# The Origin of Aromaticity: Aromatic Compounds as Intrinsic Topological Superconductors with Majorana Fermion

Kyoung Hwan Choi[1], and Dong Hack Suh[1†]

*1 Advanced Materials & Chemical Engineering Building 311, 222 Wangsimni-ro, Seongdong-Gu, Seoul, Korea, E-mail: dhsuh@hanyang.ac.kr*

**Abstract**

Topological superconductors have been discovered with recent advances in understanding the topological properties of condensed matters. These states have a full pairing gap in the bulk and gapless counter-propagating Majorana states at the boundary. A pair of Majorana zero modes is associated with each vortex. This understanding had a great influence on the theory of superconductivity and their following experiments, but its relevantce to organic compounds was not closely observed. Here, we analyze the topological states of various polyaromatic hydrocarbons (PAHs), including benzene, and reveal that they are topological superconductors. We have analyzed the momentum vectors of benzene and other PAHs through a semi-classical approach to confirm their non-trivial state. Their unique properties might be originated from the odd number of Kramers doublets in PAHs. The Hückel rule describing aromaticity can be reinterpreted with a topological viewpoint. It suggests that the (4n+2) rule can be split into two pairs of (2n+1) electrons each, namely, electrons and holes with spin up and down. Therefore, it always forms an odd number of Kramers` doublet. Moreover, n in the Hückel rule can be interpreted as the winding number in global next-nearest-neighbor(NNN) hopping. This work will re-establish the definition of aromaticity that has been known so far and extend the use of aromatic compounds as topological superconductors to quantum computers.

**Introduction**

The quantum state of matter is characterized not only by the structure of the energy spectrum but also by the nature of wave functions.[1] Of particular importance are topological properties of wave functions, i.e., properties that are invariant under small adiabatic deformation of the Hamiltonian. These properties can be specified by some topological invariant.

The quantum Hall (QH) effect[2] provides the first example of a non-trivial state of matter, where the quantized Hall conductance is a topological invariant and characterized by Chern number[3]. The quantum spin Hall (QSH) state has been theoretically predicted and experimentally observed in HgTe quantum wells.[4-7] The QSH state is characterized by a $Z_2$ topological number[8], gapless helical edge states[9,10], and winding number[11]. Through these perceptions, the $^3$He-B and the planar phase of the triplet superfluid superconductor belong to the time-reversal invariant (TRI) topological state of matter.[12,13]

Following the generalization of the QSH state, the generalization of the TRI topological superconducting state can be obtained in a straightforward manner, and has gapless Majorana surface states protected by TR symmetry.[14-16] The TRI superconductor can be classified by a $Z_2$ topological number in 2D. Furthermore, a TRI topological defect of a $Z_2$ non-trivial superconductor is accompanied by a Kramers' pair of Majorana fermions, an emergent quasi-particle with zero energy state.[17,18] Because fundamental aspects of Majoranas and their non-Abelian braiding properties[19,20] offer possible application in quantum computation[21,22], many examples of leading candidates to find Majoranas in superconductors are: the Moore-Read-type state in the fractional quantum Hall effect[23]; vortices in two-dimensional (2D) p+ip spinless superconductors[24]; and domain walls in 1D p-wave superconductors[25].

On the other hands, aromaticity is considered as one of the most important concepts in modern organic chemistry.[26] The term aromaticity/aromatic has a long history dating back to its first use in a chemical sense by Kekulé[27] and Erlenmeyer[28] in the 1860s. Similarly, as in those days, in a modern sense the term is related either to some typical properties or to a specific structure.[29] In the last few decades it has been almost generally accepted that both terms, aromaticity and aromatic character, are associated with the ground-state properties of cyclic π-electron compounds which (i) are more stable than their chain analogues, (ii) have bond lengths between those typical of the single and double ones, and (iii) have a π-electron ring current that is induced when the system is exposed to external magnetic fields.

The recent study of topological aspects of PAHs has provided a more profound understanding of a non-trivial phase of PAHs. Organic crystal with pyrene derivatives shows topological photonics through Pancharatnam-Berry and wave retardation phase merging, wavefront shaping and waveguide on edges.[30] Magnetic properties and results of transport spectroscopy of pelletized organic crystal also demonstrate the intrinsic quantized spin waves on the pyrene derivative, coherent quantum phase slip (CQPS), and Majorana hinge and corner modes which are signatures of topological superconductors.[31] However, the its origin remains unclear. Despite the the widespread use of "aromatic" and "aromaticity" in the current scientific literature, it has been defined on the empirical basis of experimental observables and quantities obtained from theoretical computations.[32] Aromatic hydrocarbons generally refer to molecules that have (4n+2) π electrons and follow the Hückel's rule, which is composed of a benzenoid ring.

Here, we propose to investigate the momentum space of aromaticity by using a semi-classical approach. The momentum vectors of up- and down-spin of electrons and holes in various aromatic hydrocarbons are analyzed by the recent theoretical suggestion about time-reversal-invariant topological superconductors.[17] In momentum space, it has been known that one pair of particles and the other pair of antiparticles must exist in zero energy state, and that all aromatic compounds have winding numbers, a topological invariant.[33] Given that, the physical meaning of the Hückel (4n+2) rule should be reinterpreted as the odd number of

Kramers doublet for (2n+1) up-spin and (2n+1) down-spin particles and antiparticles, respectively and its n as closely to do with the winding number.

**Results and discussion**

**Benzene as an exception compound to Peierls distortion**

The geometry and energy of 1,3,5-cyclohexatriene has been suggested as a reference molecule for evaluating the extra stabilization energy of benzene.[34] Among various differences between two molecules, their bond length and binding energy seem to be the key factors. 1,3,5-cyclohexatriene has three alternating sequences of double and single bonds (A-B type), while benzene has six uniform bonds (A type). Benzene could be obtained without the existence of 1,3,5-cyclohexatriene. It is very contradictory because evenly spaced chains of interacting atoms with each one contributing an electron are more unstable than those with an alternating spaced chain, as in polyacetylene.[35,36] It indicates that benzene is considered robust against Peierls distortion while 1,3,5-cyclohexatriene is not. (Fig. 1a) Although the Peierls distortion was perceived as a natural direction of stabilizing materials, the odd number of Kramers doublets resulting can make a system robust against it.[10,37]

**Momentum space and winding number of benzene**

Almost no research has been done experimentally on the superconductivity of aromatic hydrocarbons (PAHs), but some theoretical research has been done.[38-41] Recently, their various topological properties, such as Pancharatnam-Berry and wave retardation phase merging, coherent quantum phase slip, and Majorana hinge and edge modes have been reported through experimental observations.[30,31] These results allow PAHs including benzene to be considered as topological superconductors. Thus, the TRI topological superconductor model can be used for various PAHs. The BdG Hamiltonian is a block diagonal with only equal spin pairing and spin-up and -down electrons that form Cooper pairs with momentum pairing of $p_x + ip_y$ and $p_x$

- $ip_y$, respectively.[17] In this situation, the quasiparticle operators $\psi_\uparrow$ and $\psi_\downarrow$ can be expressed in terms of the eigenstates of the BdG Hamiltonian as; $\psi_\uparrow(n) = [u(n)c_\uparrow(n) + v(n)c_\uparrow^\dagger(n)]$, $\psi_\downarrow(m) = [u^*(m)c_\downarrow(m) + v^*(m)c_\downarrow^\dagger(m)]$ where n and m are odd and even carbon numbers of PAHs, respectively, u and v and $u^*$ and $v^*$ are momentum vectors of electrons and holes, respectively. Additionally, [$\psi_\uparrow(n)$, $\psi_\downarrow(m)$] transforms as a Kramers' doublet which forbids a gap in the edge state spectrum once TRS is preserved by preventing the mixing of the spin-up and spin-down modes.

To explain benzene with the above situation, the odd-numbered carbons have the up-spin fermions and the even-number carbons have the down-spin fermions. (Fig. 2a) The $u_k$ and $v_k$ of benzene can be calculated from the position vectors of each of the electrons. (See Supplementary information). For instance, the up-spin electron and hole initially positioned at C1 moves to C3 for the electron and C5 for the hole by the quasiparticle operator [$\psi_\uparrow(1)$]. Therefore, $\vec{u}(1)$ and $\vec{v}(1)$ is positioned at (-0.313, -2.315) and (-2.310, -0.896), respectively. Contrary to this, the down-spin electron paired with the up-spin electron should have the opposite sign of the y axis momentum.[17] To satisfy this, $u_\downarrow^*$ and $v_\downarrow^*$ must be matched with (-0.313, 2.315) and (-2.231, 0.896). This momentum is displayed when the electron and the hole positioned at C4 move to C2 for the electron [$\vec{u}^*(4)$] and C6 for the hole [$\vec{v}^*(4)$]. Therefore, [$\psi_\uparrow(1)$, $\psi_\downarrow(4)$] can be interpreted as a Kramers doublet. In the same way, two different quasiparticle operators make Kramers doublets: one operator pairs which move up- and down-spin electrons from C3 to C5 (C2 to C6) is [$\psi_\uparrow(3)$, $\psi_\downarrow(2)$] and the other which moves up- and down-spin electrons from C5 to C1 (C6 to C4) is [$\psi_\uparrow(5)$, $\psi_\downarrow(6)$]. Therefore, benzene can have three different Kramers doublets. It is the reason why benzene can be robust against Peierls distortion.

The relationship betweene these momentum vectors in the momentum space allows intuitive perception of the non-trivial state of benzene through the winding number. The connections among $\vec{u}(1)$, $\vec{u}(3)$ and $\vec{u}(5)$ momentum vectors can create a single closed loop containing the position (0,0). (Fig. 2b) Therefore, the spin electrons rotating clockwise have the winding number as 1. The those among $\vec{u}^*(2)$, $\vec{u}^*(4)$ and $\vec{u}^*(6)$ can so make a single closed loop containing the (0,0) position counter-clockwise (Fig. 2b). As a result, down-spin electrons has the winding number as -1. Similarly, the winding numbers up- and down-spin holes are -1 and 1, respectively. (Fig. 2c) It is equivalent to helical superconductors predicted theoretically[14,42] and its state is called "helical Majorana liquid". It reflects the topological properties of PAHs well. It comes from the boundary of $Z_2$ topological superconductors and the mass term for the edge states is forbidden by TR symmetry. Up to this point, benzene has inherently a non-trivial state, and three pairs of quasiparticle operators create an odd number of Kramers doublets, causing chiral superconducting states with the chiral Majorana edge state.

The new interpretation of benzene could be extended to the entire PAHs governed by Hückel (4n+2) rule, a symbol of aromaticity. Two concepts are used to interpret benzene as a non-trivial state: One is the up-and down-spin electrons separated by odd- and even-number carbons, respectively. The other is odd numbers of Kramer doublets that enable their NNN hopping. Therefore, the Hückel (4n+2) rule can be split into two (2n+1) electrons for up-spin and down-spin, respectively. In this case, the up- and down-spin electrons are separated and the paired quasiparticle operators have odd numbers of Kramers doublets.

**Momentum space and winding number of anthracene**

Applying the new suggestion and extending it to PAHs with linear, bent and circular types the momentum vectors and calculated in naphthalene, anthracene, pyrene, phenanthrene, chrysene, and coronene. (Tables. S2-S16) During this process, two differences can be found compared to benzene. First, all momentum vectors that appear similarly located in momentum space have slightly different values. This trend was also observed in anthracene [$\vec{u}(1)$, $\vec{u}(3)$ and $\vec{u}(5)$] and pyrene [$\vec{u}(1)$, $\vec{u}(3^b)$, $\vec{u}(13^b)$ and $\vec{u}(16^b)$], respectively. (Table. S5 and S7) It is very different from the momentum vectors in the crystals where two or three bases describe the whole system. Second, the Kramers doublet is not paired with in the same benzenoid ring. In anthracene, $\vec{u}(1)$ is paired with $\vec{u}^*(8)$, and $\vec{u}^*(3^a)$ is entangled with $\vec{u}^*(6^b)$ in pyrene. (Fig. 3a and b) It can be formed in the geometric relationship of point symmetry. Although the values of all up-and down-spin momentum vectors are slightly different, there are always momentum vectors that forms a pair of Kramers doublet in their molecules.

**Global NNN hopping for exception molecules to the Hückel rule**

PAH's topological interpretation also offers a new direction for the elusive parts of the Hückel rule. For example, pyrene is difficult to descrive with the Hückel rule. Since pyrene is made up of 16 carbons, pyrene has n as 3.5. To explain this, several qualitative models (e.g., Platt's ring perimeter model[43], Clar's model[44] and Randic's conjugated circuits model[45]) have been suggested to explain the aforementioned. The exception can be clear up by the extension of n based on the winding number of pyrene through global NNN hopping. Here, all fermions with identical spins return to their original position by hopping via the locations.

Anthracene's global NNN hopping involves seven up-spin and down-spin electrons. Therefore, the odd number of Kramers doublets was always guaranteed. On the other hand, anthracene's fermions create an Eulerian circuit because all anthracene vertices have an even degree. (Fig. 4a and b) Although momentum vectors in it are slightly different, the non-trivial state is guaranteed when fermions hop and eventually return to their initial position. Up-spin electrons rotate counter-clockwise by the same path in momentum space. Hence, the winding number is -1, while that of down-spin electrons is 1. (Fig. 4c and d)

However, for pyrene, there are eight up-and down-spin electrons. So, it can lead to Kramers doublets that is not robust to Peierls distortion. Two electrons should be excluded for robustness.

(Supplementary Fig. 53) Various methods are proposed to remove the electrons of each up-and down-spin fermion, but only in certain case, global NNN hopping can be preserved. Two odd vertices need to be an Eulerian trail and satisfy global NNN hopping. Thus, the starting momentum vector of global NNN hopping, $\vec{v}(13^a)$, is surely determined. It has a momentum vector of hole. Interestingly, it is excluded from the second run. (Fig. 5a) Although connections of momentum vectors of anthracene in momentum space make a closed-loop including the (0,0) position, a momentum vector of pyrene, $\vec{v}(13^a)$, is outside the closed loop of pyrene. It forms a half-integer winding number.[46]

In other words, the number n in the Hückel rule represents the winding number formed in the process of global NNN hopping of the electronic state. Therefore, the topological interpretation of the Hückel rule helps to understand electronic states that are completely different from the previous ones of aromaticity. In aromatic molecules, electrons exist in a static state for Hückel rule, and the electrons are persistent circulating in a dynamic state.

To probe the reinterpretation of the Hückel rule, coronene, with the half integer n, with another exception of the Hückel rule, is investigated. It has 12 up- and down-spin electrons. For global NNN hopping, one up-spin electron and one down-spin electron should be excluded to obtain the odd number of Kramers doublet. In addition, the Eulerian trail is built to connect 11 up-spin electrons and one momentum vector. It is isolated in the second rotation of global NNN hopping of up-spin electrons having the momentum vector of down-spin. (Fig. 5c) Like pyrene, the unique momentum vector forms an exotic Kramer doublet between the up-spins, giving it a half-integer winding number. (Fig. 5d) However, the most significant difference between pyrene and coronene is that the winding number is 0.5 for pyrenein the global NNN hopping excluding this momentum vector, that is 1.5 for coronene. It was confirmed that the half-integer winding number in the Hückel rule could occur for forming an Euler trail during the global NNN hopping.

**Conclusion**

In conclusion, the electronic states of aromaticity, one of the most famous concepts in chemistry, has been empirically determined by many properties without the definition. Topological phenomena of PAHs led to reinterpret the aromaticity and the Hückel rule. We further tune the concepts of TRI topological superconductors to aromatic compounds. As a result, the new suggestion for aromatics implied that the up-and down-spins must form the odd number of Kramers doublets, and the number n represents the winding numbers during the global NNN hopping of all PAHs. It is not only means that the n value expands to a half integer, but also indicates that it is a topological invariant that represents a non-trivial state. In conclusion, all aromatic compounds are thought to be intrinsic topological superconductors.

and its characterization of topological phases in one-dimensional chiral non-Hermitian systems. *Physical Review A* **97**, 052115, doi:10.1103/PhysRevA.97.052115 (2018).

**Figures.**

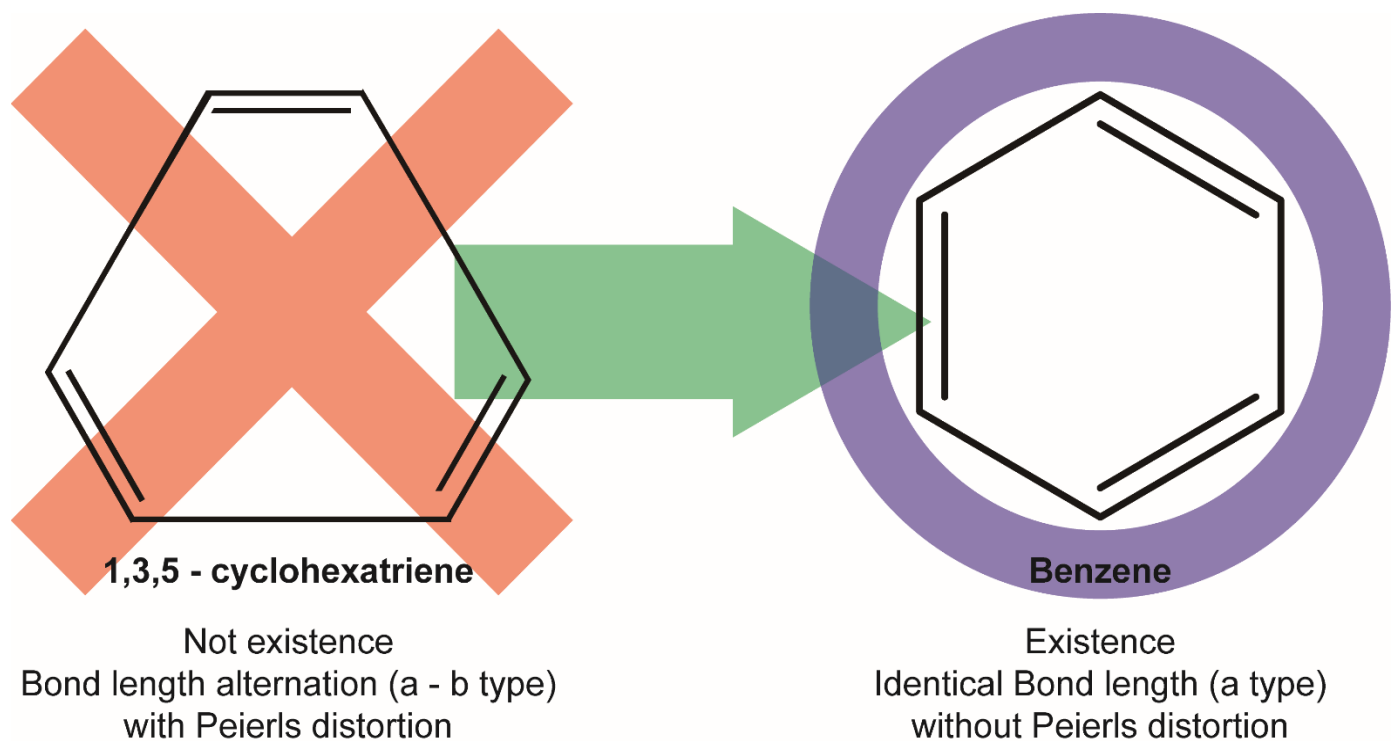

**Fig. 1. Differences between 1,3,5-cyclohexatriene and benzene.** Although Peierls distortion is a very important theory explaining the stability of molecules, benzene does not follow it. Recent topological studies show that odd number of Kramers doublet guarantees robustness of Peierls distortion.

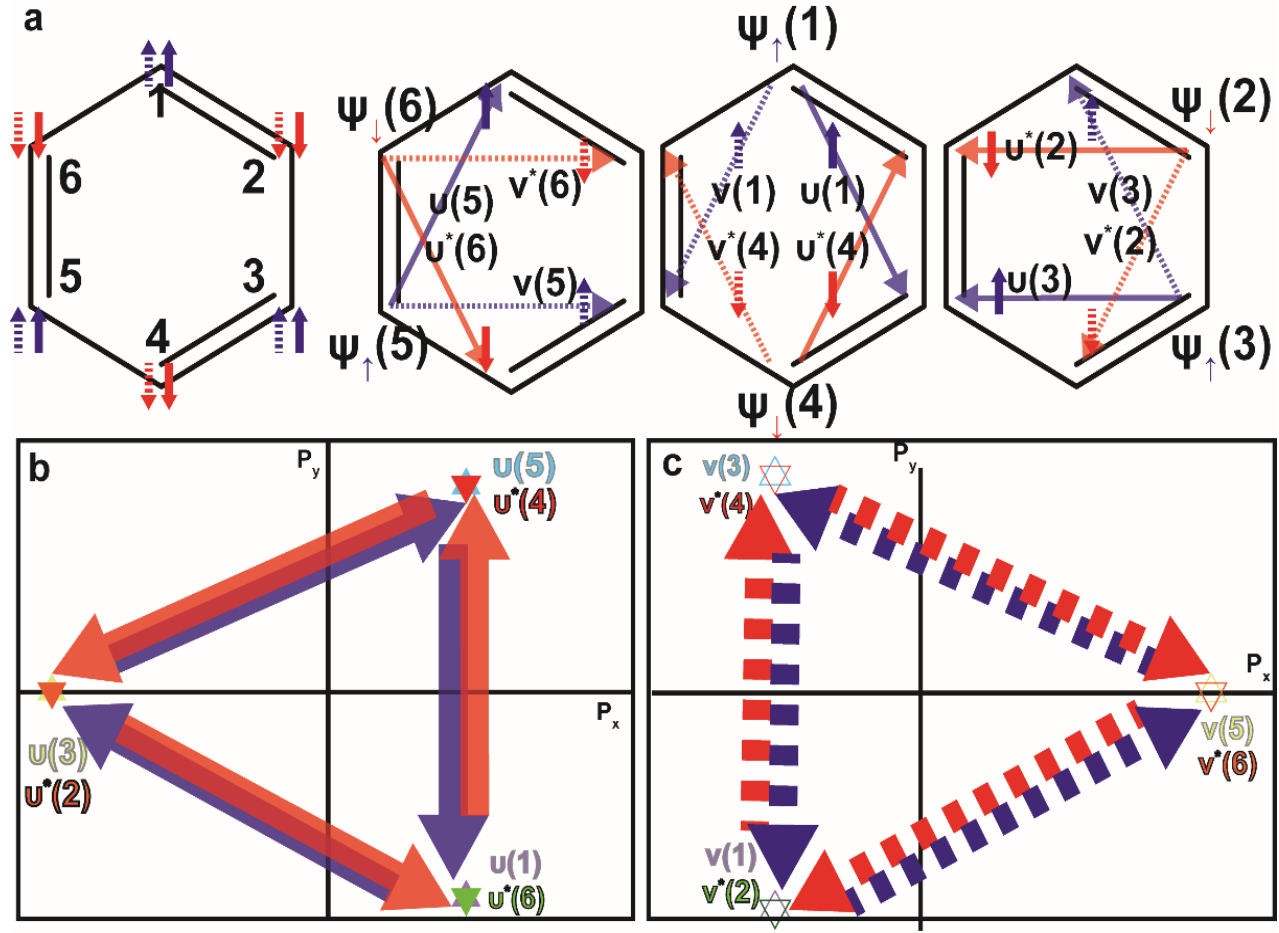

**Fig. 2. Next Nearest Neighbor(NNN) Hopping scheme and momentum space of benzene. a.** Benzene has three different momentum vectors for each particle and hole. ($\psi_\uparrow(n) = u(n)c_\uparrow(n) + u^*(n)c^\dagger_\uparrow(n)$, $\psi_\downarrow = v(m)c_\downarrow(m) + v^*(m)c^\dagger_\downarrow(m)$). ($\psi_{k1,\uparrow}$, $\psi_{-k1,\downarrow}$) consists Kramers` doublets and totally odd number of Kramers doublets exists in benzene. Therefore, benzene is robust against Peierls distortion. **b and c.** Momentum space for electron and hole. Up-spin and down-spin particles have chirality and up-spin particles and antiparticles also have chirality like quantum spin Hall model. It shows NNN hopping of benzene has a non-trivial phase.

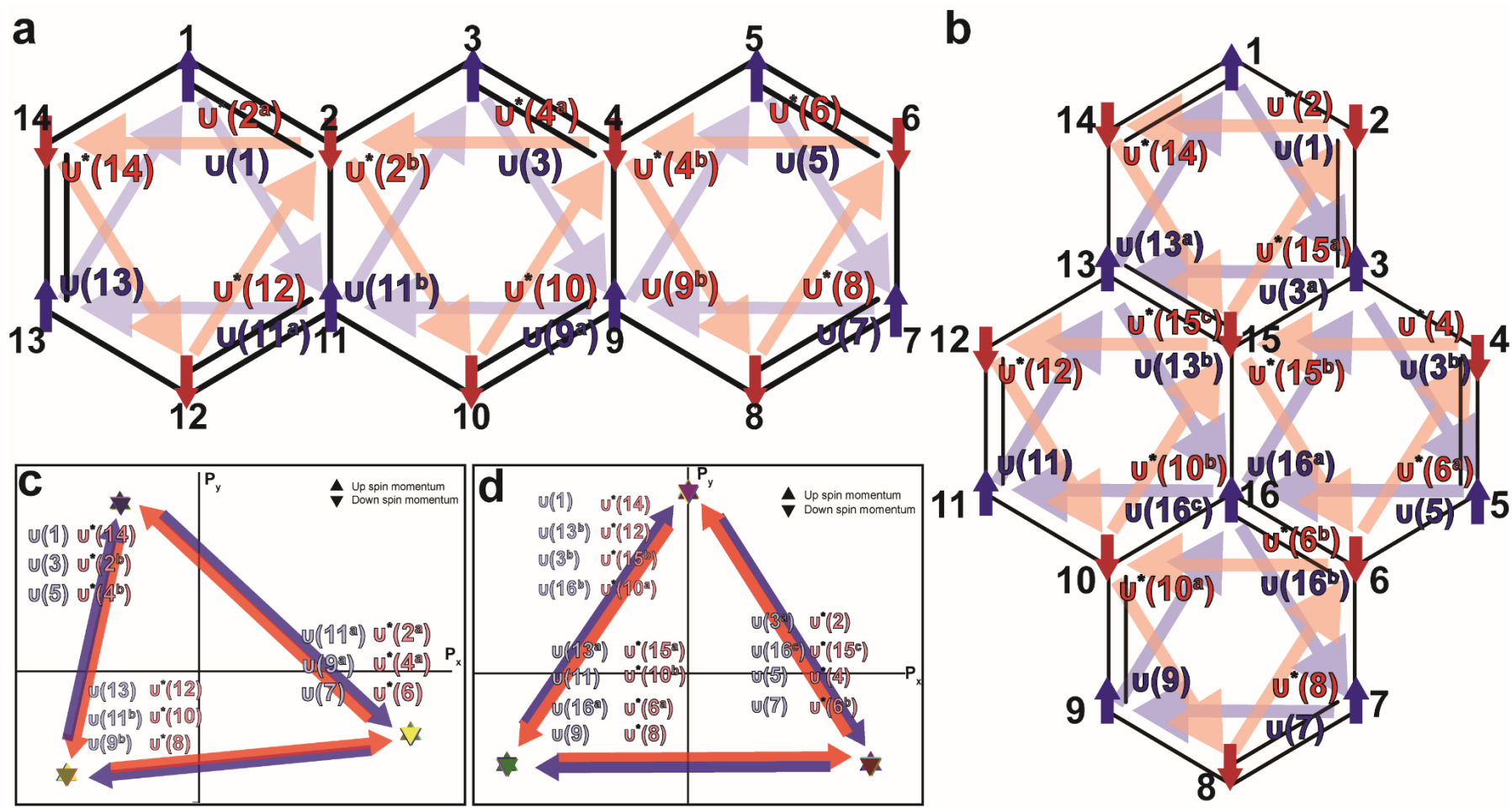

**Fig. 3. Non-trivial state for anthracene and pyrene. a and b**. NNN hopping scheme for anthracene(a) and pyrene(b). **c. and d.** Momentum space of anthracene(c) and pyrene(d).

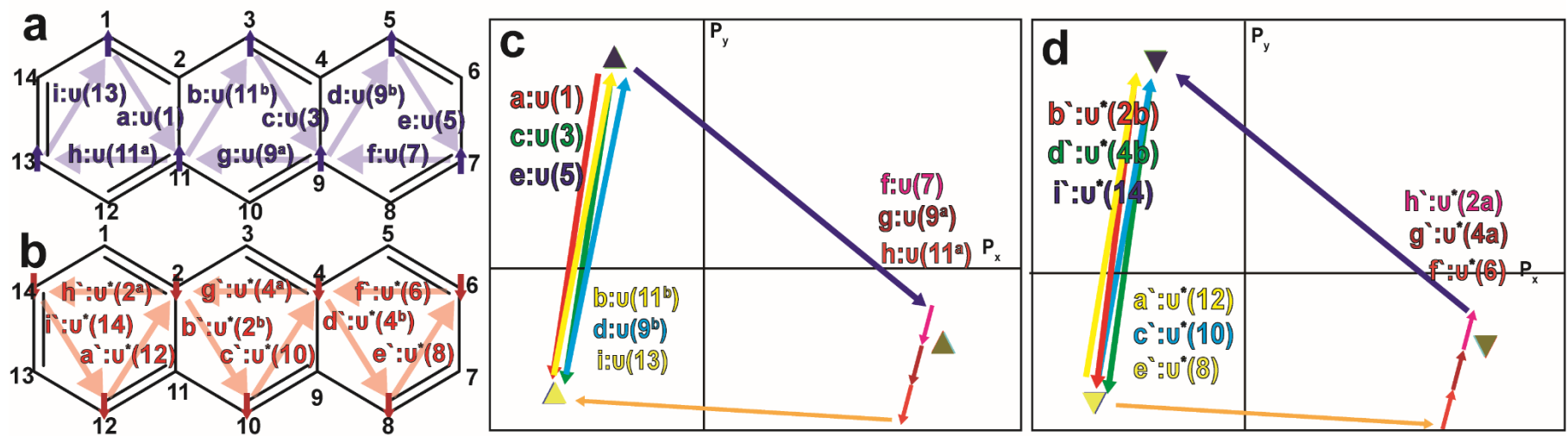

**Fig. 4. Scheme and momentum space of Global NNN hopping for anthracene**. **a and b.** Scheme of NNN hopping model of up-spin(a) and down-spin(b) for anthracene. **c. and d.** Momentum vectors of up-spin(c) and down-spin(d) of anthracene for global NNN hopping

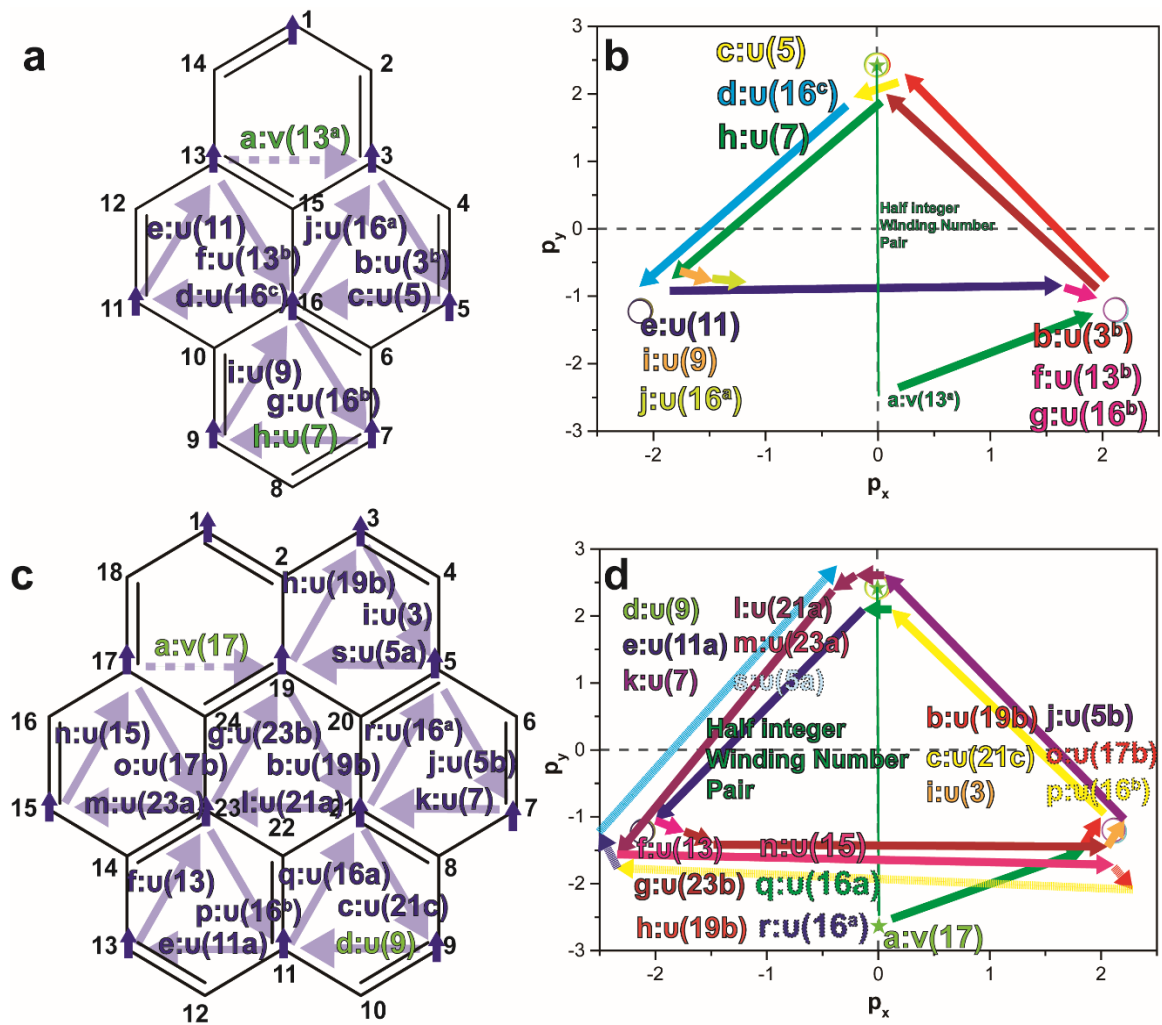

**Fig. 5. Scheme and momentum space of Global NNN hopping for pyrene and coronene. a** Scheme of NNN hopping model of up-spin for pyrene. **b.** Momentum vectors of up-spin of pyrene for global NNN hopping **c.** Scheme of NNN hopping model of up-spin for coronene **d.** Momentum vectors of up-spin of coronene for global NNN hopping

**Methods**

All molecules can be either drawn in Chemdraw. The Chemdraw drawing needs to be checked for stereochemical accuracy and the file is saved as a CDX file. The file is either opened in Chem3D or the structure is simply copied and pasted in Chem3D side window. The most stable structure of all molecules can be obtained by using a MM2 force field method which is optimized structure with little steric interactions. The carbon coordinates were then used as the position vector. A detail method for calculation of bond characteristics and modified bond dissociation energy in PAHs are discussed in supplementary information.

**Supplementary information** is available in the online version of the paper

**Acknowledgements**

This research was supported by Basic Science Research Program through the National Research Foundation of Korea(NRF) funded by the Ministry of Education (2018R1D1A1A02047853).

**Author contribution**

Dong Hack Suh designed, initiated and directed this research. Kyoung Hwan Choi performed the experiment. Dong Hack Suh and Kyoung Hwan Choi discussed the experiment, the data, the results and co-wrote the manuscript.

**Competing interest.** The authors declare no competing interests.

**Supplementary information** is available for this paper

**Reprints and permissions information is available**

Supplementary information for

The origin of aromaticity: Aromatic compound as the Intrinsic Topological Superconductor with a Majorana Fermion.

**Supplementary Note 1**: According to the definition of momentum (P=mv), the difference between the two position vectors has the same physical meaning as the momentum vector when the mass is constant. Therefore, momentum vectors of each PAHs are calculated from their position vector. For instance, benzene has six carbons, and each has their position (Supplementary table 2) and momentum vector for NNN hopping can be calculated (Supplementary table 3).

| Carbon number | X position | Y position |
|---|---|---|
| 1 | 0.0000 | 1.3990 |
| 2 | 1.2116 | 0.6995 |
| 3 | 1.2116 | -0.6995 |
| 4 | 0.0000 | -1.3990 |
| 5 | -1.2116 | -0.6995 |
| 6 | -1.2116 | 0.6995 |

Table S1. Position vector of carbons in benzene

| Particle/antiparticle | Up / down spin | Momentum vector | x | y | z |
|---|---|---|---|---|---|
| Electron | Up spin | u(1) | 1.2116 | -2.0985 | 0 |
| | | u(3) | -2.4231 | 0.0000 | 0 |
| | | u(5) | 1.2116 | 2.0985 | 0 |
| | Down spin | u*(2) | -2.4231 | 0.0000 | 0 |
| | | u*(4) | 1.2116 | 2.0985 | 0 |
| | | u*(6) | 1.2116 | -2.0985 | 0 |
| Hole | Up spin | v(1) | -1.2116 | -2.0985 | 0 |
| | | v(3) | -1.2116 | 2.0985 | 0 |
| | | v(5) | 2.4231 | 0.0000 | 0 |
| | Down spin | v*(2) | -1.2116 | -2.0985 | 0 |
| | | v*(4) | -1.2116 | 2.0985 | 0 |
| | | v*(6) | 2.4231 | 0.0000 | 0 |

Table S2. Momentum vectors of fermions in benzene

| Carbon number | X position | Y position | Z position |
|---|---|---|---|
| 1 | 2.513 | 0.304 | -1.24 |
| 2 | 1.258 | 0.159 | -0.626 |
| 3 | 0.194 | 1.03 | -0.918 |
| 4 | -1.061 | 0.884 | -0.304 |
| 5 | -2.125 | 1.754 | -0.596 |
| 6 | -3.366 | 1.597 | 0.023 |
| 7 | -3.56 | 0.568 | 0.939 |
| 8 | -2.513 | -0.304 | 1.24 |
| 9 | -1.258 | -0.159 | 0.626 |
| 10 | -0.194 | -1.03 | 0.918 |
| 11 | 1.61 | -0.884 | 0.304 |
| 12 | 2.124 | -1.754 | 0.596 |
| 13 | 3.366 | -1.597 | -0.023 |
| 14 | 3.56 | -0.569 | -0.94 |

**Table S3. Position vector of carbons in anthracene**

| Particle/antiparticle | Up / down spin | Momentum vector | x | y | z |
|---|---|---|---|---|---|
| Electron | Up spin | u(1) | -1.452 | -1.188 | 1.544 |
| | | u(11a) | 2.305 | -0.713 | -0.327 |
| | | u(13) | -0.853 | 1.901 | -1.217 |
| | | u(3) | -1.452 | -1.189 | 1.544 |
| | | u(9a) | 2.319 | -0.725 | -0.322 |
| | | u(11b) | -0.867 | 1.914 | -1.222 |
| | | u(5) | -1.435 | -1.186 | 1.535 |
| | | u(7) | 2.302 | -0.727 | -0.313 |
| | | u(9b) | -0.867 | 1.913 | -1.222 |
| | Down spin | u*(2a) | 2.302 | -0.728 | -0.314 |
| | | u*(14) | -1.436 | -1.185 | 1.536 |
| | | u*(12) | -0.866 | 1.913 | -1.222 |
| | | u*(4a) | 2.319 | -0.725 | -0.322 |
| | | u*(2b) | -1.452 | -1.189 | 1.544 |
| | | u*(10) | -0.867 | 1.914 | -1.222 |
| | | u*(6) | 2.305 | -0.713 | -0.327 |
| | | u*(4b) | -1.452 | -1.188 | 1.544 |
| | | u*(8) | -0.853 | 1.901 | -1.217 |
| Hole | Up spin | v(1) | 0.853 | -1.901 | 1.217 |
| | | v(13) | -2.305 | 0.713 | 0.327 |
| | | v(11a) | 1.452 | 1.188 | -1.544 |
| | | v(3) | 0.867 | -1.914 | 1.222 |
| | | v(11b) | -2.319 | 0.725 | 0.322 |
| | | v(9a) | 1.452 | 1.189 | -1.544 |
| | | v(5) | 0.867 | -1.913 | 1.222 |
| | | v(9b) | -2.302 | 0.727 | 0.313 |
| | | v(7) | 1.435 | 1.186 | -1.535 |
| | Down spin | v*(14) | -2.302 | 0.728 | 0.314 |
| | | v*(2a) | 0.866 | -1.913 | 1.222 |
| | | v*(12) | 1.436 | 1.185 | -1.536 |
| | | v*(2b) | -2.319 | 0.725 | 0.322 |
| | | v*(4a) | 0.867 | -1.914 | 1.222 |
| | | v*(10) | 1.452 | 1.189 | -1.544 |
| | | v*(4b) | -2.305 | 0.713 | 0.327 |
| | | v*(6) | 0.853 | -1.901 | 1.217 |
| | | v*(8) | 1.452 | 1.188 | -1.544 |

**Table S4. Momentum vectors of fermions in anthracene**

| Carbon number | X position | Y position | Z position |
| --- | --- | --- | --- |
| 1 | -3.513 | -0.007 | -0.106 |
| 2 | -2.814 | -1.214 | -0.1 |
| 3 | -1.412 | -1.228 | -0.058 |
| 4 | -0.693 | -2.43 | -0.051 |
| 5 | 0.704 | -2.427 | -0.009 |
| 6 | 1.418 | -1.222 | 0.028 |
| 7 | 2.82 | -1.202 | 0.07 |
| 8 | 3.513 | 0.007 | 0.106 |
| 9 | 2.814 | 1.214 | 0.1 |
| 10 | 1.412 | 1.228 | 0.058 |
| 11 | 0.693 | 2.43 | 0.051 |
| 12 | -0.704 | 2.428 | 0.009 |
| 13 | -1.418 | 1.222 | -0.028 |
| 14 | -2.82 | 1.203 | -0.07 |
| 15 | -0.708 | -0.001 | -0.021 |
| 16 | 0.708 | 0.001 | 0.021 |

**Table S5. Position vector of carbons in pyrene**

| Particle/antiparticle | Up / down spin | Momentum vector | x | y | z |
|---|---|---|---|---|---|
| Electron | Up spin | u(1) | 2.101 | -1.221 | 0.048 |
| | | u(3a) | -0.006 | 2.45 | 0.03 |
| | | u(13a) | -2.095 | -1.229 | -0.078 |
| | | u(13b) | 2.126 | -1.221 | 0.049 |
| | | u(16c) | -0.015 | 2.429 | 0.03 |
| | | u(11) | -2.111 | -1.208 | -0.079 |
| | | u(3b) | 2.116 | -1.199 | 0.049 |
| | | u(5) | 0.004 | 2.428 | 0.03 |
| | | u(16a) | -2.12 | -1.229 | -0.079 |
| | | u(16b) | 2.112 | -1.203 | 0.049 |
| | | u(7) | -0.006 | 2.416 | 0.03 |
| | | u(9) | -2.106 | -1.213 | -0.079 |
| | Down spin | u*(2) | -0.006 | 2.417 | 0.03 |
| | | u*(14) | 2.112 | -1.204 | 0.049 |
| | | u*(15a) | -2.106 | -1.213 | -0.079 |
| | | u*(15c) | 0.004 | 2.429 | 0.03 |
| | | u*(12) | 2.116 | -1.2 | 0.049 |
| | | u*(10b) | -2.12 | -1.229 | -0.079 |
| | | u*(4) | -0.015 | 2.429 | 0.03 |
| | | u*(15b) | 2.126 | -1.221 | 0.049 |
| | | u*(6a) | -2.111 | -1.208 | -0.079 |
| | | u*(6b) | -0.006 | 2.45 | 0.03 |
| | | u*(10a) | 2.101 | -1.221 | 0.048 |
| | | u*(8) | -2.095 | -1.229 | -0.078 |

**Table S6. Momentum vectors of particles in pyrene**

| Particle/antiparticle | Up / down spin | Momentum vector | x | y | z |
|---|---|---|---|---|---|
| Hole | Up spin | v(1) | 2.095 | 1.229 | 0.078 |
| | | v(13a) | 0.006 | -2.45 | -0.03 |
| | | v(3a) | -2.101 | 1.221 | -0.048 |
| | | v(13b) | 2.111 | 1.208 | 0.079 |
| | | v(11) | 0.015 | -2.429 | -0.03 |
| | | v(16a) | -2.126 | 1.221 | -0.049 |
| | | v(3b) | 2.12 | 1.229 | 0.079 |
| | | v(16b) | -0.004 | -2.428 | -0.03 |
| | | v(5) | -2.116 | 1.199 | -0.049 |
| | | v(16c) | 2.106 | 1.213 | 0.079 |
| | | v(9) | 0.006 | -2.416 | -0.03 |
| | | v(7) | -2.112 | 1.203 | -0.049 |
| | Down spin | v*(2) | 0.006 | -2.417 | -0.03 |
| | | v*(15a) | 2.106 | 1.213 | 0.079 |
| | | v*(14) | -2.112 | 1.204 | -0.049 |
| | | v*(12) | -0.004 | -2.429 | -0.03 |
| | | v*(15b) | 2.12 | 1.229 | 0.079 |
| | | v*(10a) | -2.116 | 1.2 | -0.049 |
| | | v*(15c) | 0.015 | -2.429 | -0.03 |
| | | v*(4) | 2.111 | 1.208 | 0.079 |
| | | v*(6a) | -2.126 | 1.221 | -0.049 |
| | | v*(10b) | 0.006 | -2.45 | -0.03 |
| | | v*(6b) | 2.095 | 1.229 | 0.078 |
| | | v*(8) | -2.101 | 1.221 | -0.048 |

**Table S7. Momentum vectors of antiparticles in pyrene**

| Carbon number | X position | Y position | Z position |
|---|---|---|---|
| 1 | 3.575 | -0.022 | 0.005 |
| 2 | 2.75 | 1.099 | 0.001 |
| 3 | 1.351 | 0.971 | 0 |
| 4 | 0.556 | 2.12 | -0.004 |
| 5 | -0.833 | 2.027 | -0.006 |
| 6 | -1.468 | 0.782 | -0.004 |
| 7 | -2.872 | 0.723 | -0.006 |
| 8 | -3.541 | -0.498 | -0.004 |
| 9 | -2.81 | -1.677 | 0 |
| 10 | -1.413 | -1.631 | 0.002 |
| 11 | -0.695 | -0.408 | 0 |
| 12 | 0.743 | -0.311 | 0.002 |
| 13 | 1.618 | -1.428 | 0.006 |
| 14 | 3.009 | -1.287 | 0.008 |

**Table S8. Position vector of carbons in phenanthrene**

| Particle/antiparticle | Up / down spin | Momentum vector | x | y | z |
|---|---|---|---|---|---|
| Electron | Up spin | u(1) | -2.224 | 0.993 | -0.005 |
| | | u(3a) | 0.267 | -2.399 | 0.006 |
| | | u(13) | 1.957 | 1.406 | -0.001 |
| | | u(3b) | -2.184 | 1.056 | -0.006 |
| | | u(5) | 0.138 | -2.435 | 0.006 |
| | | u(11a) | 2.046 | 1.379 | 0 |
| | | u(11b) | -2.177 | 1.131 | -0.006 |
| | | u(7) | 0.062 | -2.4 | 0.006 |
| | | u(9) | 2.115 | 1.269 | 0 |
| | Down spin | u*(2) | 0.259 | -2.386 | 0.007 |
| | | u*(14) | -2.266 | 0.976 | -0.006 |
| | | u*(12a) | 2.007 | 1.41 | -0.001 |
| | | u*(4) | 0.187 | -2.431 | 0.006 |
| | | u*(12b) | -2.211 | 1.093 | -0.006 |
| | | u*(6a) | 2.024 | 1.338 | 0 |
| | | u*(6b) | 0.055 | -2.413 | 0.006 |
| | | u*(10) | -2.128 | 1.133 | -0.006 |
| | | u*(8) | 2.073 | 1.28 | 0 |
| Hole | Up spin | v(1) | -1.957 | -1.406 | 0.001 |
| | | v(13) | -0.267 | 2.399 | -0.006 |
| | | v(3a) | 2.224 | -0.993 | 0.005 |
| | | v(3b) | -2.046 | -1.379 | 0 |
| | | v(11a) | -0.138 | 2.435 | -0.006 |
| | | v(5) | 2.184 | -1.056 | 0.006 |
| | | v(11b) | -2.115 | -1.269 | 0 |
| | | v(9) | -0.062 | 2.4 | -0.006 |
| | | v(7) | 2.177 | -1.131 | 0.006 |
| | Down spin | v*(14) | -0.259 | 2.386 | -0.007 |
| | | v*(2) | -2.007 | -1.41 | 0.001 |
| | | v*(12a) | 2.266 | -0.976 | 0.006 |
| | | v*(12b) | -0.187 | 2.431 | -0.006 |
| | | v*(4) | -2.024 | -1.338 | 0 |
| | | v*(6) | 2.211 | -1.093 | 0.006 |
| | | v*(10) | -0.055 | 2.413 | -0.006 |
| | | v*(6) | -2.073 | -1.28 | 0 |
| | | v*(8) | 2.128 | -1.133 | 0.006 |

**Table S9. Momentum vectors of fermions in phenanthrene**

| Carbon number | X position | Y position | Z position |
|---|---|---|---|
| 1 | 3.921 | 0.58 | 0.032 |
| 2 | 2.515 | 0.566 | 0.04 |
| 3 | 1.822 | 1.765 | 0.168 |
| 4 | 0.429 | 1.787 | 0.18 |
| 5 | -0.35 | 0.617 | 0.065 |
| 6 | -1.797 | 0.647 | 0.078 |
| 7 | -2.581 | 1.826 | 0.204 |
| 8 | -3.98 | 1.803 | 0.211 |
| 9 | -4.651 | 0.597 | 0.093 |
| 10 | -3.921 | -0.58 | -0.032 |
| 11 | -2.515 | -0.566 | -0.041 |
| 12 | -1.823 | -1.766 | -0.168 |
| 13 | -0.429 | -1.787 | -0.18 |
| 14 | 0.35 | -0.617 | -0.065 |
| 15 | 1.797 | -0.647 | -0.078 |
| 16 | 2.581 | -1.826 | -0.204 |
| 17 | 3.98 | -1.803 | -0.211 |
| 18 | 4.651 | -0.597 | -0.093 |

**Table S1. Position vector of carbons in chrysene**

| Particle/antiparticle | Up / down spin | Momentum vector | x | y | z |
|---|---|---|---|---|---|
| Electron | Up spin | u(1) | -2.124 | -1.227 | -0.11 |
| | | u(15a) | 2.183 | -1.156 | -0.133 |
| | | u(17) | -0.059 | 2.383 | 0.243 |
| | | u(3) | -2.172 | -1.148 | -0.103 |
| | | u(5a) | 2.147 | -1.264 | -0.143 |
| | | u(15b) | 0.025 | 2.412 | 0.246 |
| | | u(5b) | -2.165 | -1.183 | -0.106 |
| | | u(11a) | 2.086 | -1.221 | -0.139 |
| | | u(13) | 0.079 | 2.404 | 0.245 |
| | | u(7) | -2.07 | -1.229 | -0.111 |
| | | u(9) | 2.136 | -1.163 | -0.134 |
| | | u(11b) | -0.066 | 2.392 | 0.245 |
| | Down spin | u*(2a) | 2.136 | -1.163 | -0.133 |
| | | u*(18) | -2.07 | -1.229 | -0.111 |
| | | u*(16) | -0.066 | 2.392 | 0.244 |
| | | u*(4) | 2.086 | -1.221 | -0.14 |
| | | u*(2b) | -2.165 | -1.183 | -0.105 |
| | | u*(14a) | 0.079 | 2.404 | 0.245 |
| | | u*(6a) | 2.147 | -1.264 | -0.143 |
| | | u*(14b) | -2.173 | -1.149 | -0.103 |
| | | u*(12) | 0.026 | 2.413 | 0.246 |
| | | u*(8) | 2.183 | -1.156 | -0.133 |
| | | u*(6b) | -2.124 | -1.227 | -0.11 |
| | | u*(10) | -0.059 | 2.383 | 0.243 |

**Table S2. Momentum vectors of particles in chrysene**

| Particle/antiparticle | Up / down spin | Momentum vector | x | y | z |
|---|---|---|---|---|---|
| Hole | Up spin | v(1) | 0.059 | -2.383 | -0.243 |
| | | v(17) | -2.183 | 1.156 | 0.133 |
| | | v(15a) | 2.124 | 1.227 | 0.11 |
| | | v(3) | -0.025 | -2.412 | -0.246 |
| | | v(15b) | -2.147 | 1.264 | 0.143 |
| | | v(5a) | 2.172 | 1.148 | 0.103 |
| | | v(5b) | -0.079 | -2.404 | -0.245 |
| | | v(13) | -2.086 | 1.221 | 0.139 |
| | | v(11a) | 2.165 | 1.183 | 0.106 |
| | | v(7) | 0.066 | -2.392 | -0.245 |
| | | v(11b) | -2.136 | 1.163 | 0.134 |
| | | v(9) | 2.07 | 1.229 | 0.111 |
| | Down spin | v*(18) | -2.136 | 1.163 | 0.133 |
| | | v*(2a) | 0.066 | -2.392 | -0.244 |
| | | v*(16) | 2.07 | 1.229 | 0.111 |
| | | v*(2b) | -2.086 | 1.221 | 0.14 |
| | | v*(4) | -0.079 | -2.404 | -0.245 |
| | | v*(14a) | 2.165 | 1.183 | 0.105 |
| | | v*(14b) | -2.147 | 1.264 | 0.143 |
| | | v*(6a) | -0.026 | -2.413 | -0.246 |
| | | v*(12) | 2.173 | 1.149 | 0.103 |
| | | v*(6b) | -2.183 | 1.156 | 0.133 |
| | | v*(8) | 0.059 | -2.383 | -0.243 |
| | | v*(10) | 2.124 | 1.227 | 0.11 |

**Table S3. Momentum vectors of antiparticles in chrysene**

| Carbon number | X position | Y position | Z position |
|---|---|---|---|
| 1  | 0.686  | -3.685 | 0.015  |
| 2  | 1.416  | -2.453 | 0.008  |
| 3  | 2.848  | -2.437 | 0.007  |
| 4  | 3.534  | -1.249 | 0.002  |
| 5  | 2.833  | 0      | -0.004 |
| 6  | 3.534  | 1.247  | -0.006 |
| 7  | 2.849  | 2.436  | -0.012 |
| 8  | 1.417  | 2.453  | -0.018 |
| 9  | 0.686  | 3.684  | -0.021 |
| 10 | -0.686 | 3.685  | -0.024 |
| 11 | -1.417 | 2.453  | -0.021 |
| 12 | -2.849 | 2.436  | -0.018 |
| 13 | -3.534 | 1.248  | -0.008 |
| 14 | -2.833 | -0.001 | -0.003 |
| 15 | -3.534 | -1.249 | 0.008  |
| 16 | -2.848 | -2.437 | 0.017  |
| 17 | -1.416 | -2.454 | 0.013  |
| 18 | -0.685 | -3.685 | 0.019  |
| 19 | 0.716  | -1.24  | -0.001 |
| 20 | 1.432  | 0      | -0.008 |
| 21 | 0.716  | 1.24   | -0.016 |
| 22 | -0.716 | 1.24   | -0.017 |
| 23 | -1.432 | 0      | -0.008 |
| 24 | -0.716 | -1.24  | 0.001  |

**Table S4. Position vector of carbons in coronene**

| Particle/antiparticle | Up / down spin | Momentum vector | x | y | z |
|---|---|---|---|---|---|
| Electron | Up spin | u(1) | 0.03 | 2.445 | -0.016 |
| | | u(19a) | -2.132 | -1.214 | 0.014 |
| | | u(17a) | 2.102 | -1.231 | 0.002 |
| | | u(3) | -0.015 | 2.437 | -0.011 |
| | | u(5a) | -2.117 | -1.24 | 0.003 |
| | | u(19b) | 2.132 | -1.197 | 0.008 |
| | | u(7b) | -0.016 | 2.454 | -0.021 |
| | | u(23a) | -2.102 | -1.249 | 0.016 |
| | | u(15) | 2.118 | -1.205 | 0.005 |
| | | u(19c) | 0 | 2.48 | -0.015 |
| | | u(21a) | -2.148 | -1.24 | 0.008 |
| | | u(23b) | 2.148 | -1.24 | 0.007 |
| | | u(5b) | 0.016 | 2.436 | -0.008 |
| | | u(7) | -2.133 | -1.196 | -0.004 |
| | | u(21b) | 2.117 | -1.24 | 0.012 |
| | | u(23c) | 0.015 | 2.453 | -0.013 |
| | | u(11a) | -2.117 | -1.205 | 0.013 |
| | | u(13) | 2.102 | -1.248 | 0 |
| | | u(21c) | -0.03 | 2.444 | -0.005 |
| | | u(9) | -2.103 | -1.231 | 0 |
| | | u(11b) | 2.133 | -1.213 | 0.005 |
| | Down spin | u*(2a) | -2.101 | -1.232 | 0.011 |
| | | u*(18) | -0.031 | 2.445 | -0.018 |
| | | u*(24a) | 2.132 | -1.213 | 0.007 |
| | | u*(4) | -2.118 | -1.204 | 0.006 |
| | | u*(2b) | 0.016 | 2.453 | -0.016 |
| | | u*(20a) | 2.102 | -1.249 | 0.01 |
| | | u*(24b) | -2.132 | -1.197 | 0.016 |
| | | u*(16) | 0.015 | 2.436 | -0.02 |
| | | u*(14a) | 2.117 | -1.239 | 0.004 |
| | | u*(20b) | -2.148 | -1.24 | 0.009 |
| | | u*(24c) | 0 | 2.48 | -0.018 |
| | | u*(22a) | 2.148 | -1.24 | 0.009 |
| | | u*(6) | -2.102 | -1.247 | -0.002 |
| | | u*(20c) | -0.015 | 2.453 | -0.01 |
| | | u*(8a) | 2.117 | -1.206 | 0.012 |
| | | u*(22b) | -2.117 | -1.241 | 0.014 |
| | | u*(14b) | -0.016 | 2.437 | -0.015 |
| | | u*(12) | 2.133 | -1.196 | 0.001 |
| | | u*(8b) | -2.133 | -1.213 | 0.001 |
| | | u*(22c) | 0.03 | 2.445 | -0.007 |
| | | u*(10) | 2.103 | -1.232 | 0.006 |

**Table S5. Momentum vectors of particles in coronene**

| Particle/antiparticle | Up / down spin | Momentum vector | x | y | z |
|---|---|---|---|---|---|
| Hole | Up spin | v(1) | -2.102 | 1.231 | -0.002 |
| | | v(17) | 2.132 | 1.214 | -0.014 |
| | | v(19a) | -0.03 | -2.445 | 0.016 |
| | | v(3) | -2.132 | 1.197 | -0.008 |
| | | v(19b) | 2.117 | 1.24 | -0.003 |
| | | v(5a) | 0.015 | -2.437 | 0.011 |
| | | v(17b) | -2.118 | 1.205 | -0.005 |
| | | v(15) | 2.102 | 1.249 | -0.016 |
| | | v(23a) | 0.016 | -2.454 | 0.021 |
| | | v(19c) | -2.148 | 1.24 | -0.007 |
| | | v(23b) | 2.148 | 1.24 | -0.008 |
| | | v(21a) | 0 | -2.48 | 0.015 |
| | | v(5b) | -2.117 | 1.24 | -0.012 |
| | | v(21b) | 2.133 | 1.196 | 0.004 |
| | | v(7) | -0.016 | -2.436 | 0.008 |
| | | v(23c) | -2.102 | 1.248 | 0 |
| | | v(13) | 2.117 | 1.205 | -0.013 |
| | | v(11a) | -0.015 | -2.453 | 0.013 |
| | | v(121c) | -2.133 | 1.213 | -0.005 |
| | | v(11b) | 2.103 | 1.231 | 0 |
| | | v(9) | 0.03 | -2.444 | 0.005 |
| | Down spin | v*(18) | 2.101 | 1.232 | -0.011 |
| | | v*(2a) | -2.132 | 1.213 | -0.007 |
| | | v*(24a) | 0.031 | -2.445 | 0.018 |
| | | v*(2b) | 2.118 | 1.204 | -0.006 |
| | | v*(4) | -2.102 | 1.249 | -0.01 |
| | | v*(20a) | -0.016 | -2.453 | 0.016 |
| | | v*(16) | 2.132 | 1.197 | -0.016 |
| | | v*(24b) | -2.117 | 1.239 | -0.004 |
| | | v*(14a) | -0.015 | -2.436 | 0.02 |
| | | v*(24c) | 2.148 | 1.24 | -0.009 |
| | | v*(20b) | -2.148 | 1.24 | -0.009 |
| | | v*(22a) | 0 | -2.48 | 0.018 |
| | | v*(20c) | 2.102 | 1.247 | 0.002 |
| | | v*(6) | -2.117 | 1.206 | -0.012 |
| | | v*(8a) | 0.015 | -2.453 | 0.01 |
| | | v*(14b) | 2.117 | 1.241 | -0.014 |
| | | v*(22b) | -2.133 | 1.196 | -0.001 |
| | | v*(12) | 0.016 | -2.437 | 0.015 |
| | | v*(22c) | 2.133 | 1.213 | -0.001 |
| | | v*(8) | -2.103 | 1.232 | -0.006 |
| | | v*(10) | -0.03 | -2.445 | 0.007 |

**Table S6. Momentum vectors of antiparticles in coronene**

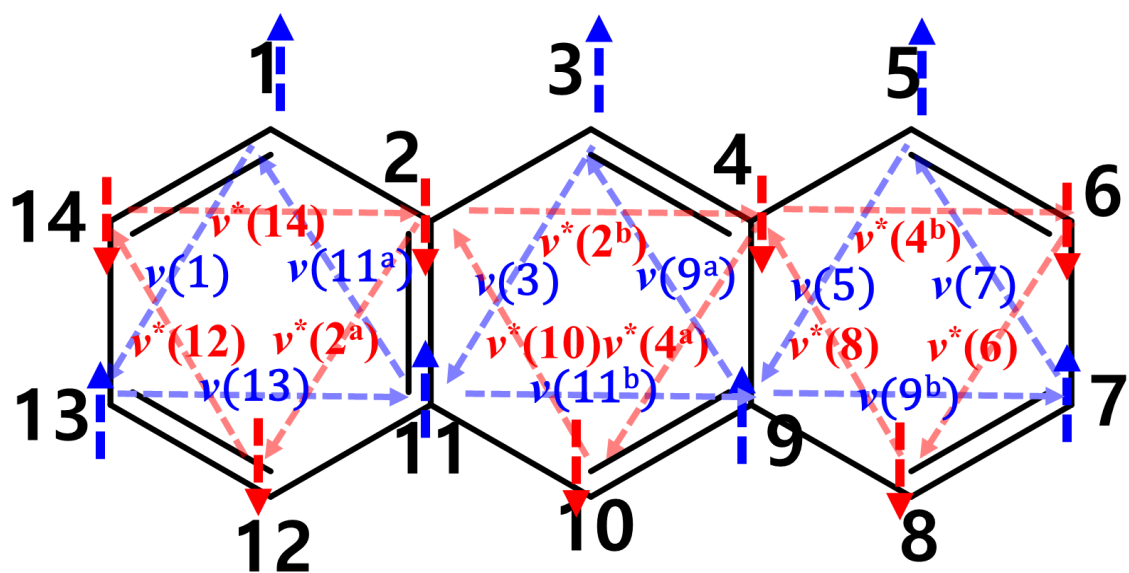

**Figure S1. Momentum vectors of antiparticle fermions in anthracene**

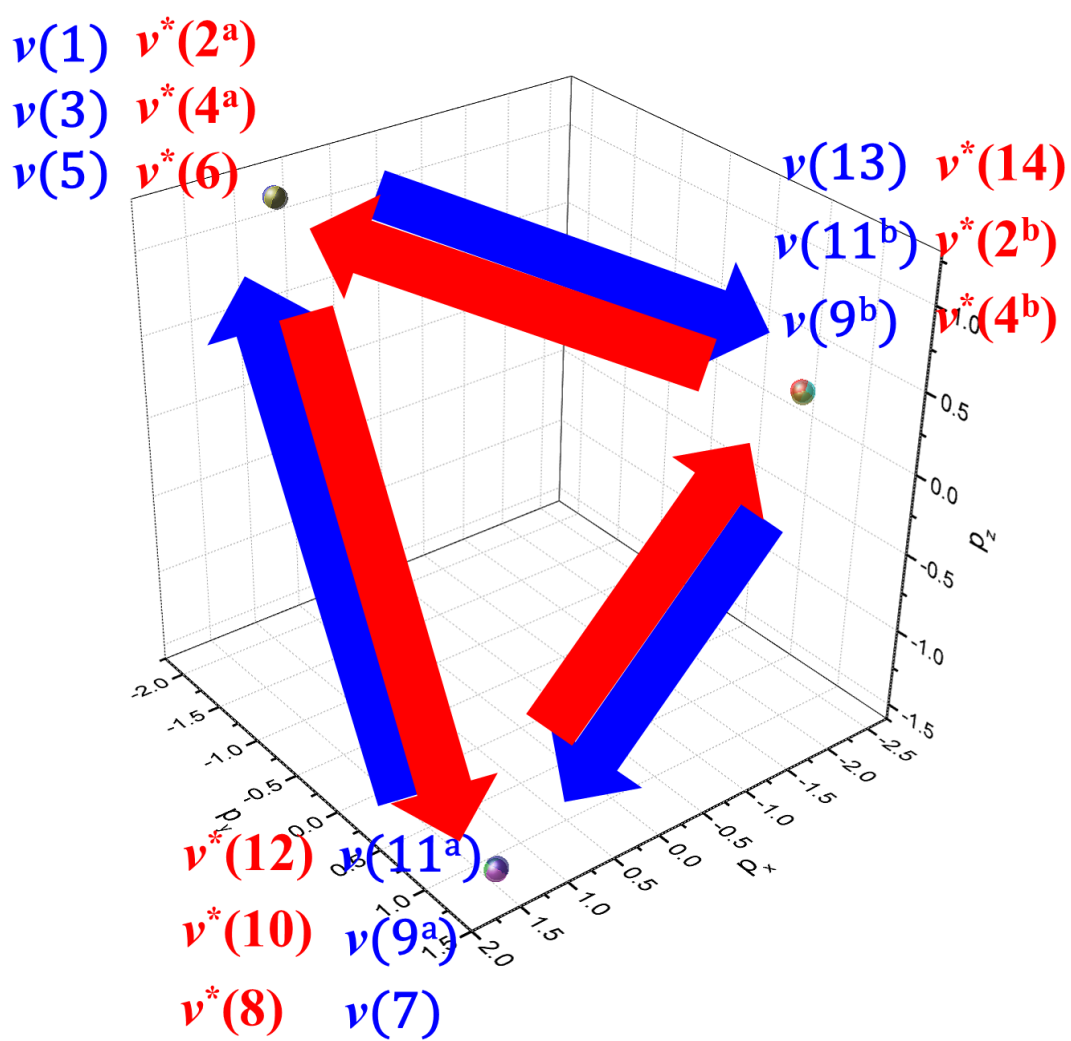

**Figure S2. Momentum space of antiparticle fermions in anthracene**

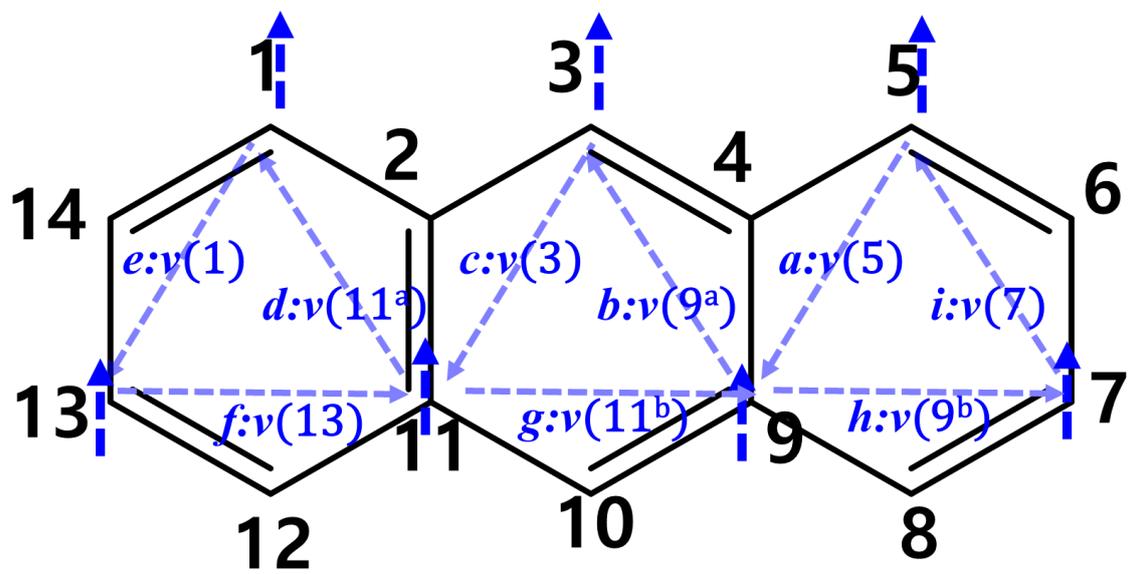

**Figure S3.** Momentum vector for global NNN hopping of up spin antiparticle of anthracene

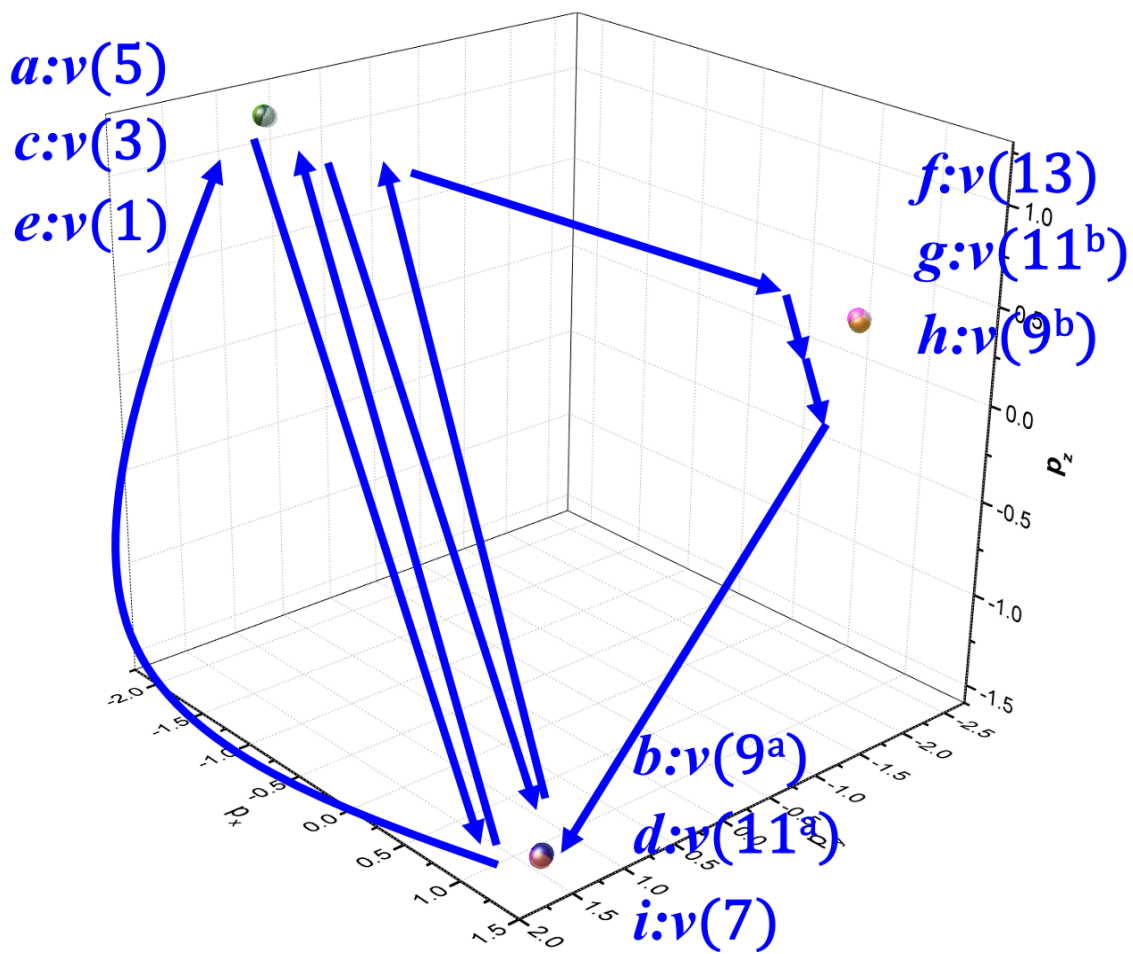

**Figure S4. Momentum space for global NNN hopping of up spin antiparticle of anthracene**

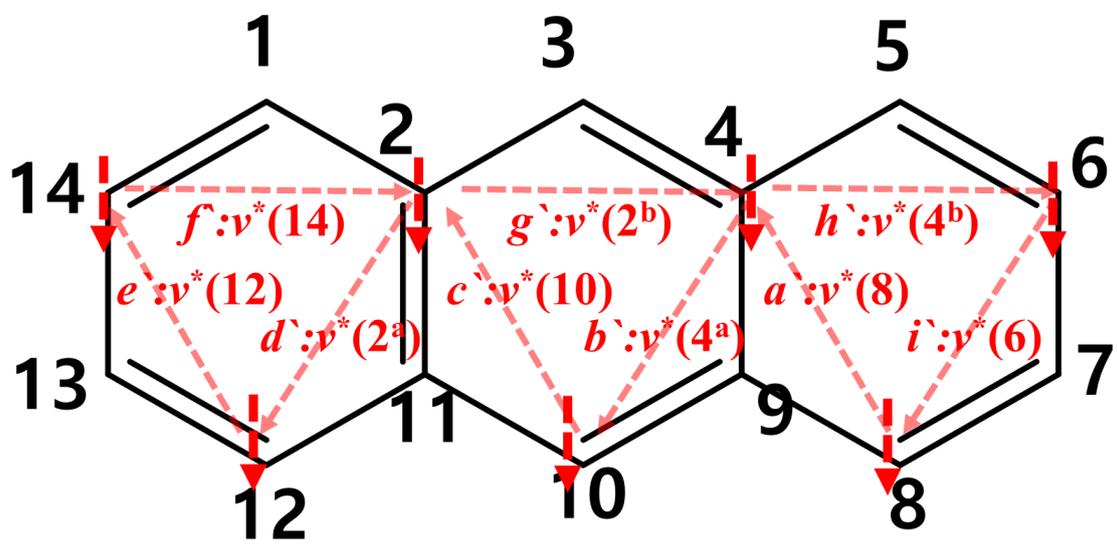

**Figure S5.** Momentum vector for global NNN hopping of down spin antiparticle of anthracene

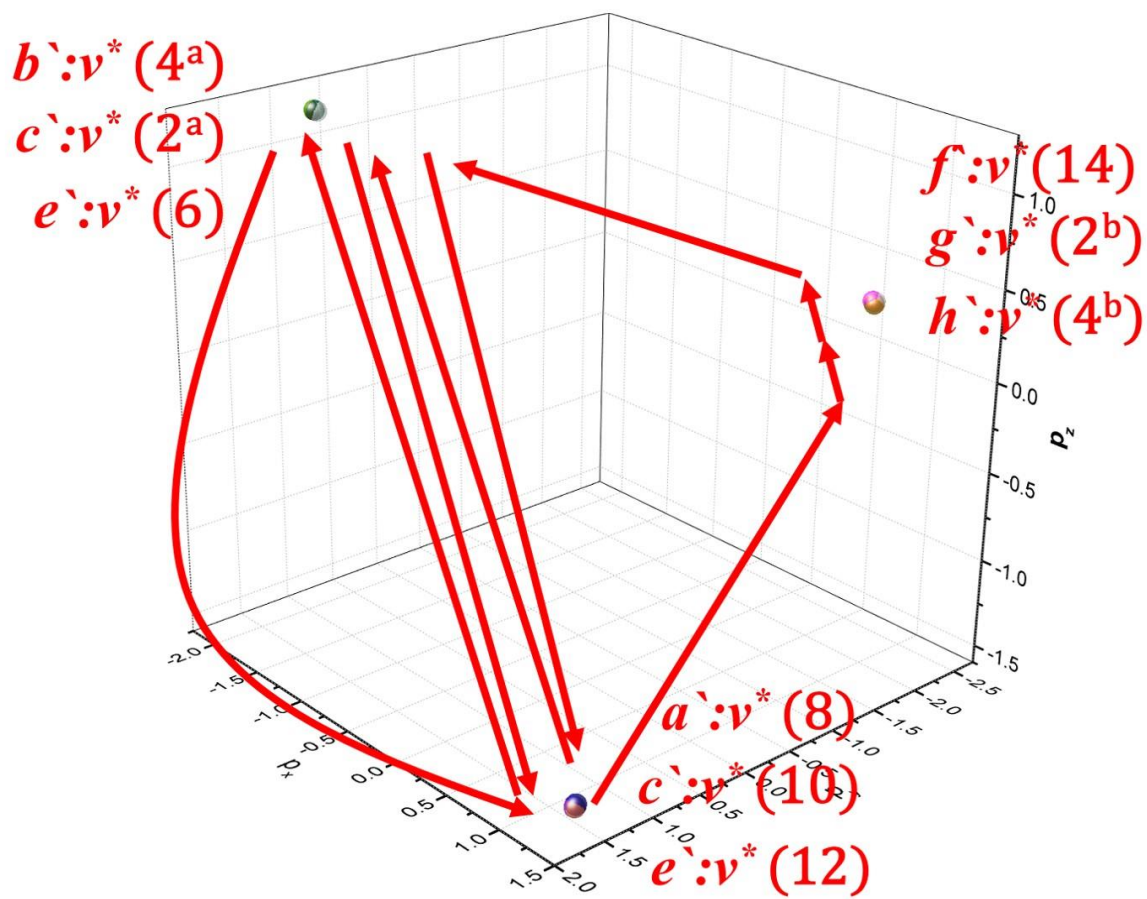

**Figure S6.** Momentum space for global NNN hopping of down spin antiparticle of anthracene

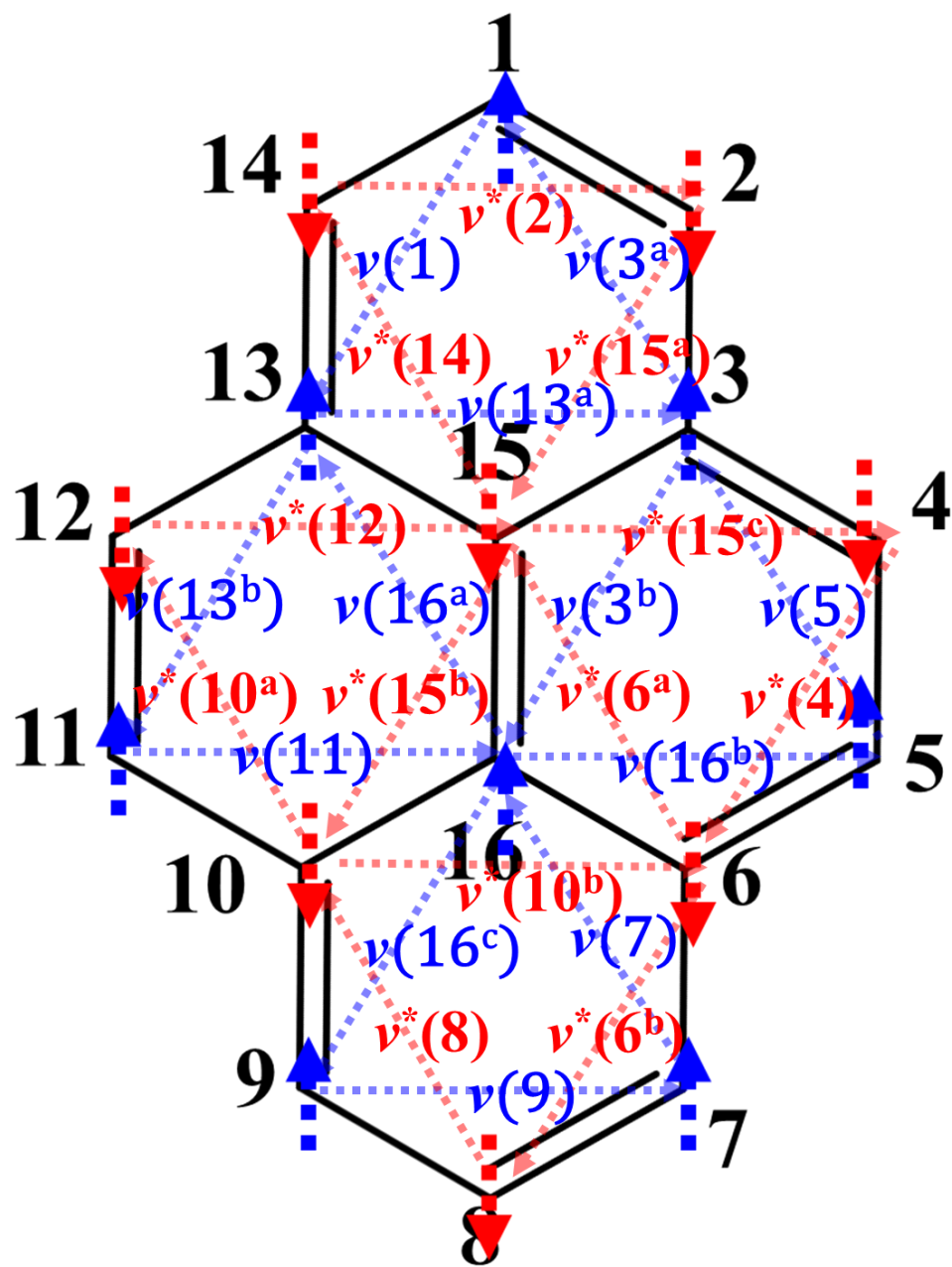

Figure S7. Momentum vectors of antiparticle fermions in pyrene.

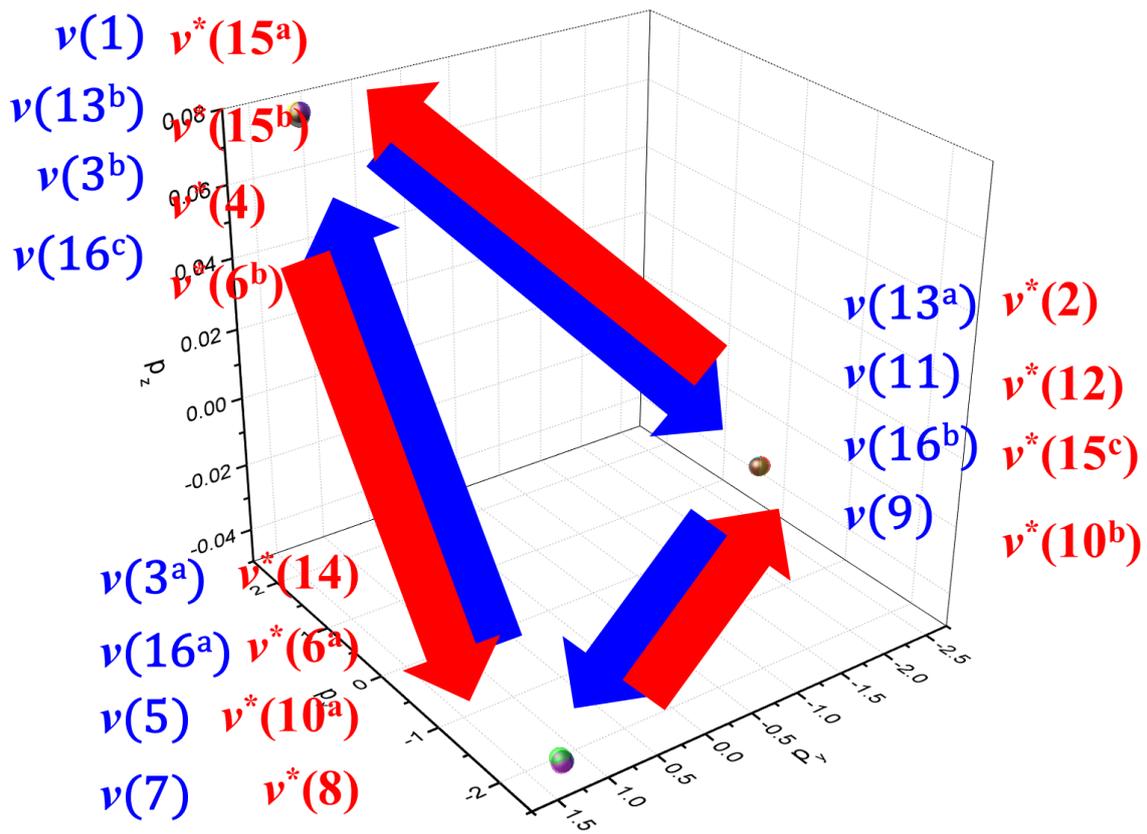

**Figure S8.** Momentum space of antiparticle fermions in pyrene

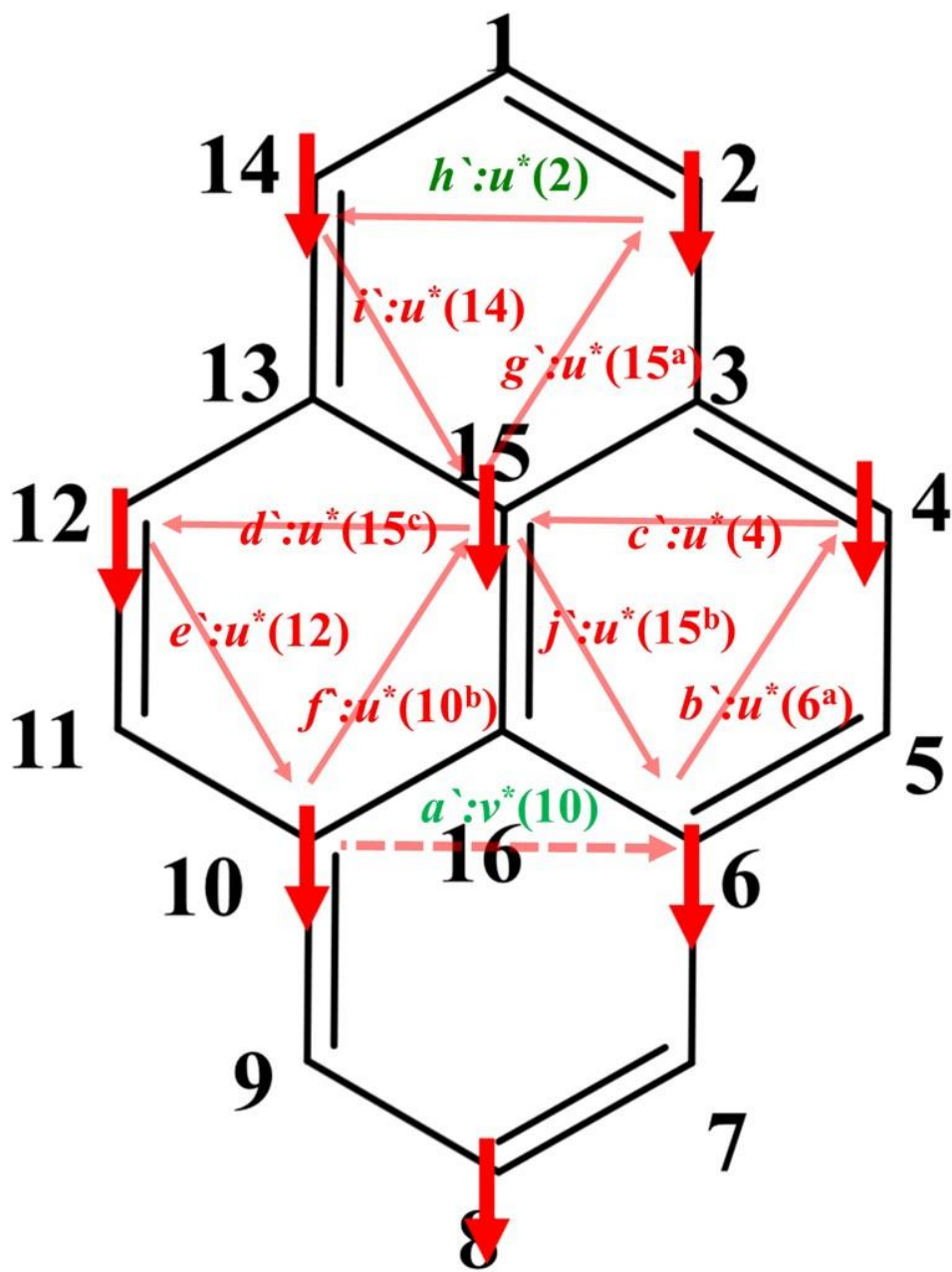

**Figure S9. Momentum vector for global NNN hopping of down spin particle of pyrene**

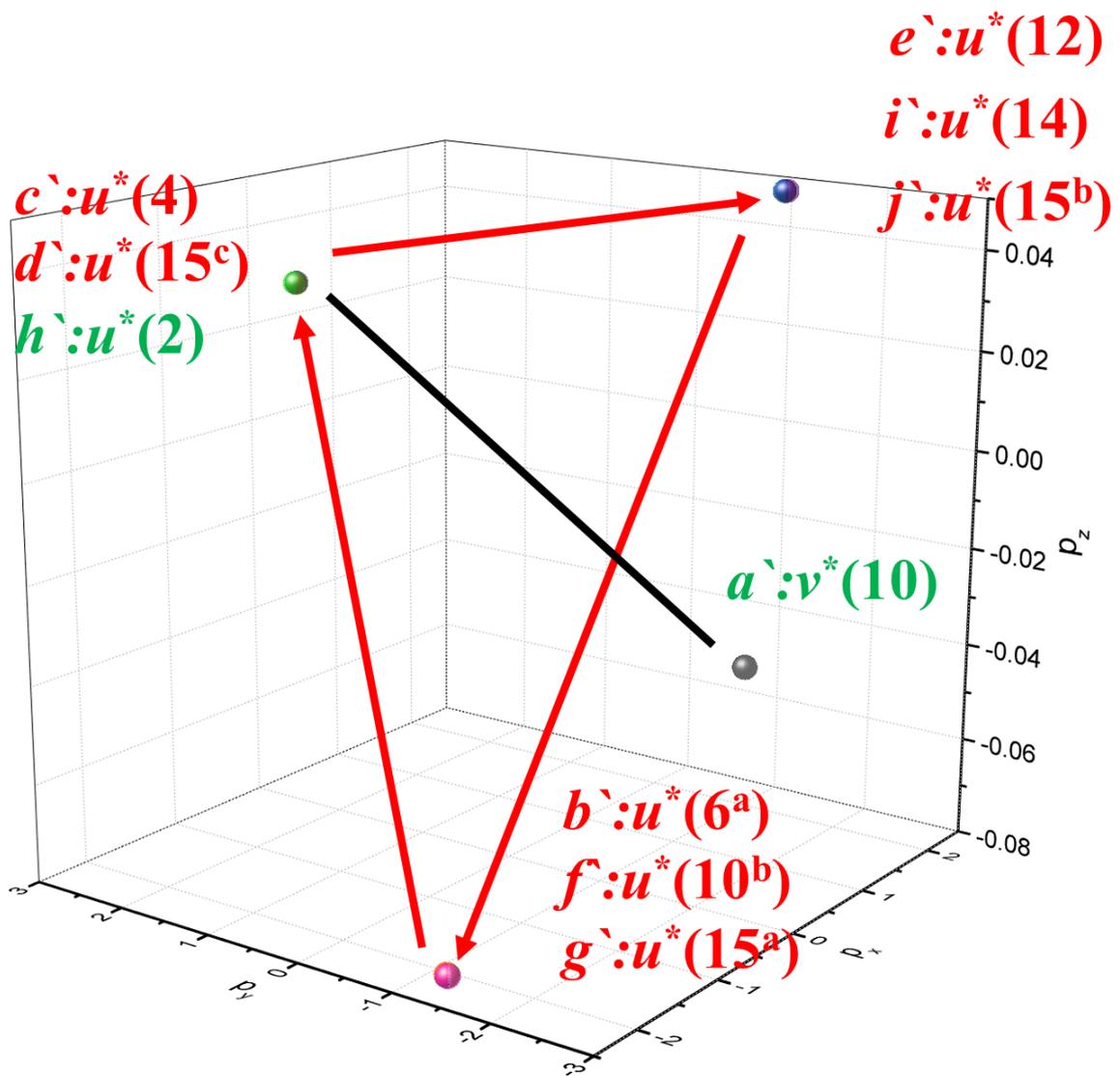

**Figure S10. Momentum space for global NNN hopping of down spin particle of pyrene.**

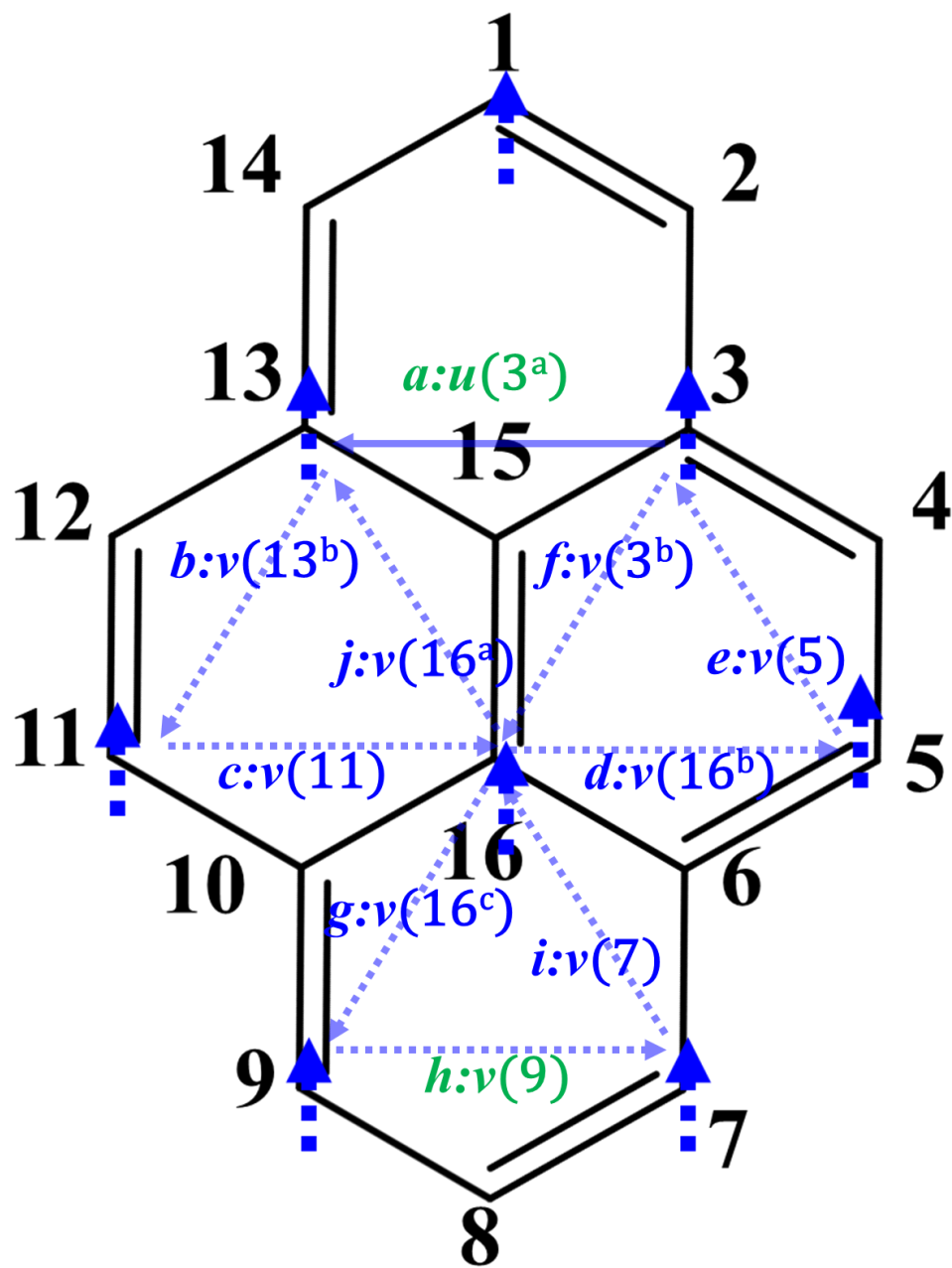

**Figure S11.** Momentum vector for global NNN hopping of up spin antiparticle of pyrene

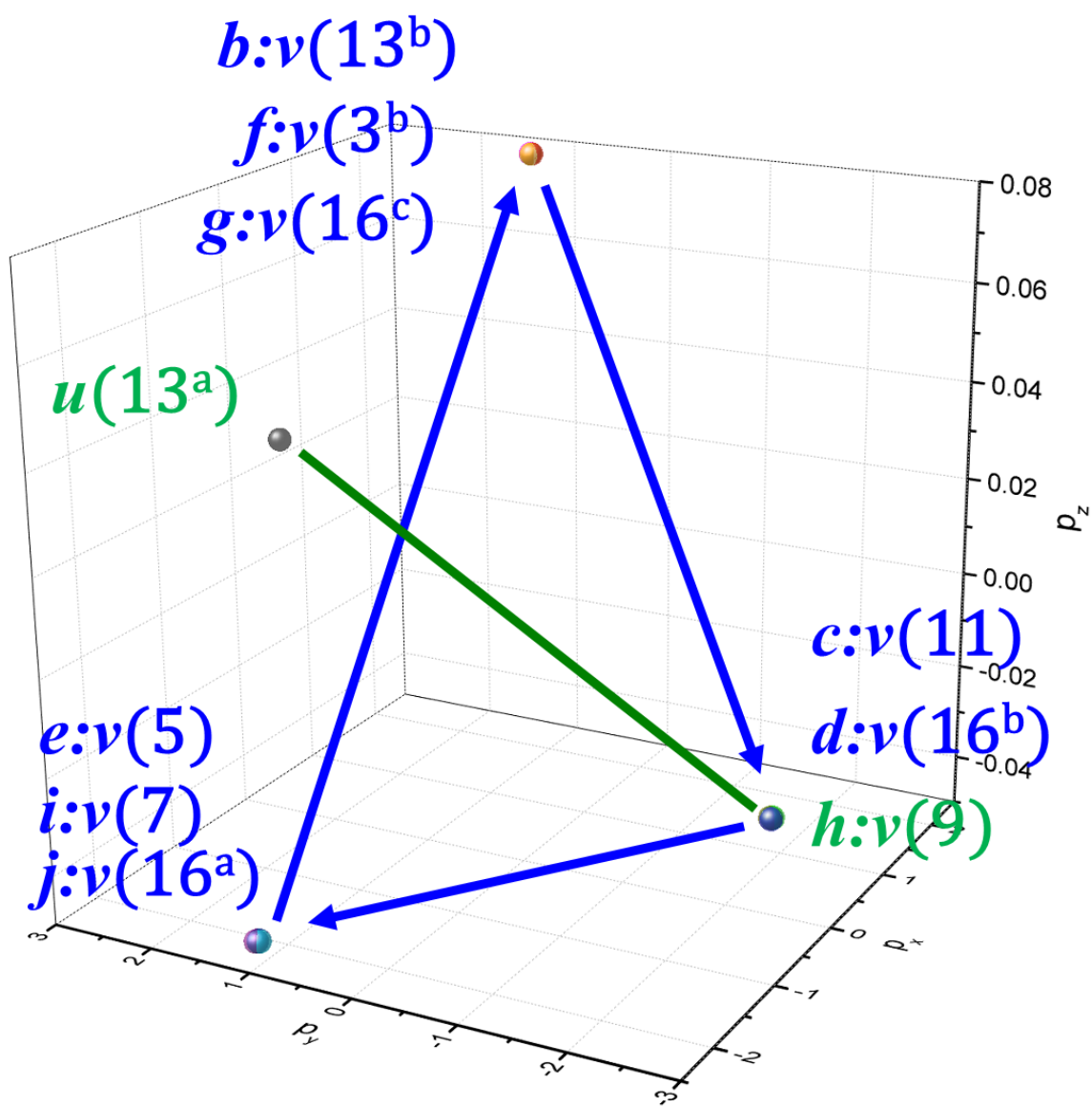

Figure S1. Momentum space for global NNN hopping of up spin antiparticle of pyrene

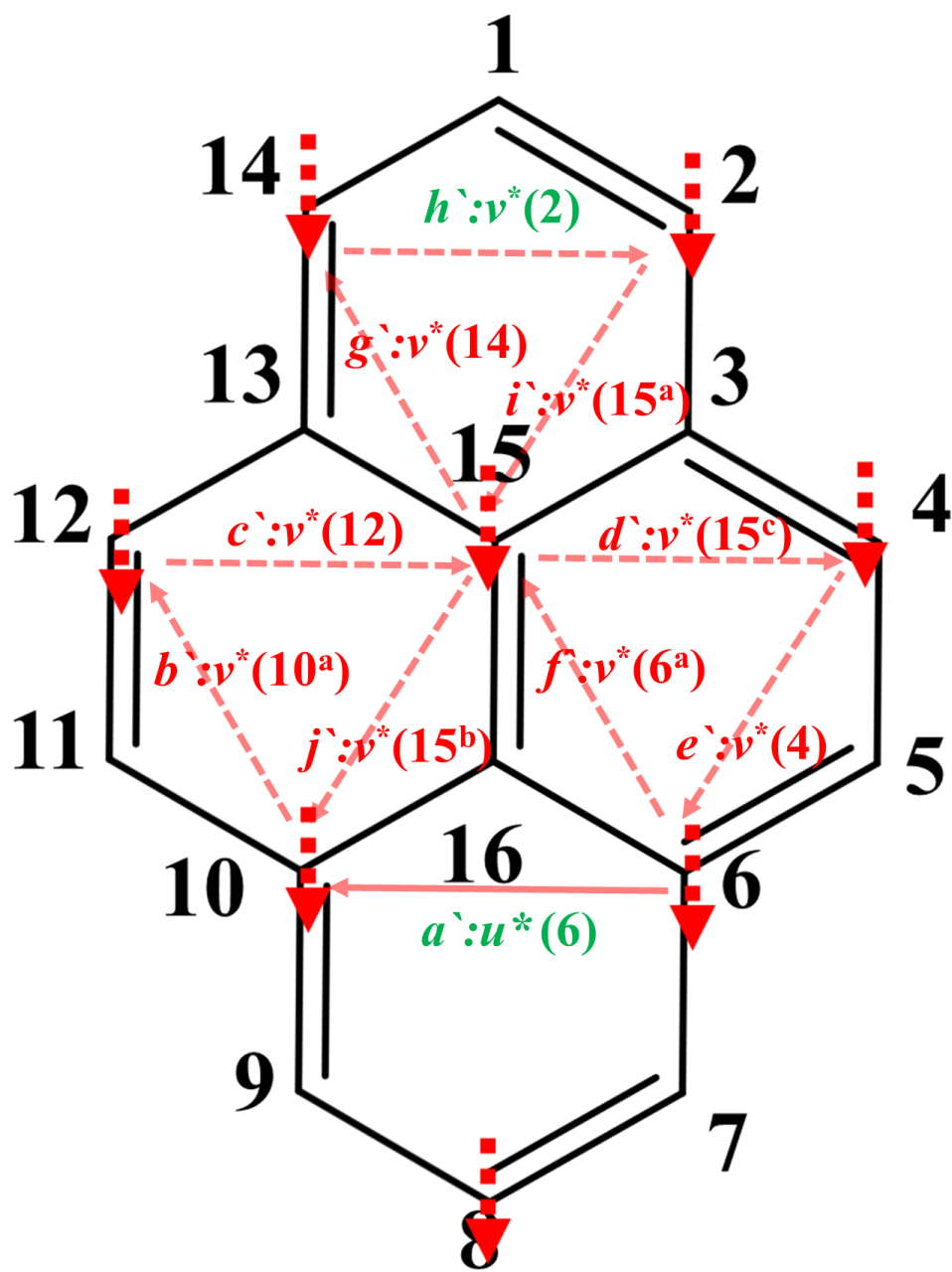

Figure S13. Momentum vector for global NNN hopping of down spin antiparticle of pyrene

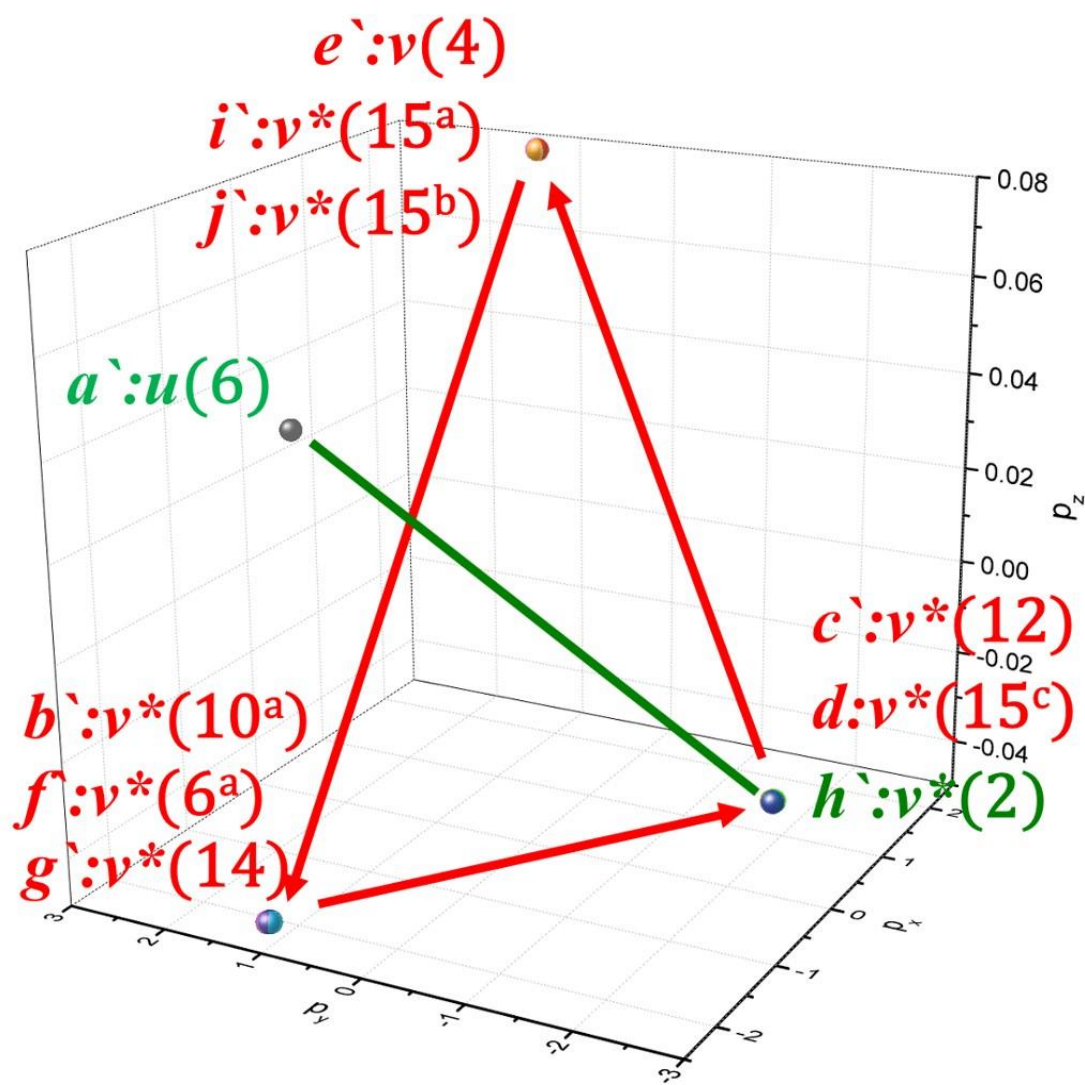

Figure S14. Momentum space for global NNN hopping of down spin antiparticle of pyrene

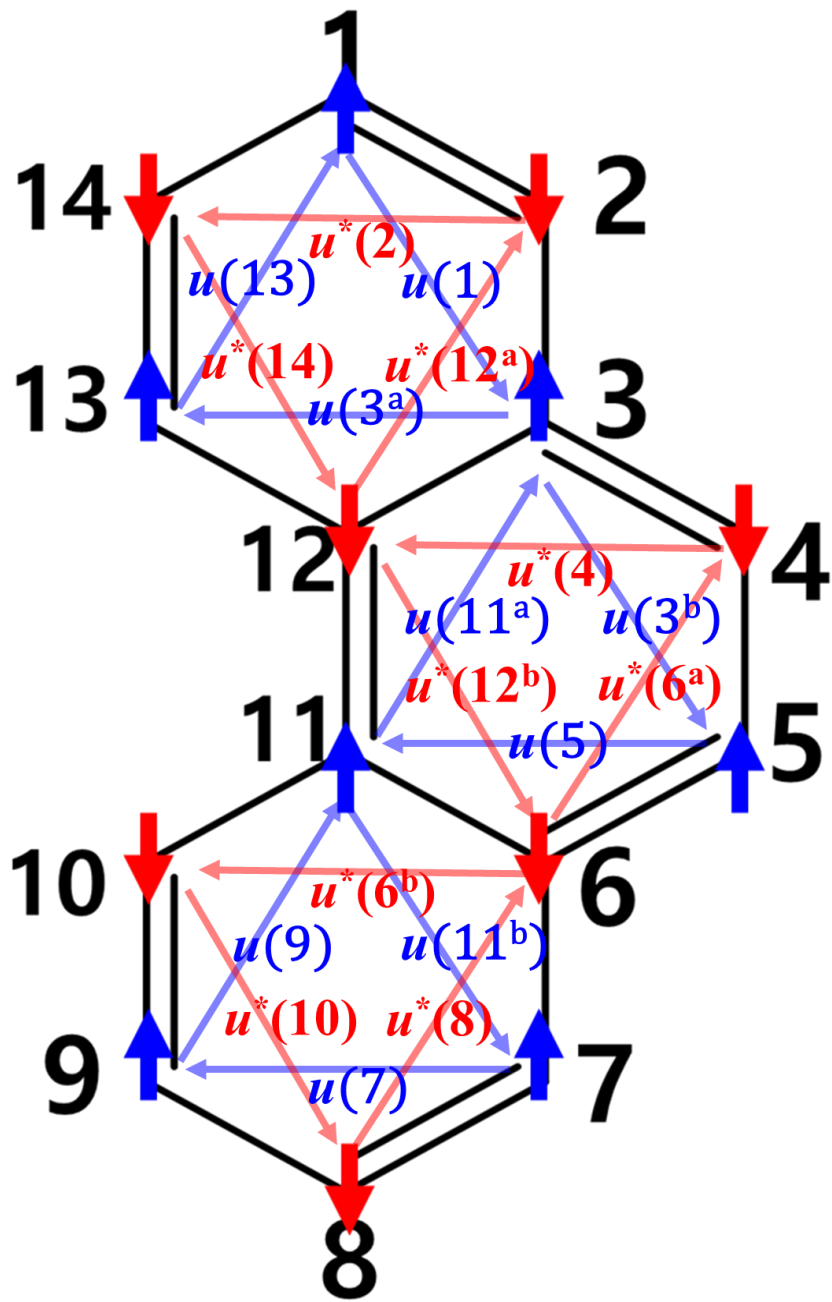

**Figure S15. Momentum vectors of particle fermions in phenanthrene**

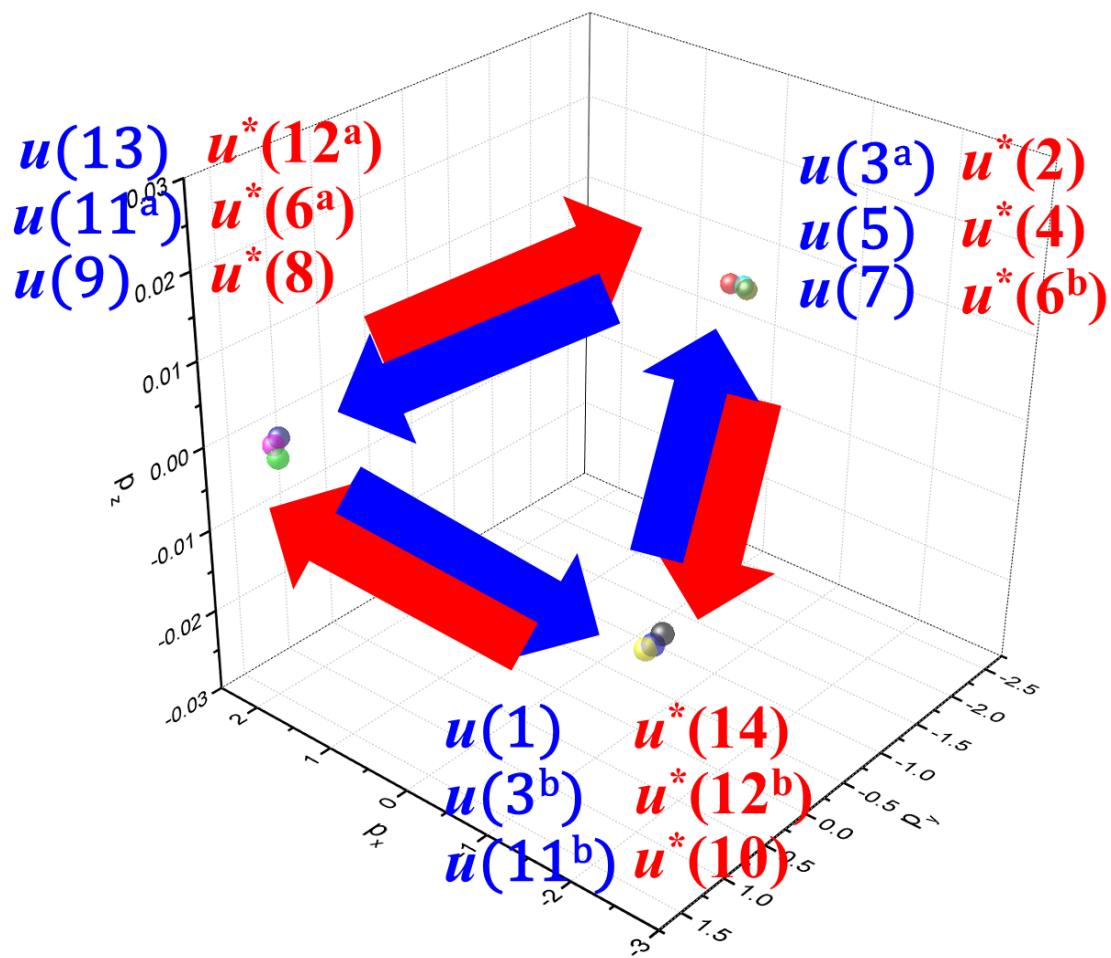

**Figure S16. Momentum space of particle fermions in phenanthrene**

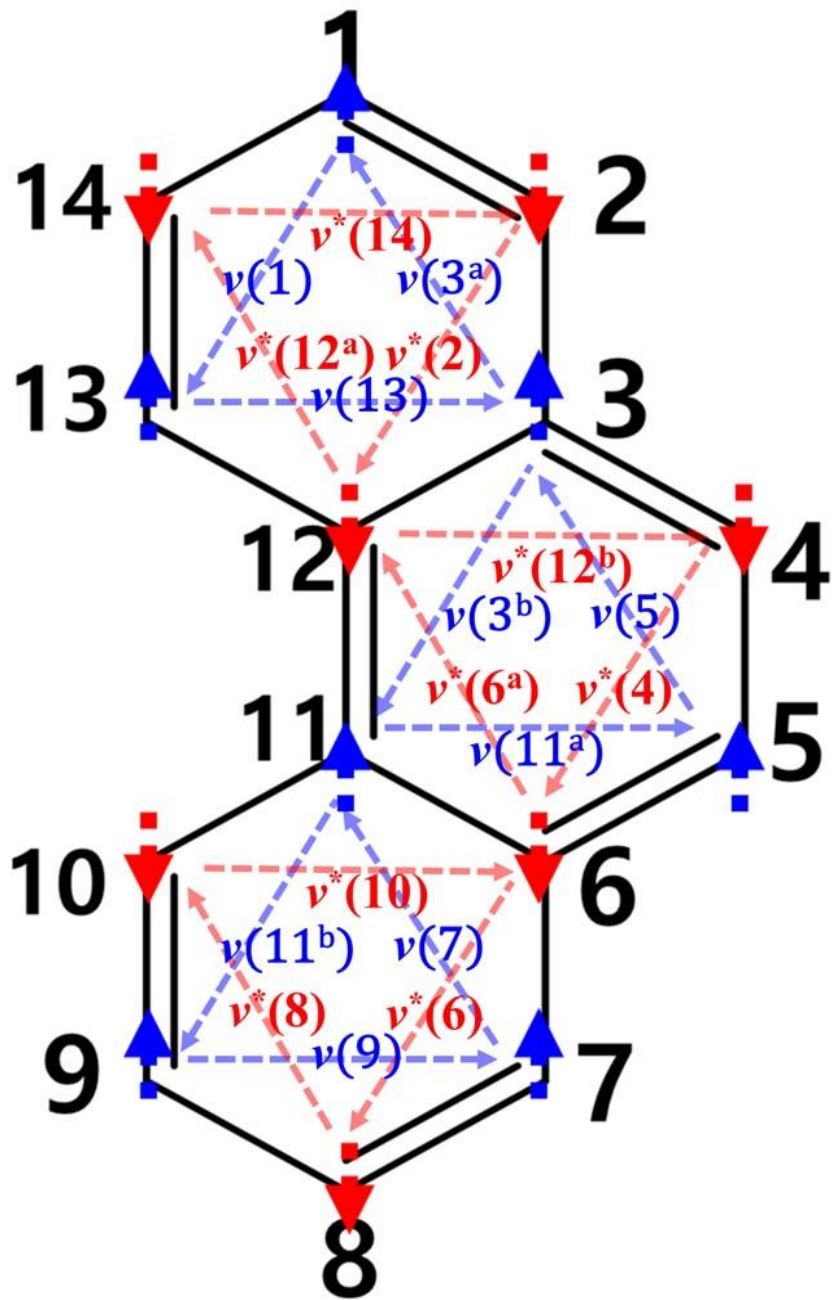

**Figure S17. Momentum vectors of antiparticle fermions in phenanthrene**

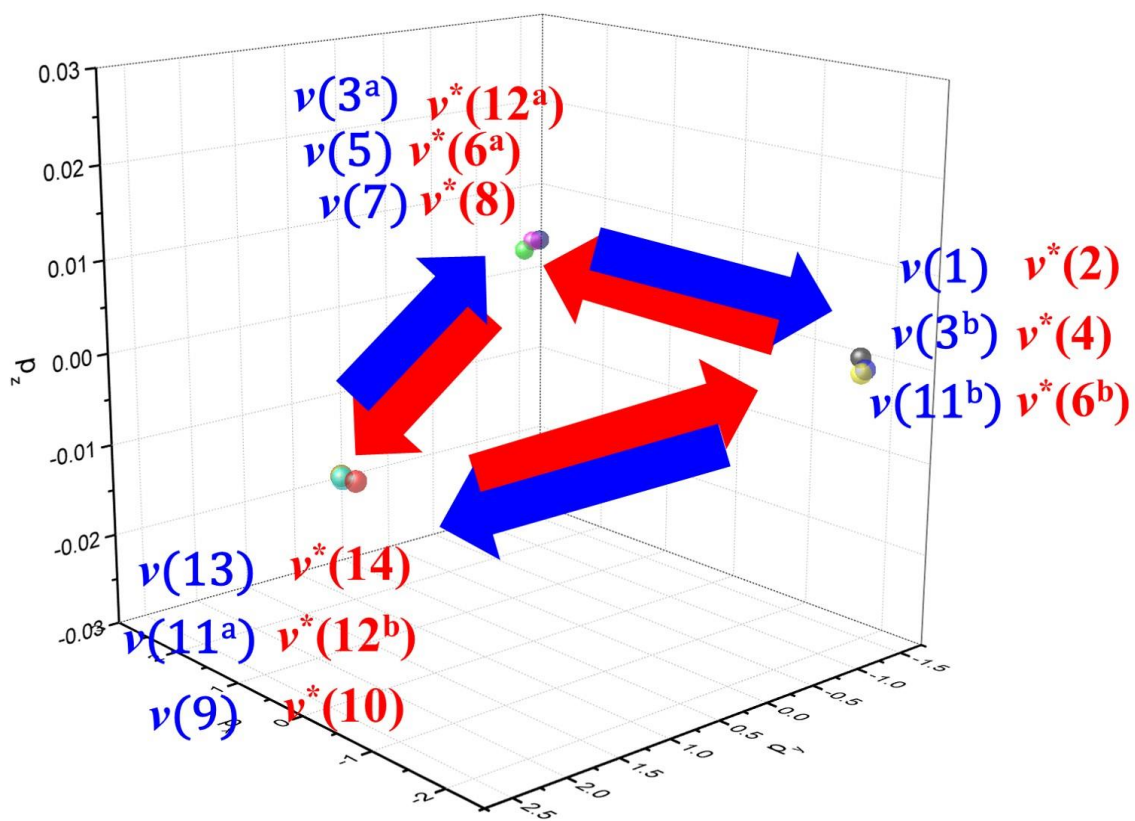

Figure S18. Momentum space of antiparticle fermions in phenanthrene

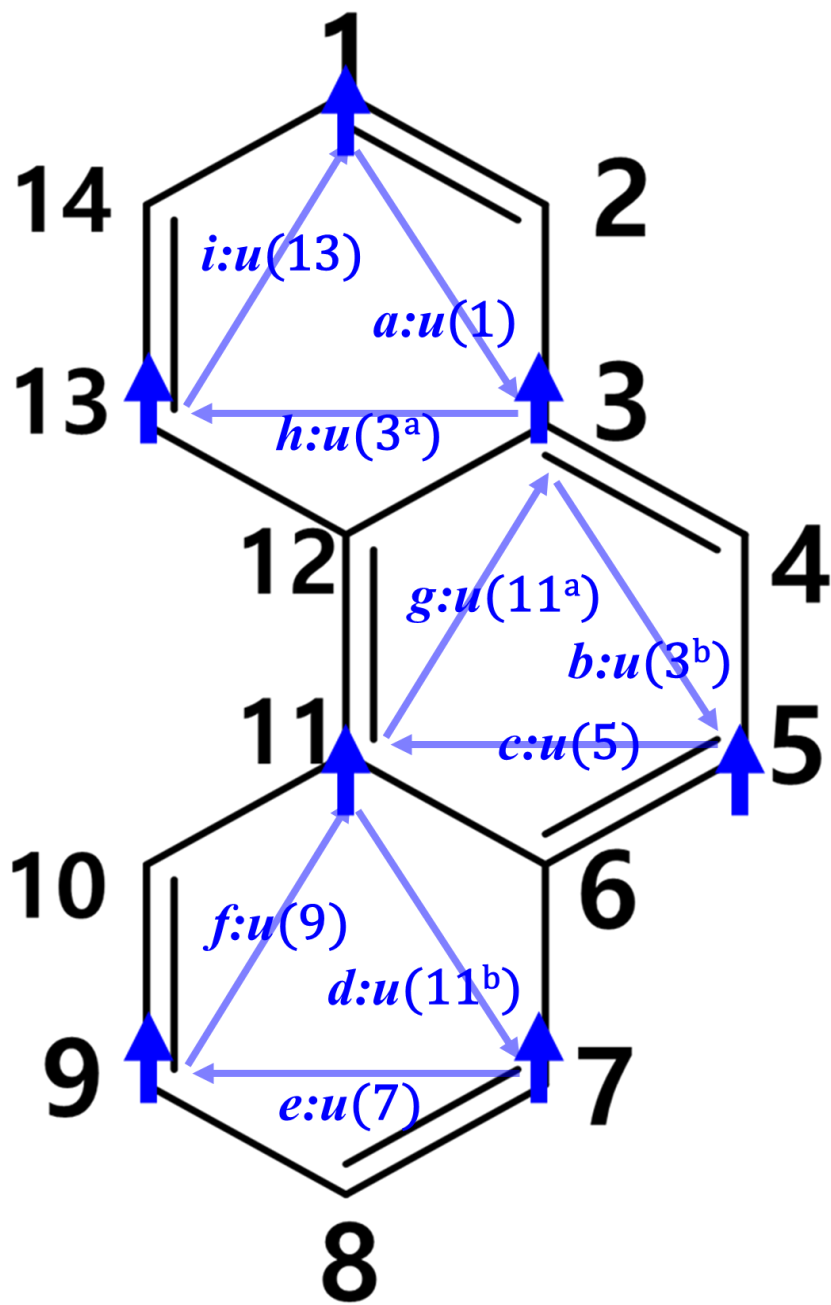

Figure S19. Momentum vector for global NNN hopping of up spin particle of phenanthrene

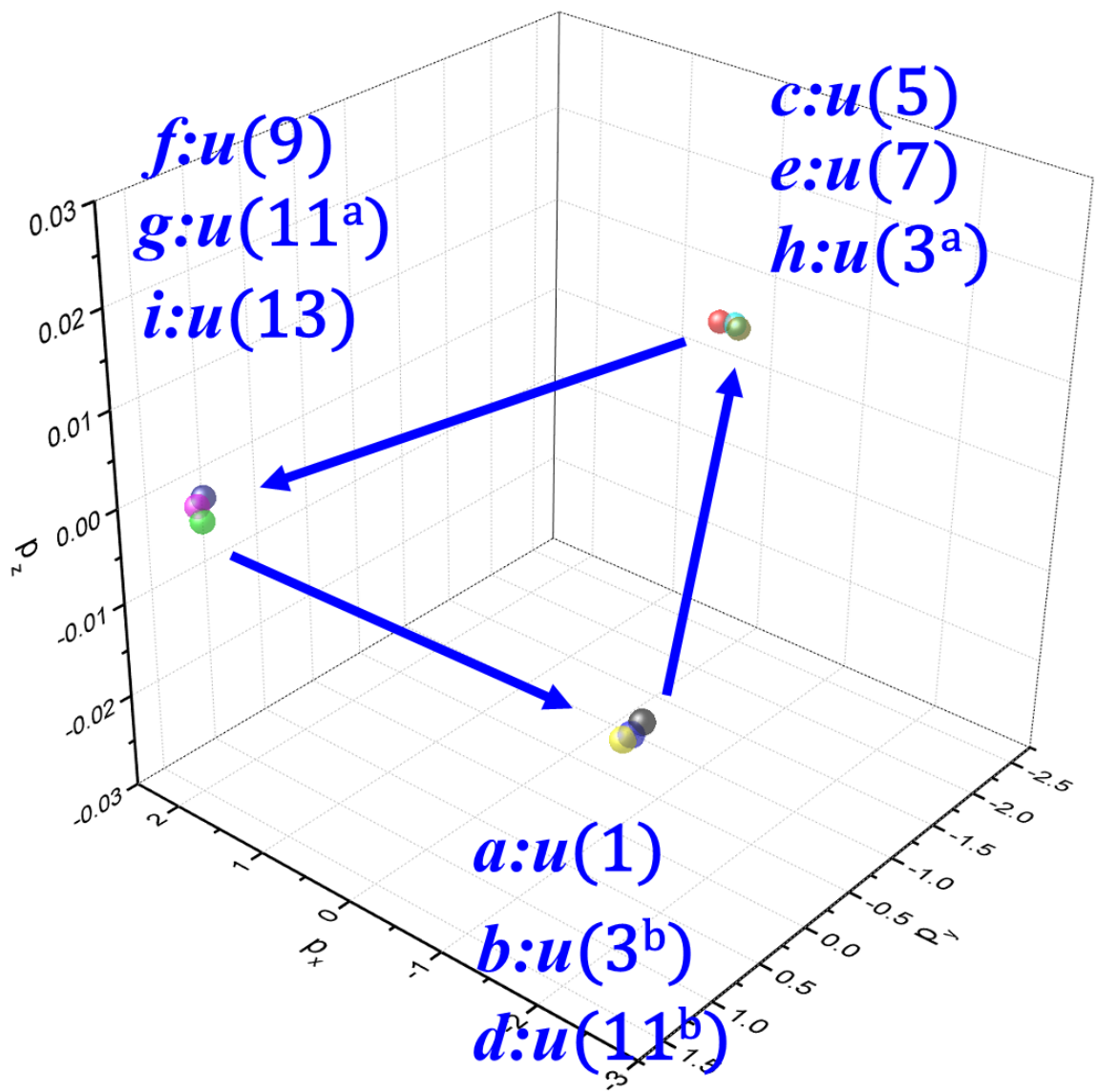

**Figure S20. Momentum space for global NNN hopping of up spin particle of phenanthrene**

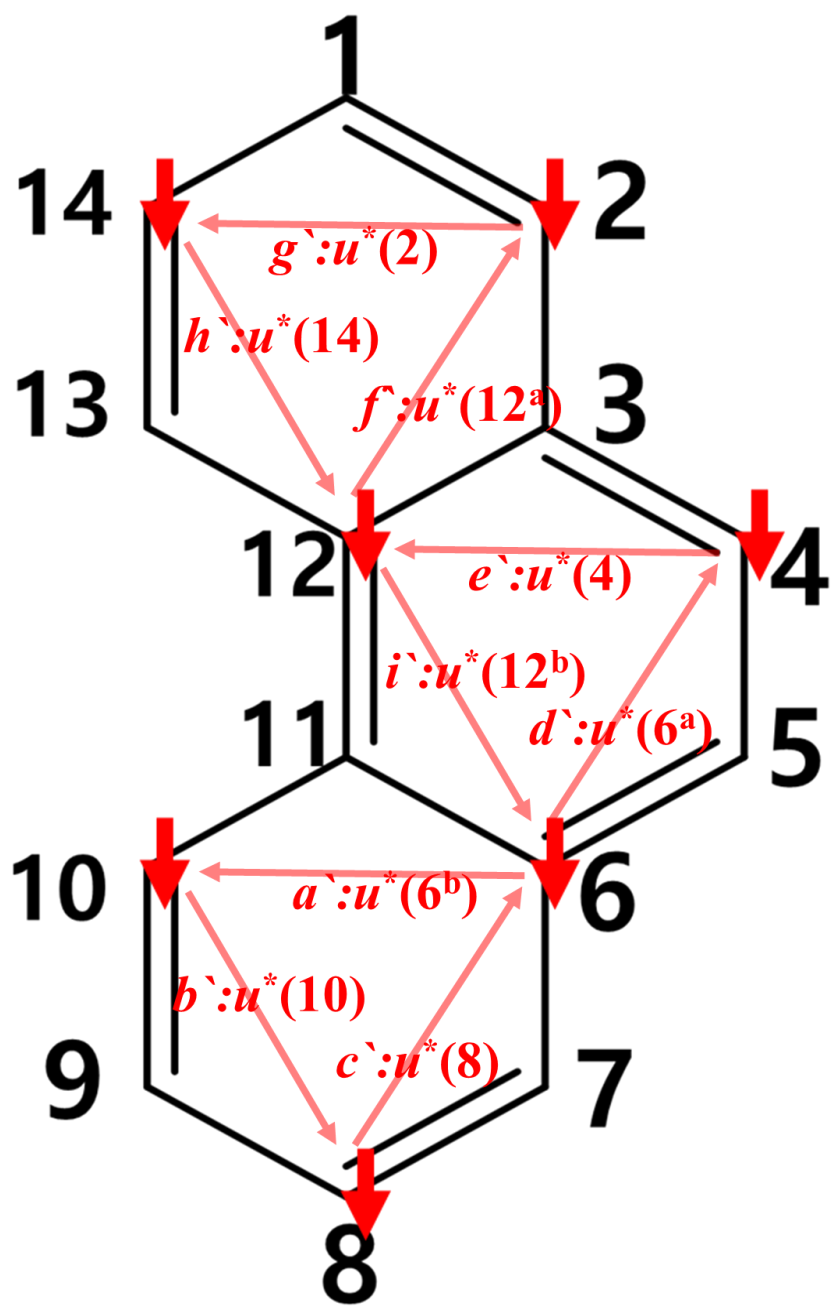

**Figure S21. Momentum vector for global NNN hopping of down spin particle of phenanthrene**

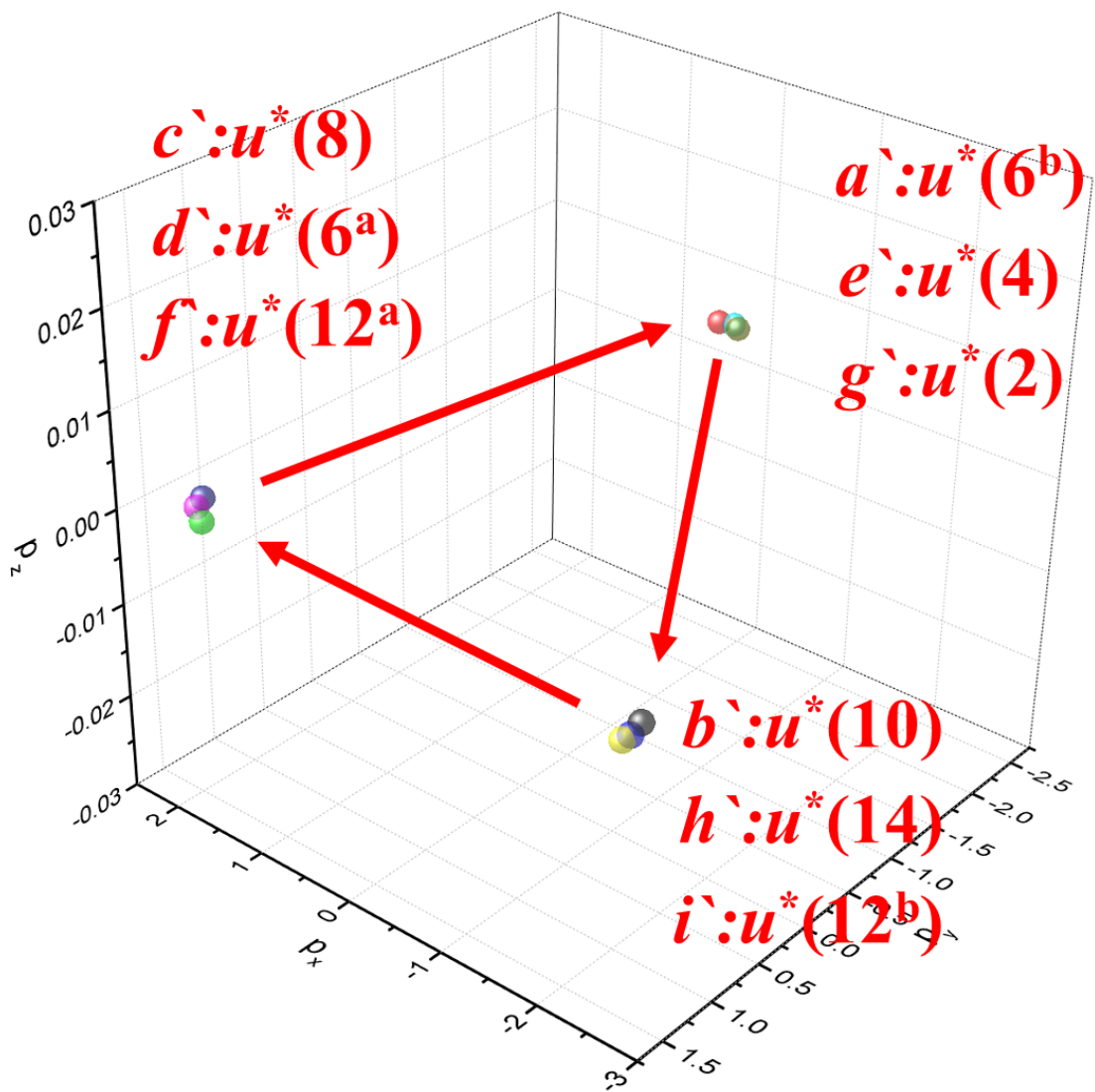

**Figure S22. Momentum space for global NNN hopping of down spin particle of phenanthrene**

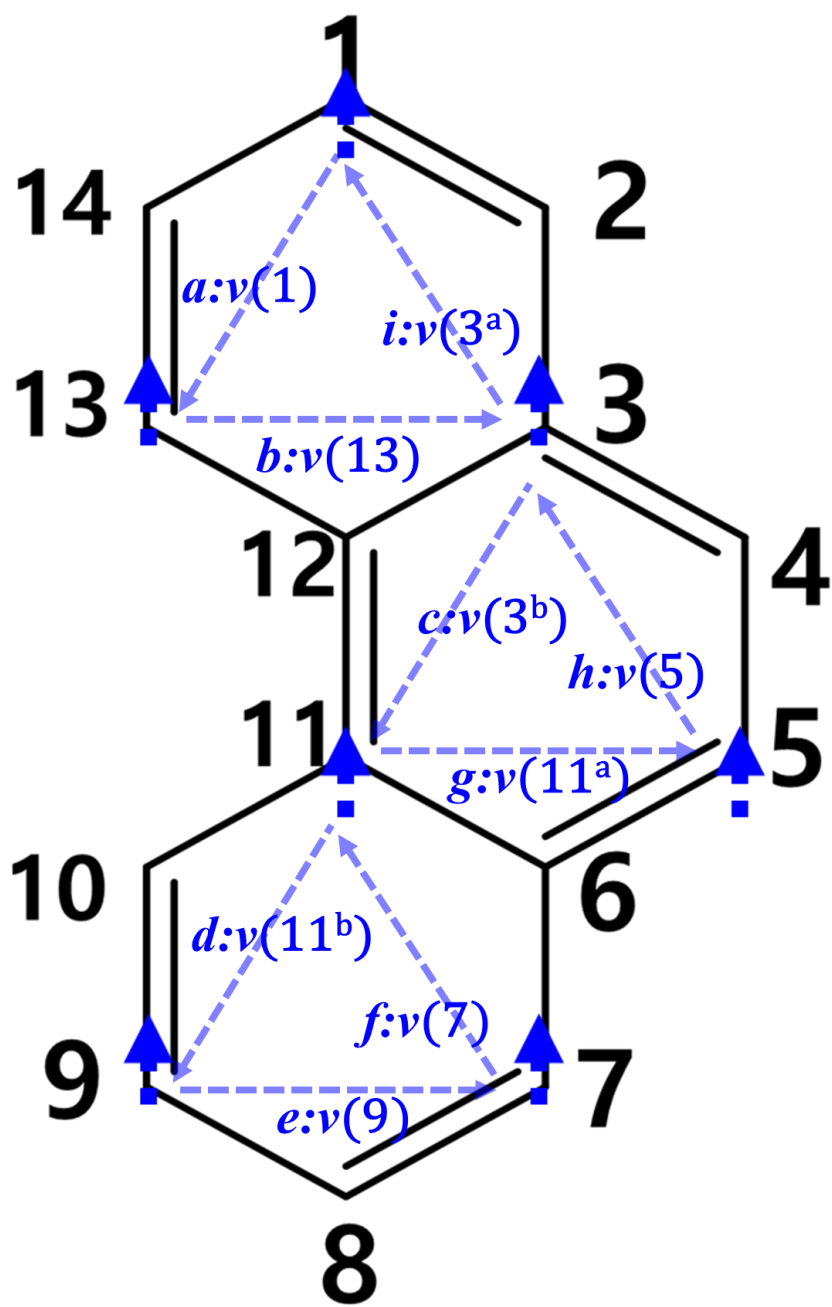

Figure S23. Momentum vector for global NNN hopping of up spin antiparticle of phenanthrene.

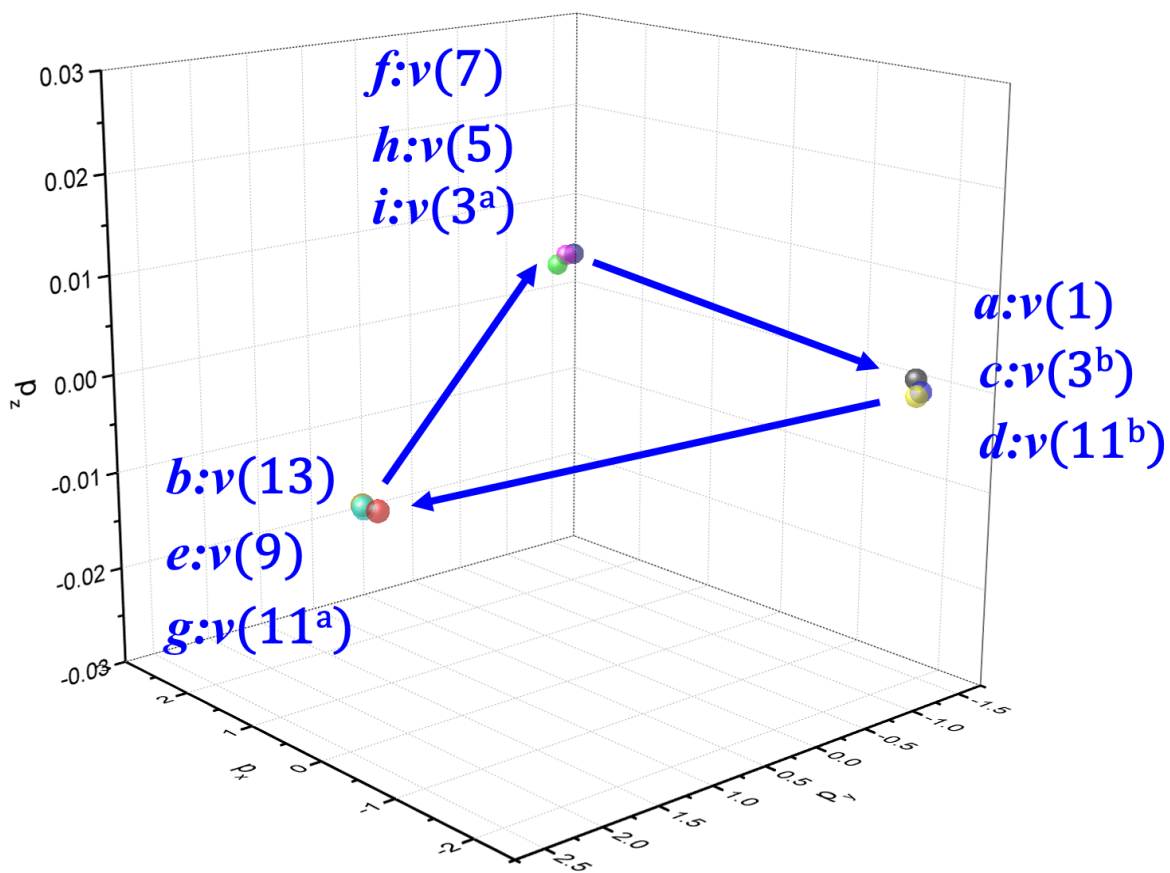

Figure S24. Momentum space for global NNN hopping of up spin antiparticle of anthracene

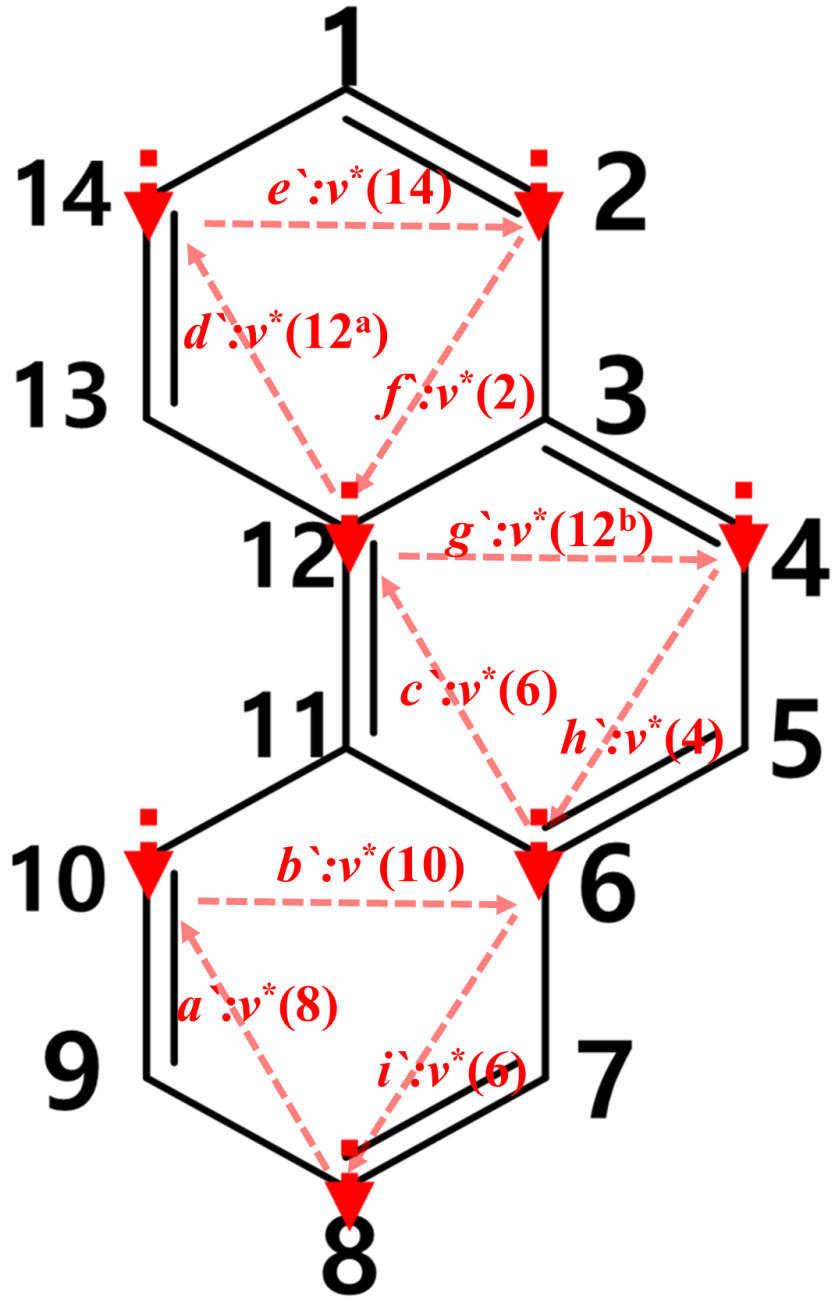

**Figure S25.** Momentum vector for global NNN hopping of down spin antiparticle of phenanthrene.

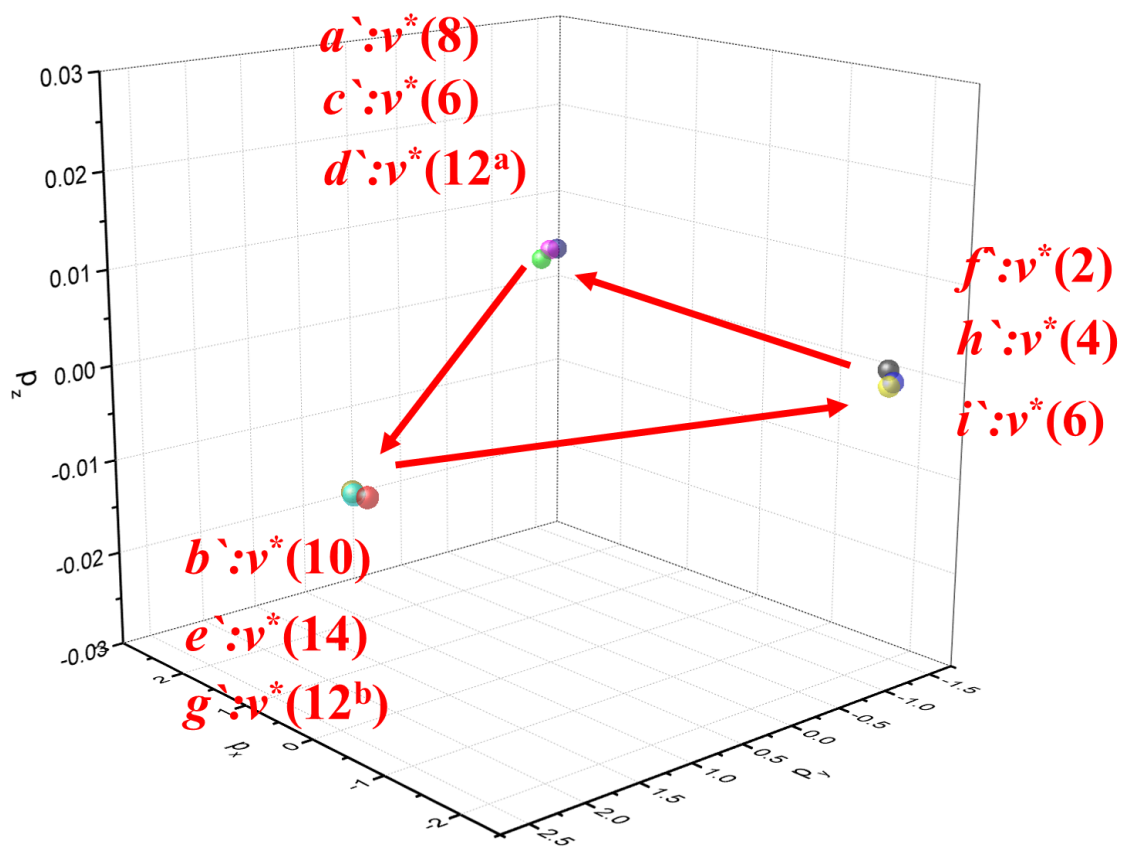

Figure S26. Momentum space for global NNN hopping of down spin antiparticle of phenanthrene.

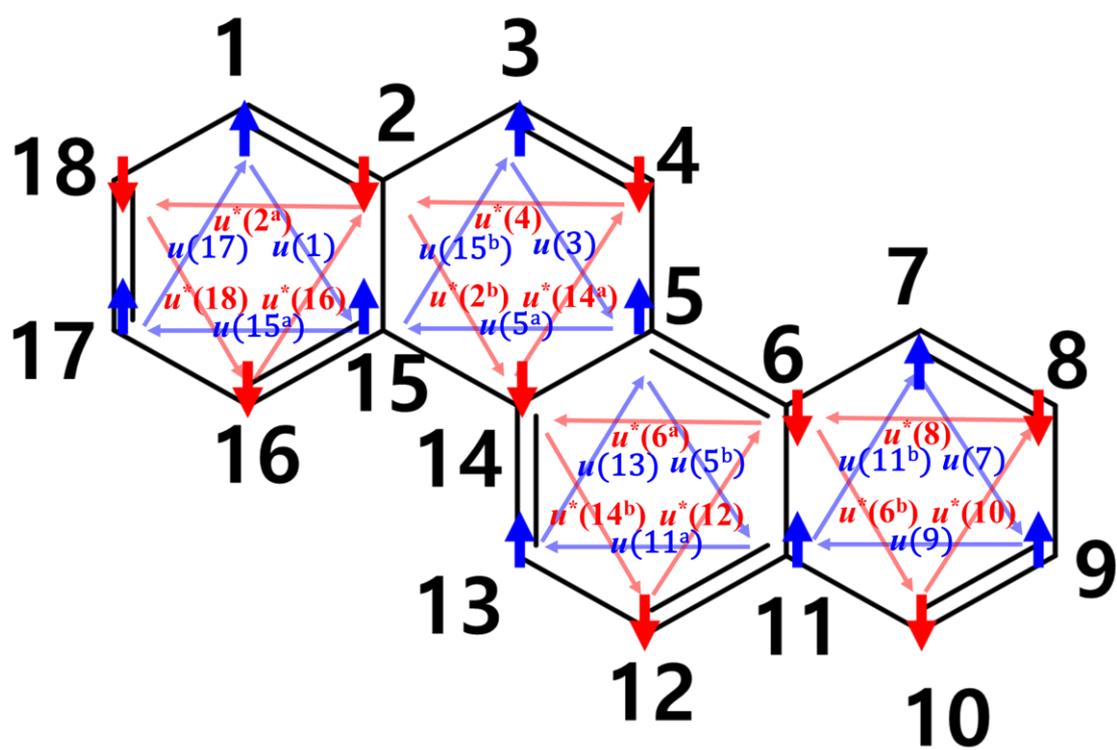

**Figure S27. Momentum vectors of particle fermions in chrysene.**

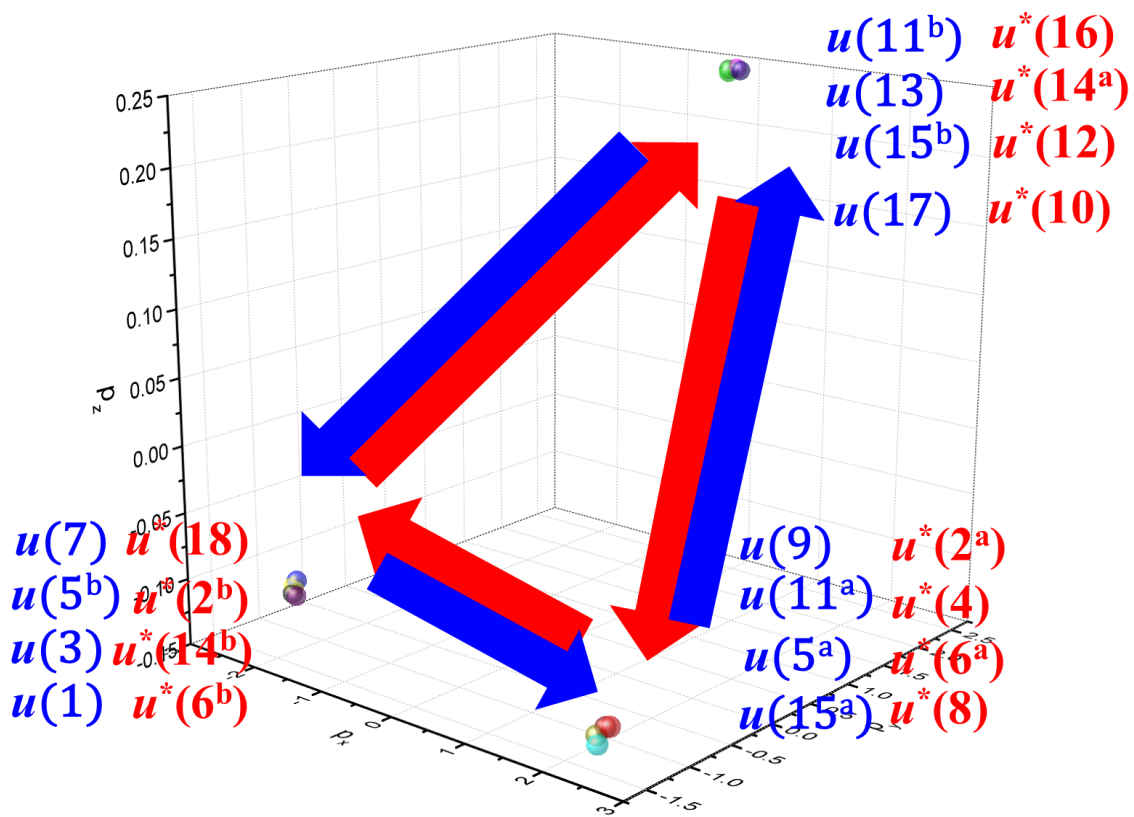

**Figure S28. Momentum space of particle fermions in phenanthrene.**

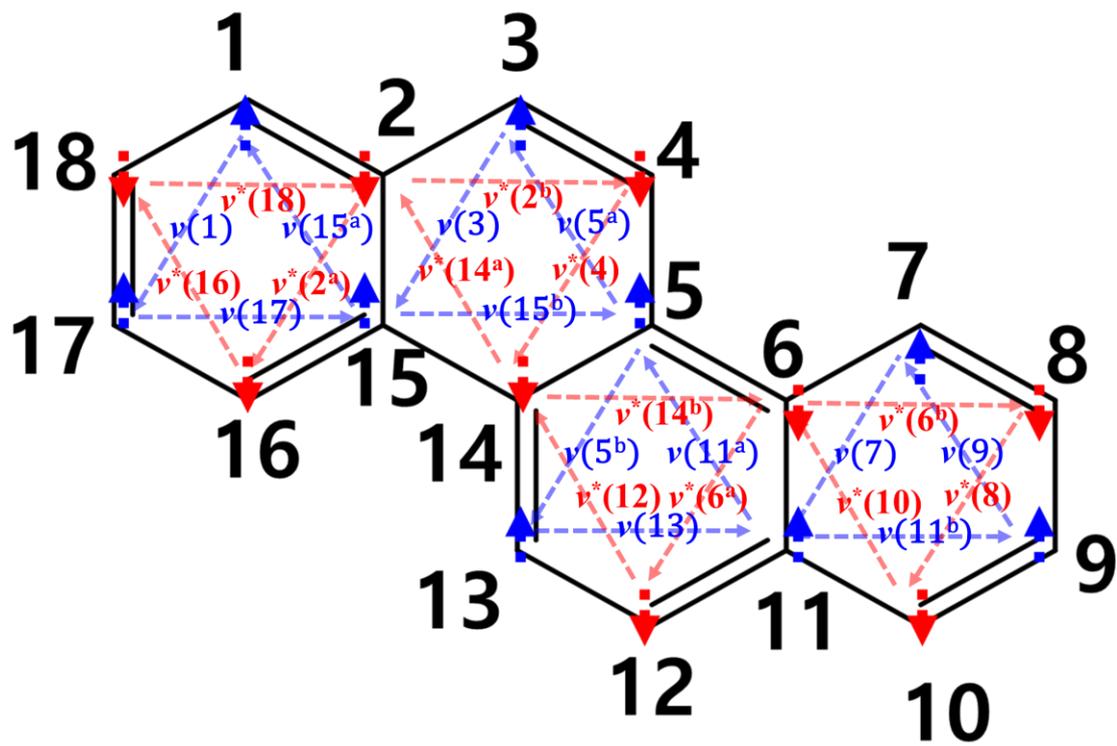

**Figure S29. Momentum vectors of antiparticle fermions in phenanthrene.**

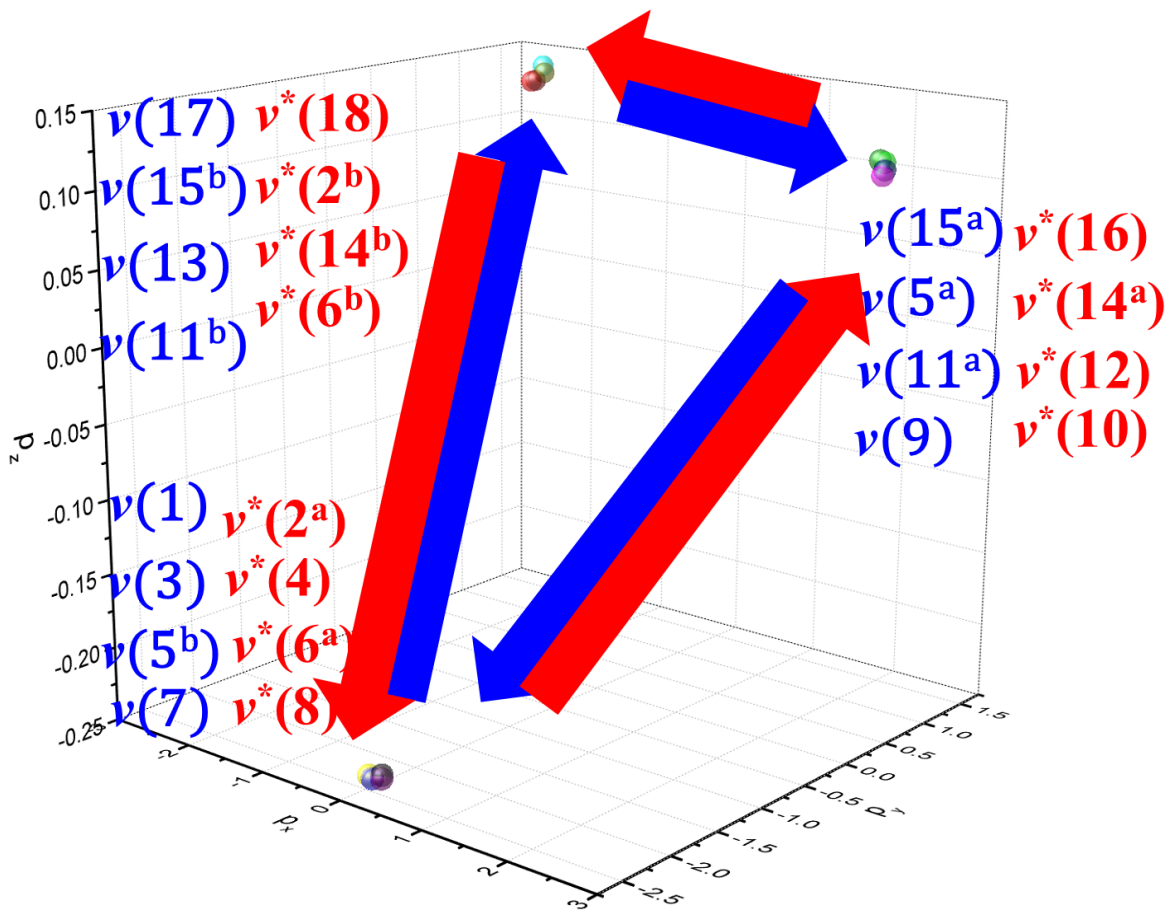

**Figure S30.** Momentum space of antiparticle fermions in chrysene

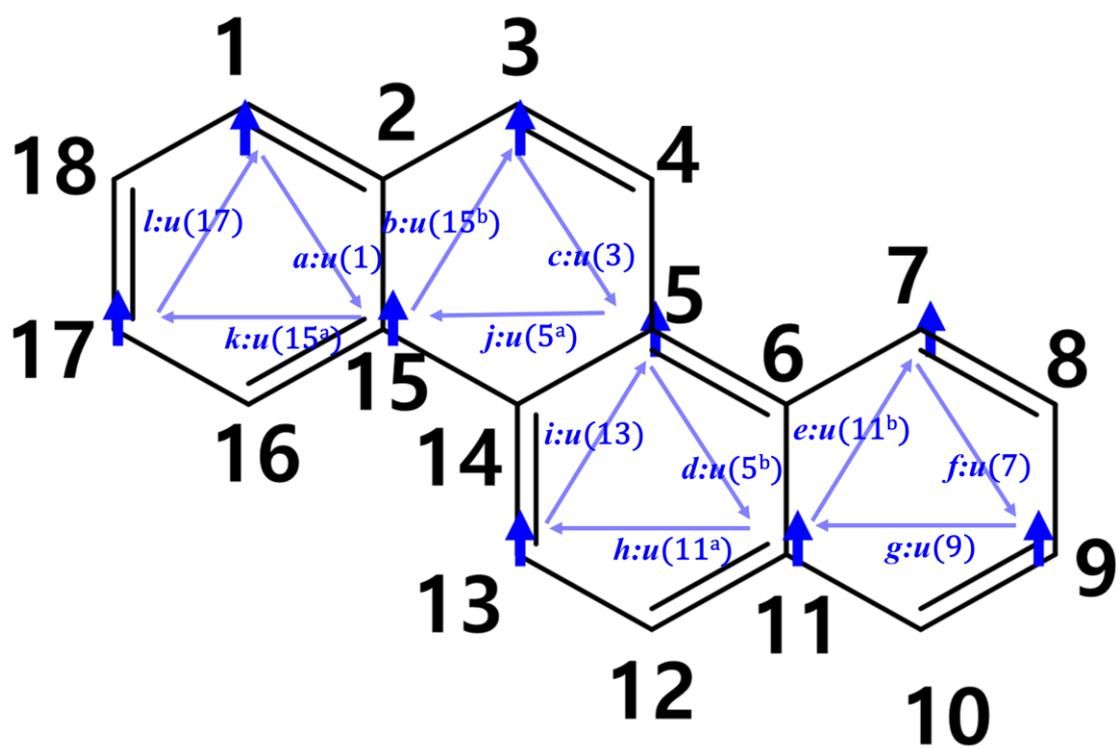

**Figure S31. Momentum vector for global NNN hopping of up spin particle of chrysene.**

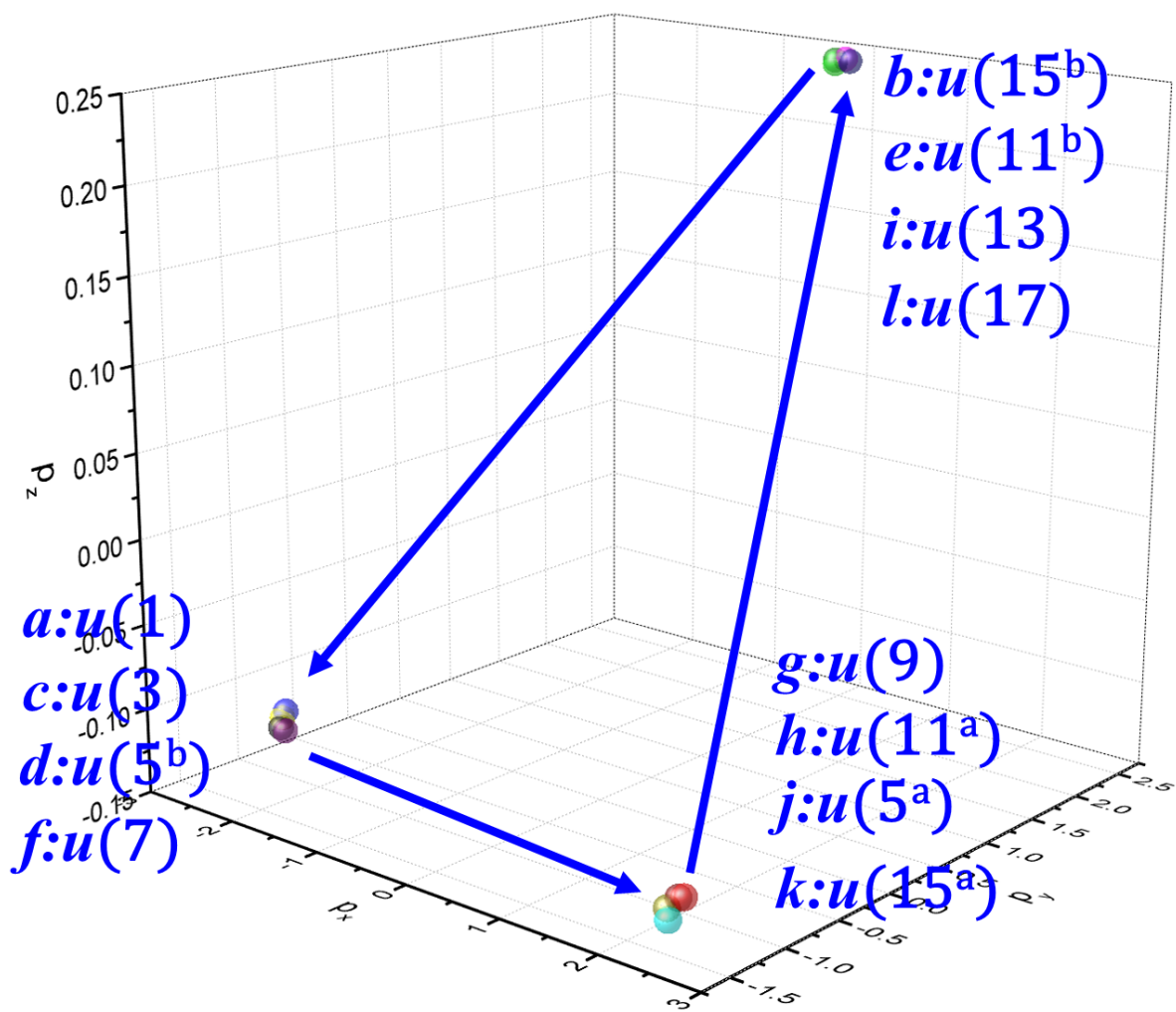

**Figure S32. Momentum space for global NNN hopping of up spin particle of phenanthrene.**

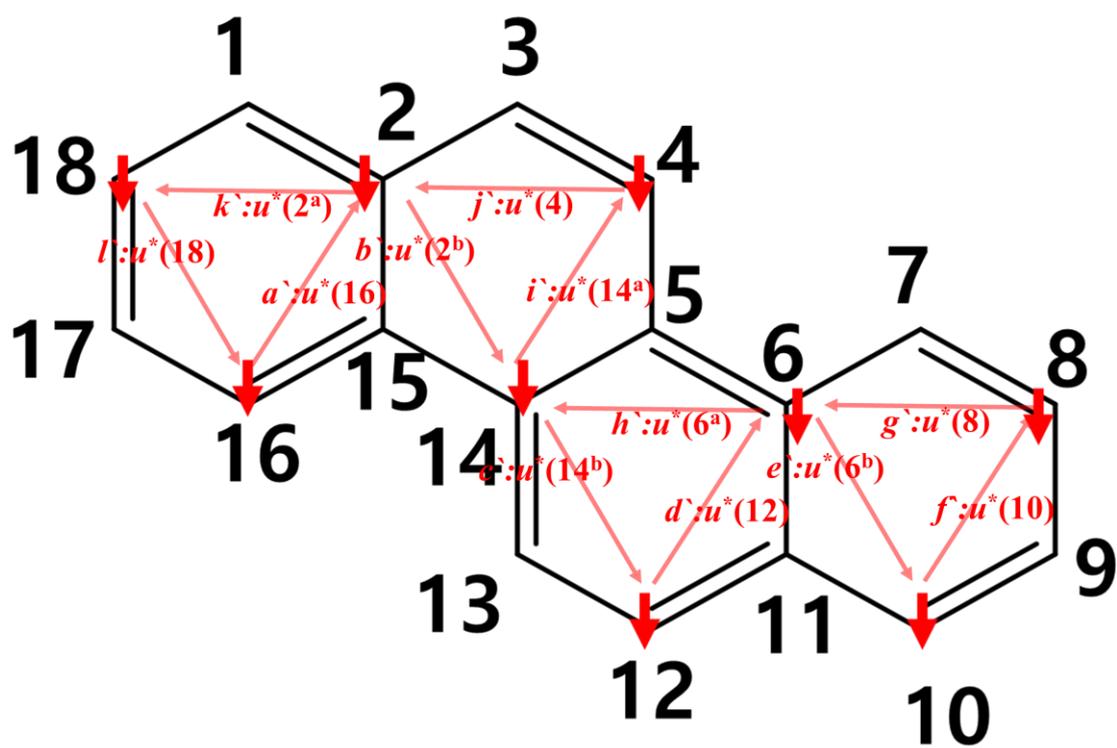

**Figure S33.** Momentum vector for global NNN hopping of down spin particle of chrysene.

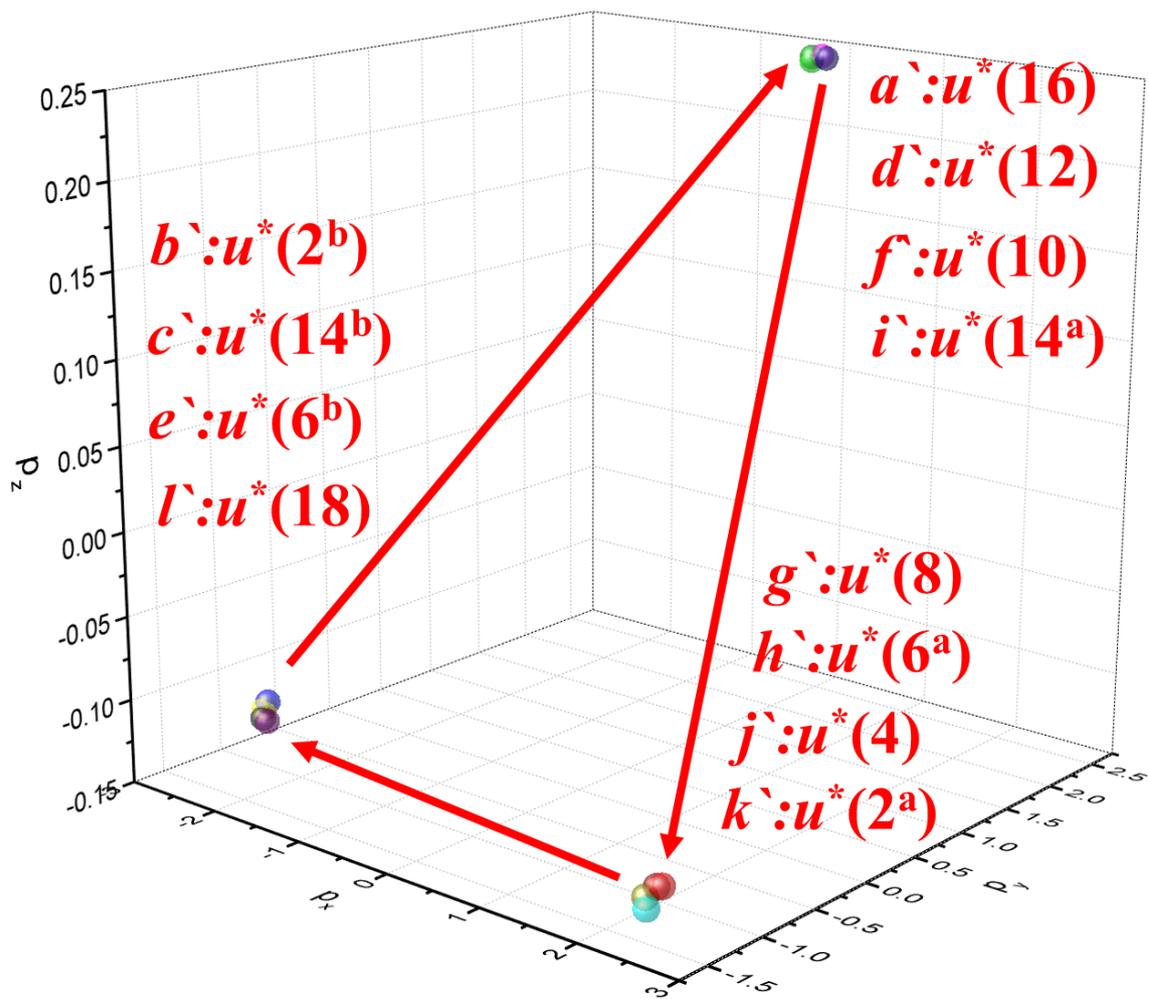

**Figure S34.** Momentum space for global NNN hopping of down spin particle of chrysene

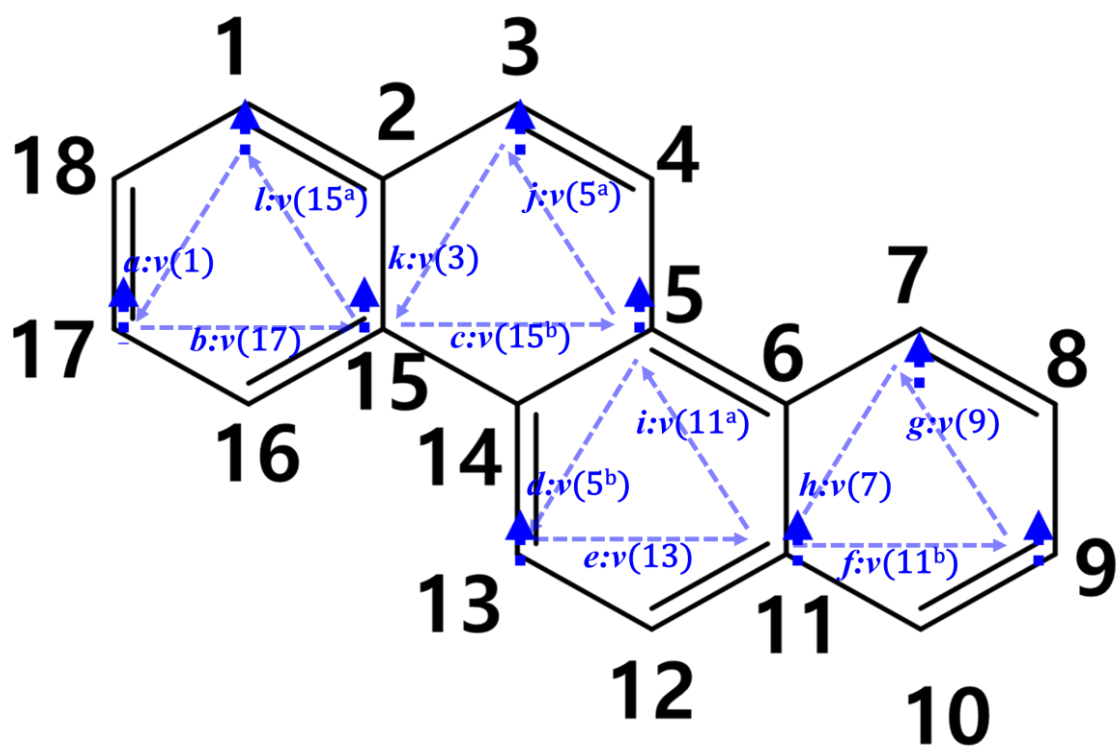

**Figure S35.** Momentum vector for global NNN hopping of up spin antiparticle of chrysene.

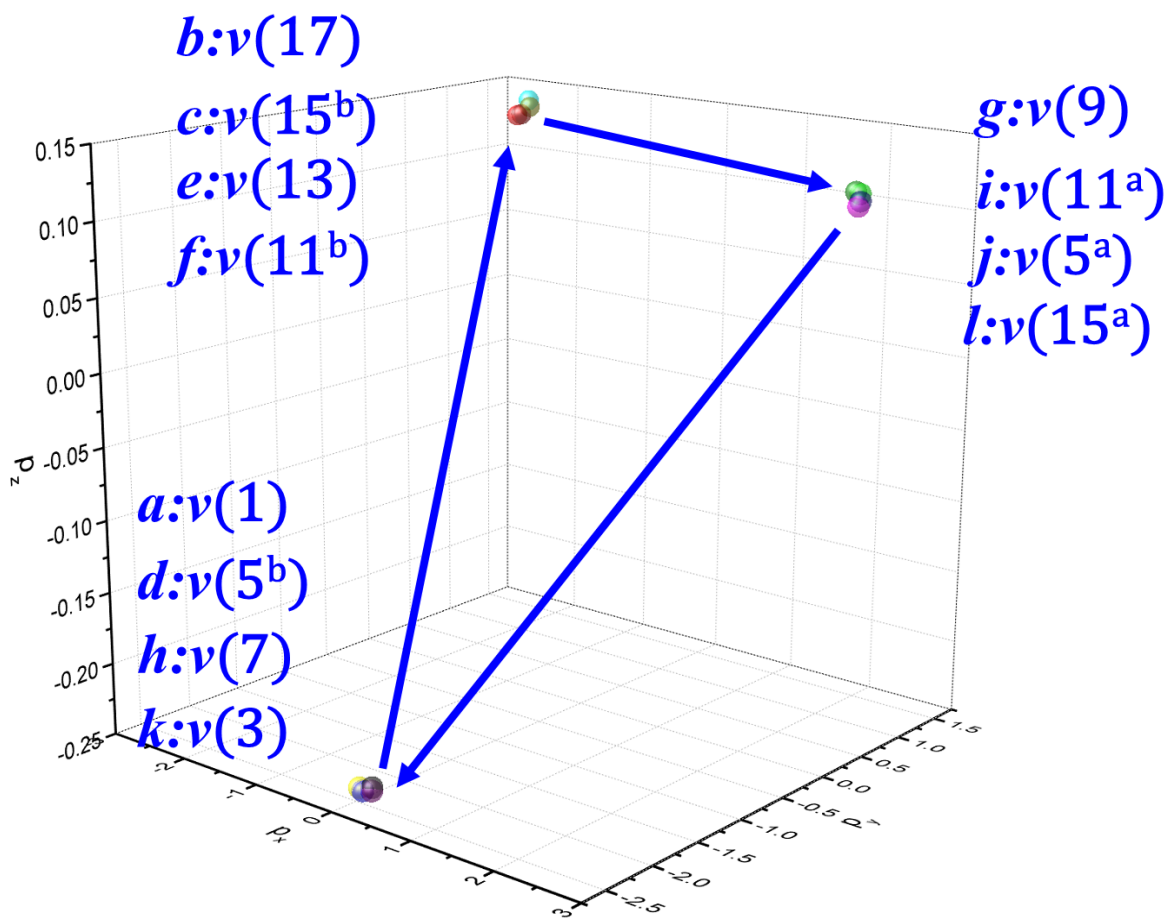

Figure S36. Momentum space for global NNN hopping of up spin antiparticle of chrysene

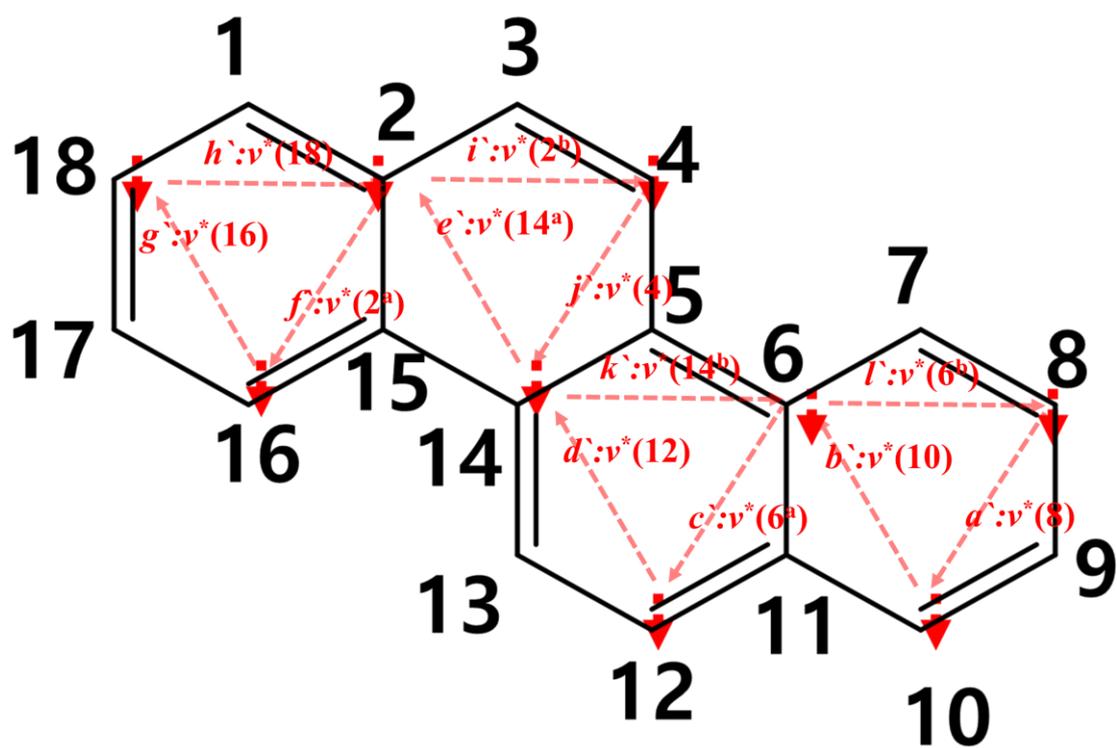

**Figure S37.** Momentum vector for global NNN hopping of down spin antiparticle of phenanthrene

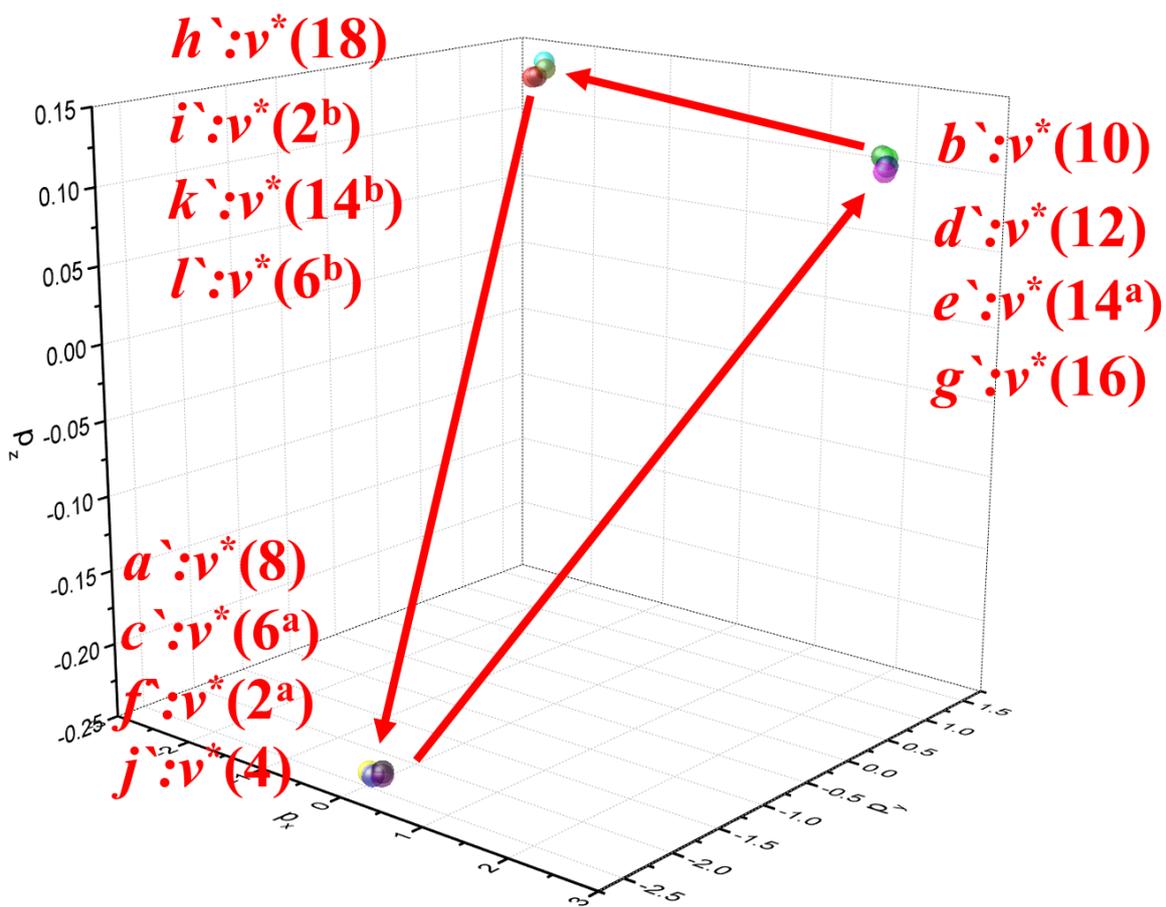

**Figure S38.** Momentum space for global NNN hopping of down spin antiparticle of phenanthrene.

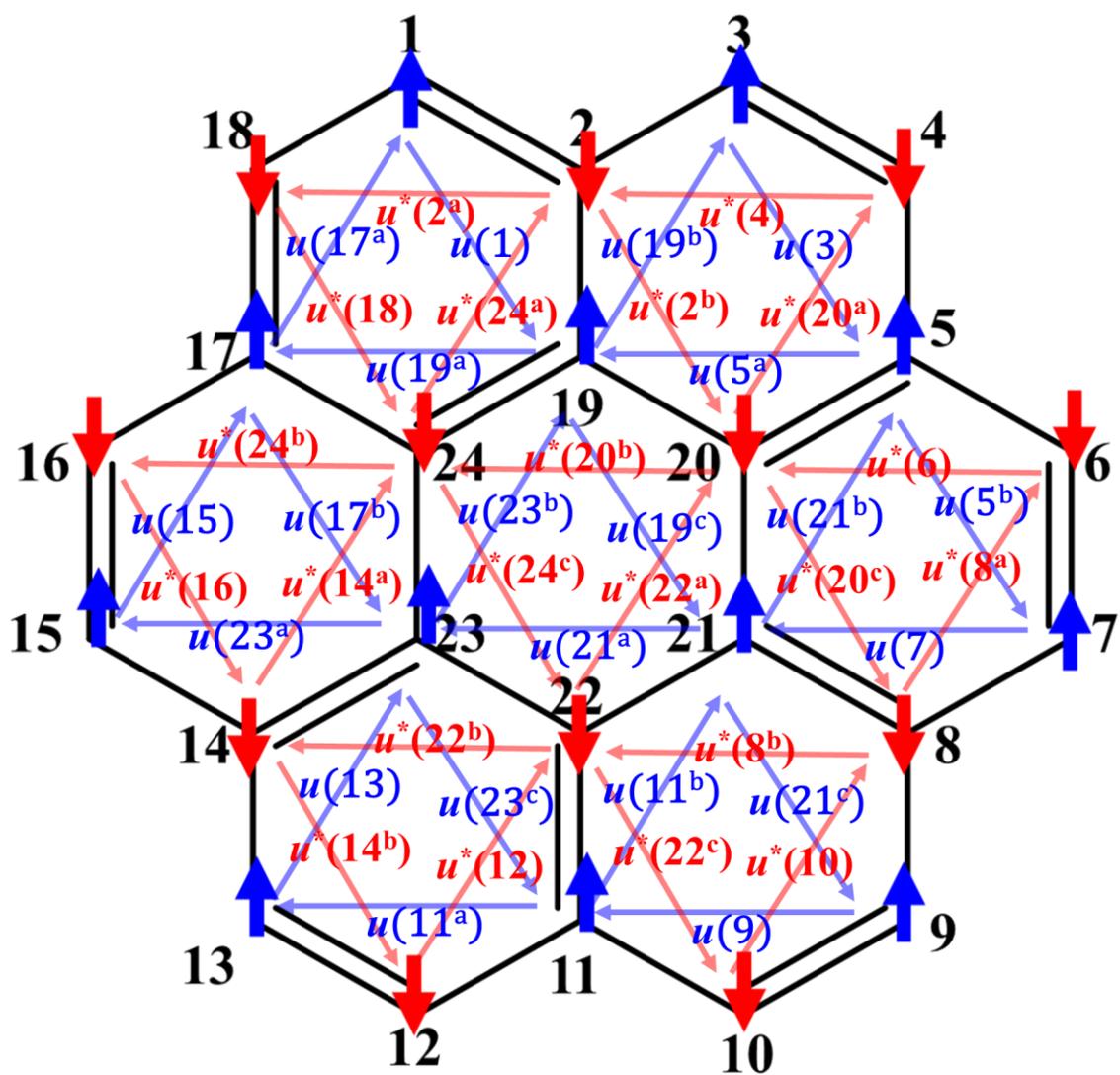

**Figure S39.** Momentum vectors of particle fermions in coronene.

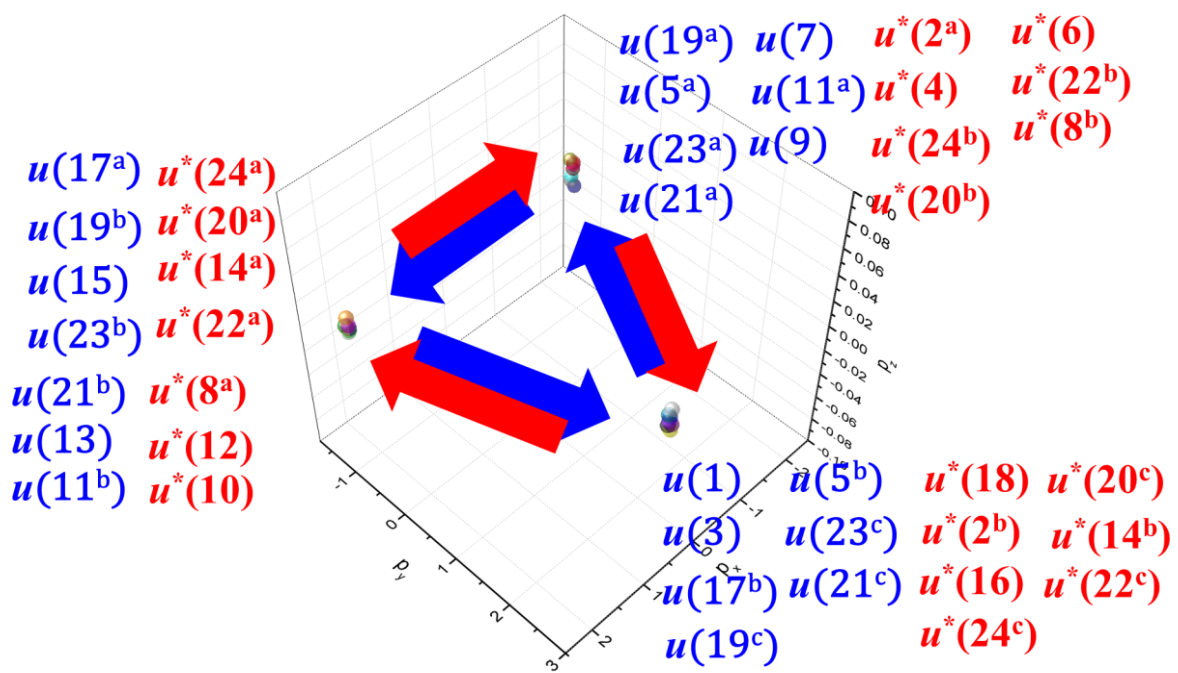

**Figure S40.** Momentum space of particle fermions in coronene.

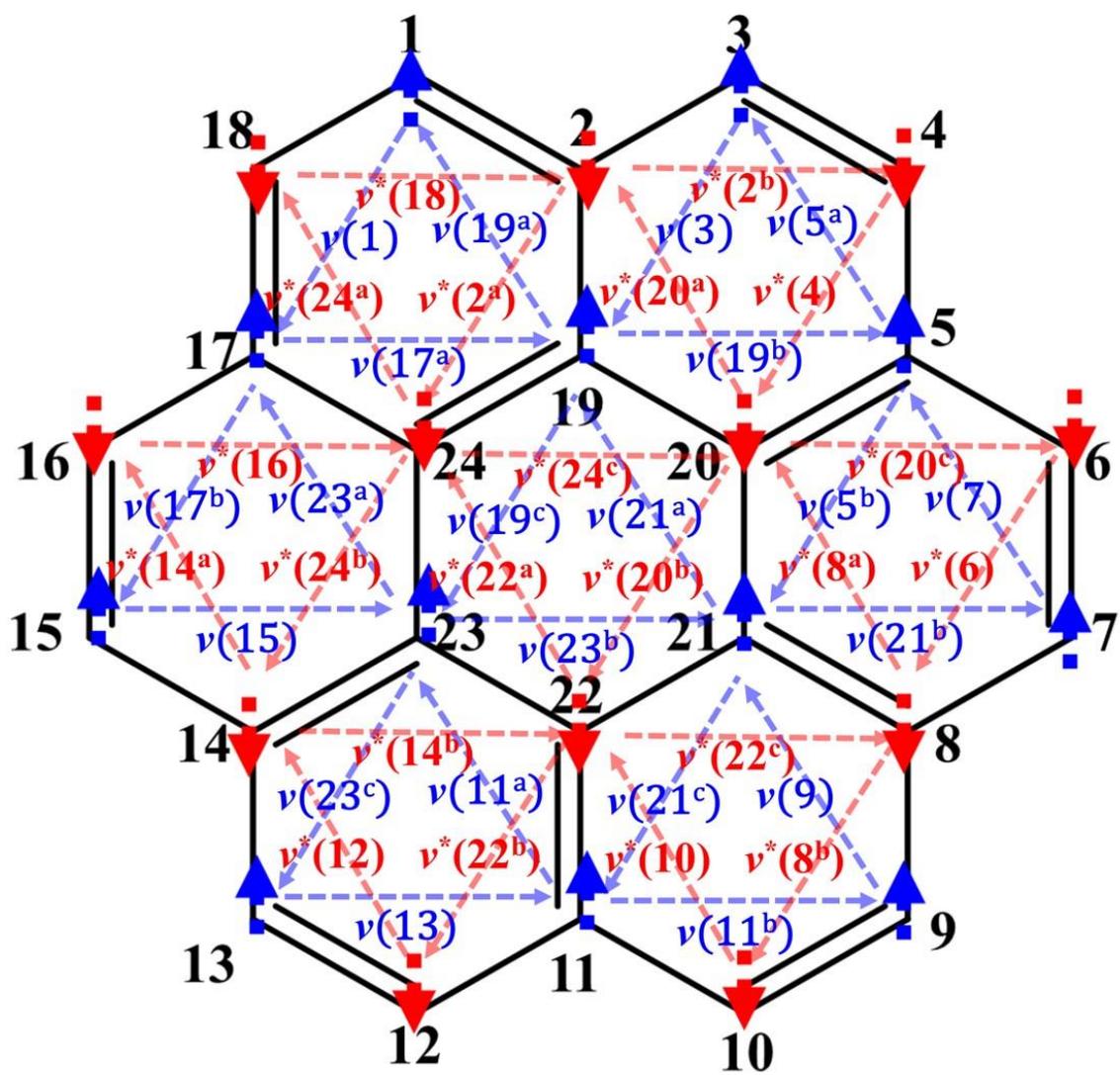

**Figure S41.** Momentum vectors of antiparticle fermions in coronene

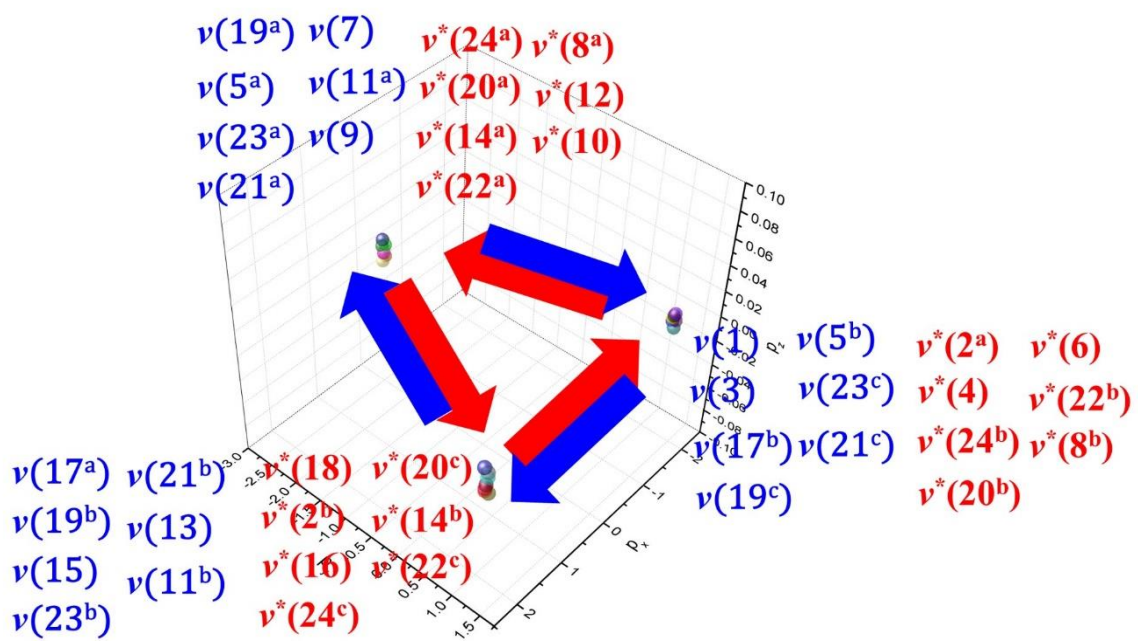

**Figure S42.** Momentum space of antiparticle fermions in coronene

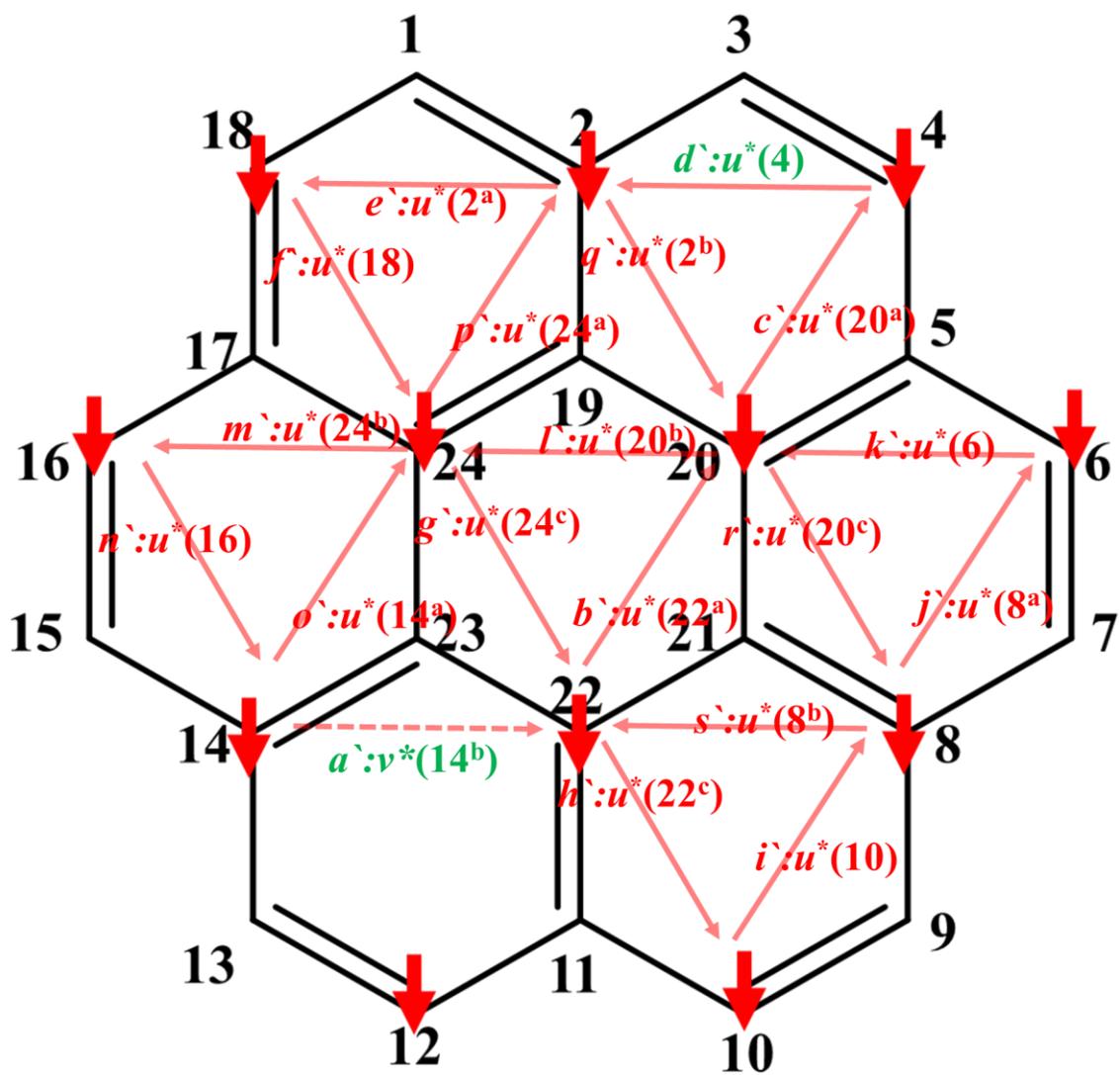

**Figure S43.** Momentum vector for global NNN hopping of down spin particle of coronene.

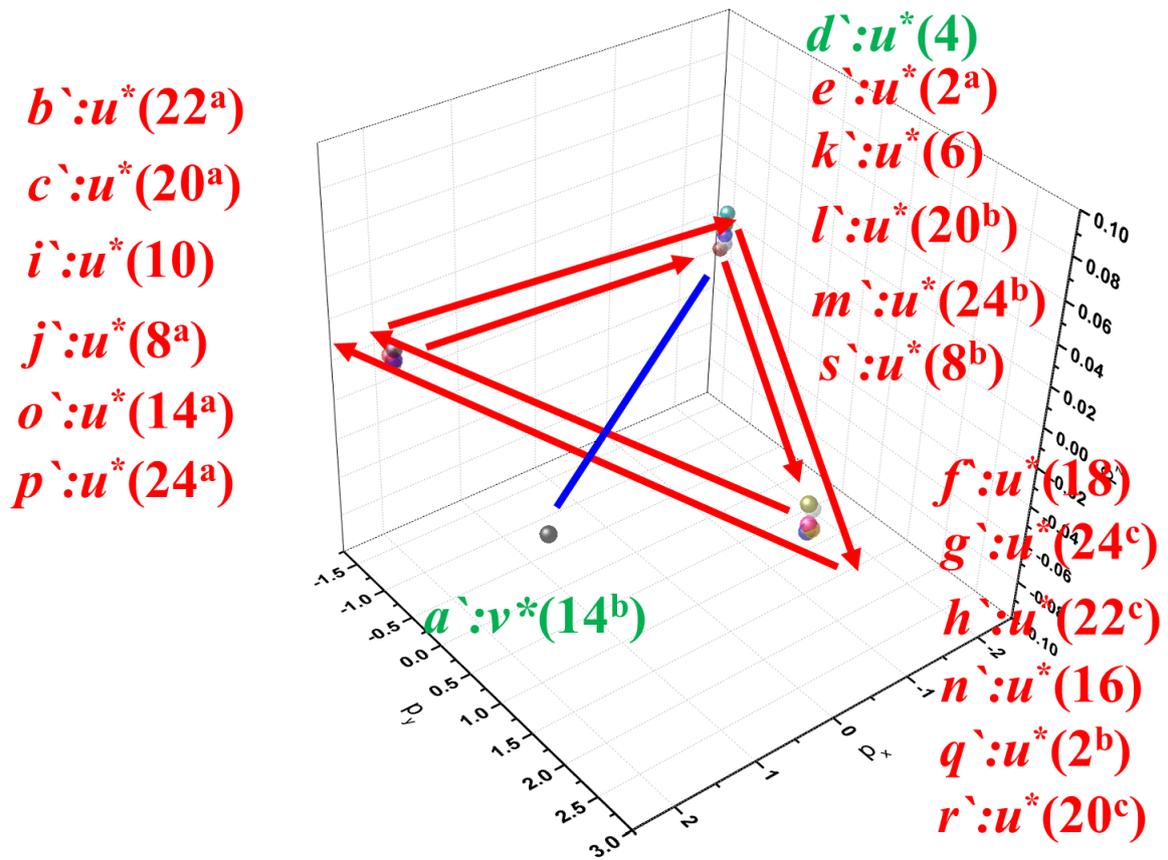

**Figure S44.** Momentum space for global NNN hopping of down spin particle of coronene.

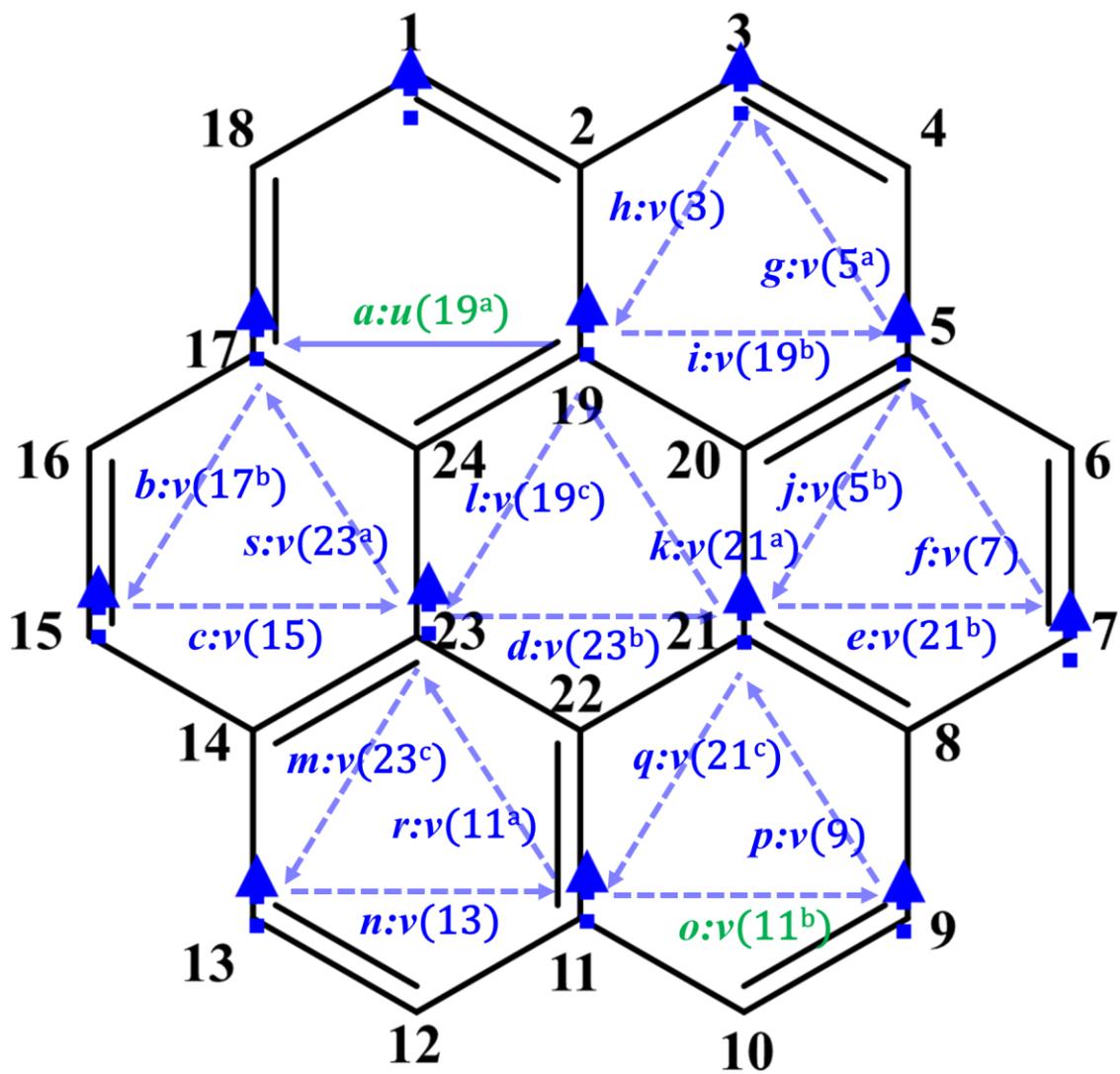

**Figure S45.** Momentum vector for global NNN hopping of up spin antiparticle of coronene.

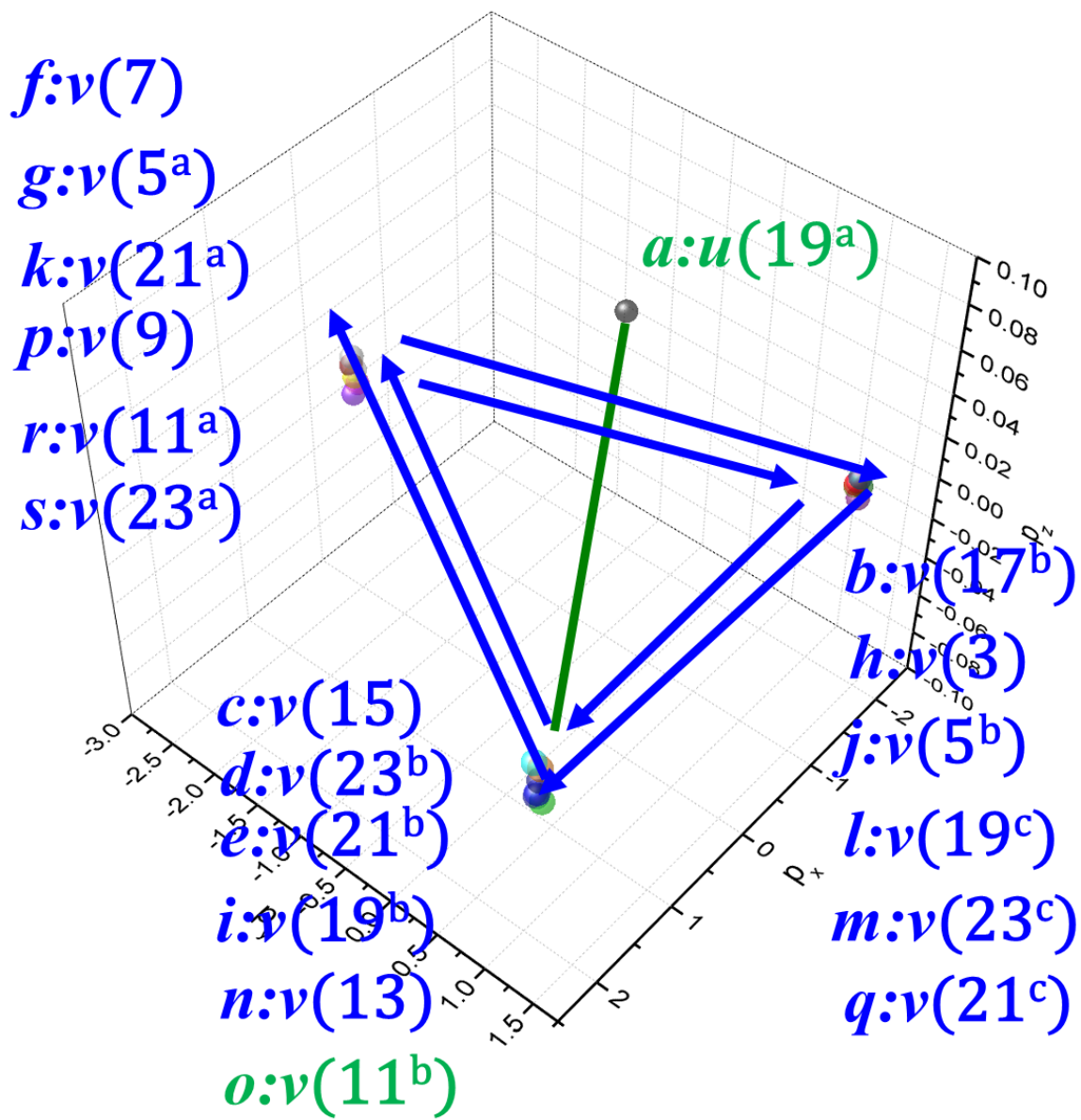

**Figure S46.** Momentum space for global NNN hopping of up spin antiparticle of coronene

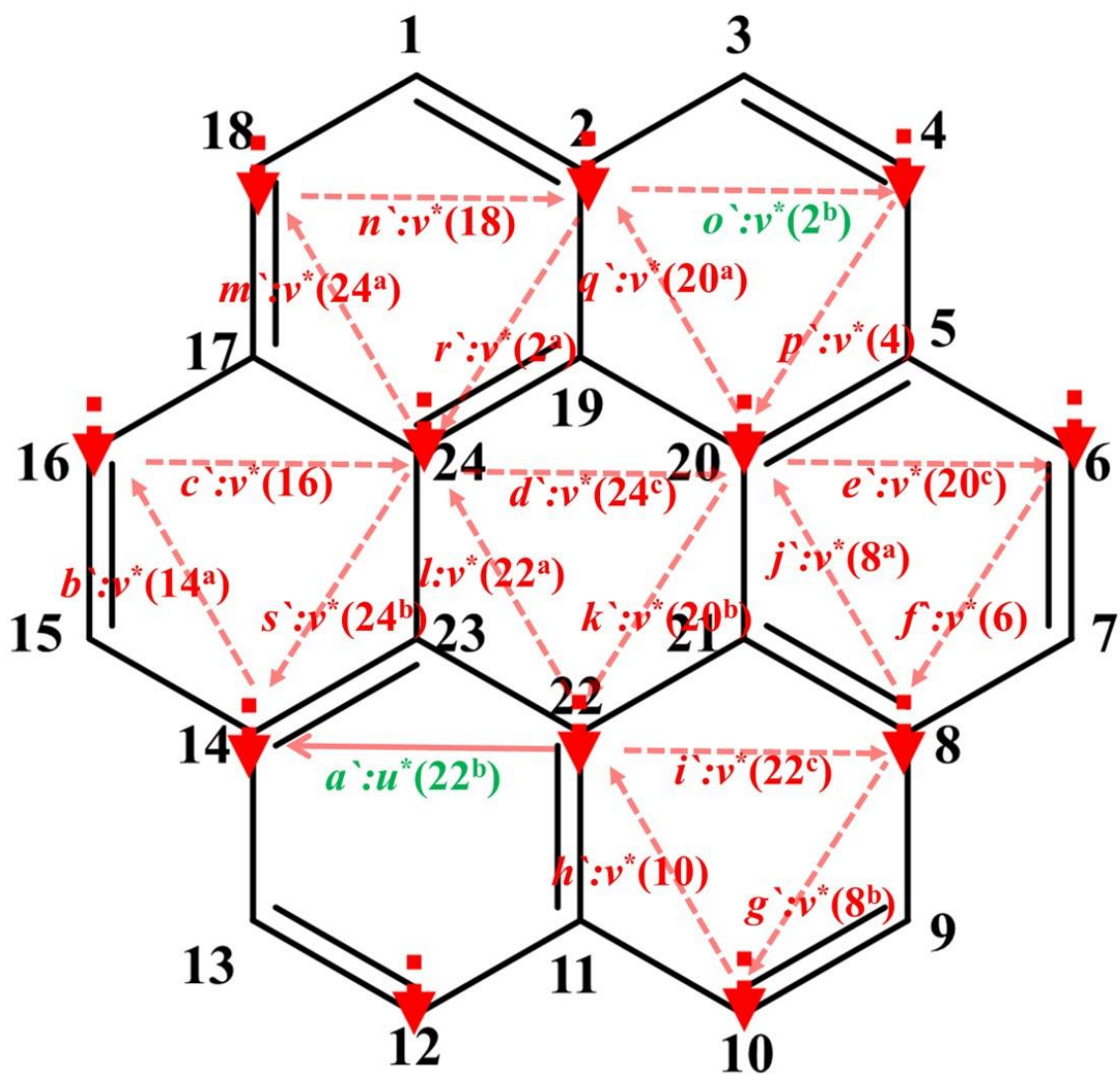

**Figure S47.** Momentum vector for global NNN hopping of down spin antiparticle of coronene

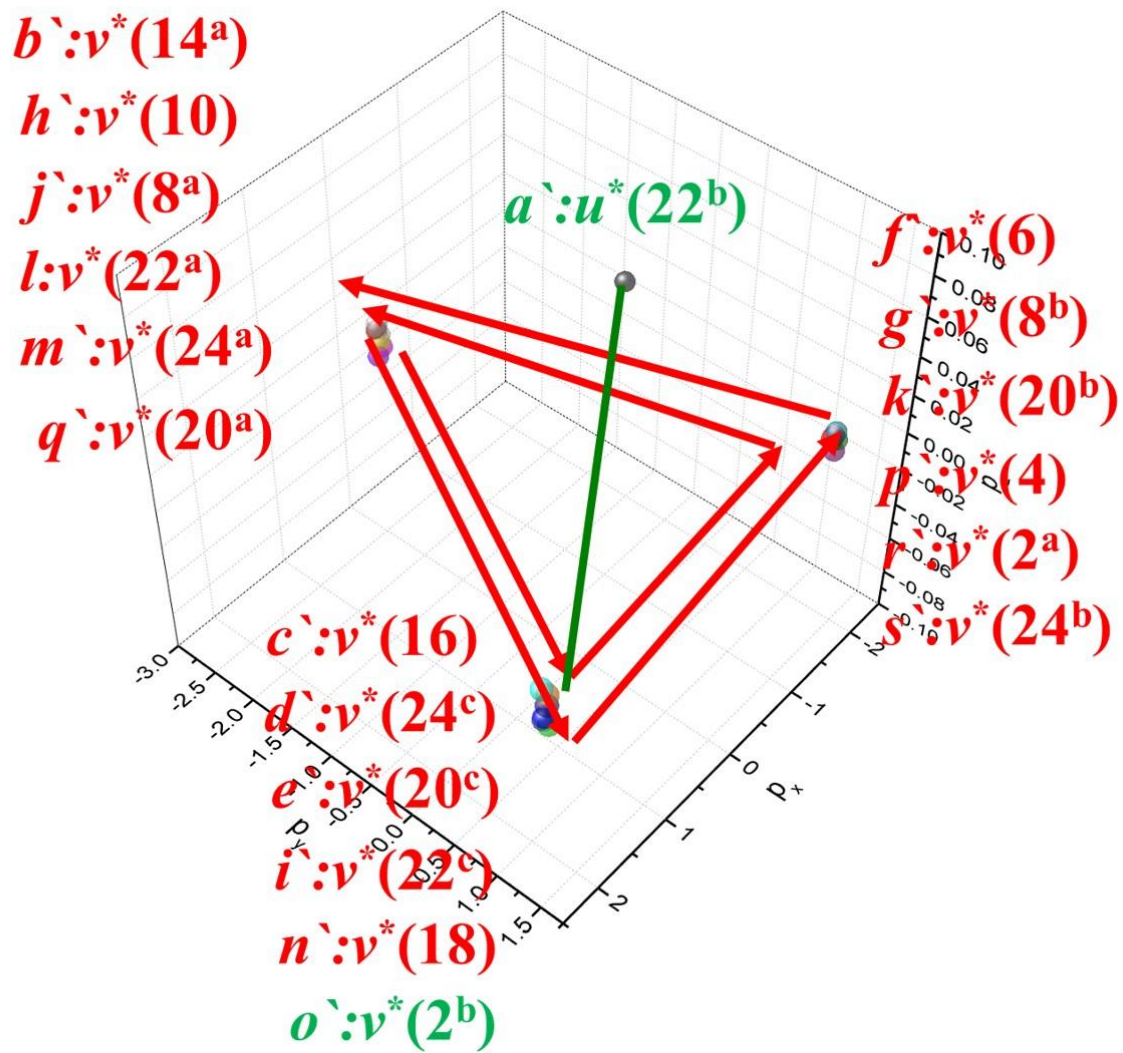

**Figure S48.** Momentum space for global NNN hopping of down spin antiparticle of coronene.

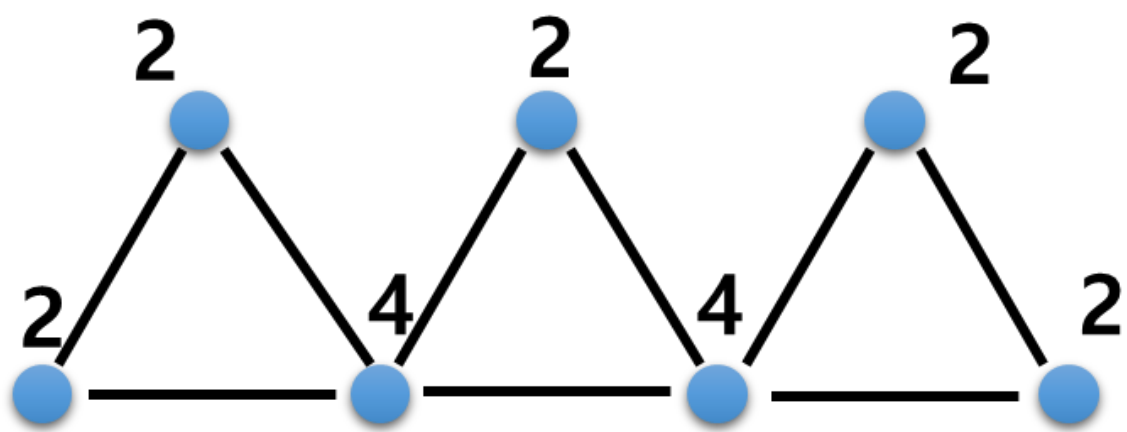

Figure S49. Eulerian graph of anthracene for up spin. Because all vertices in the graph have an even degree, it can make Eulerian circuit.

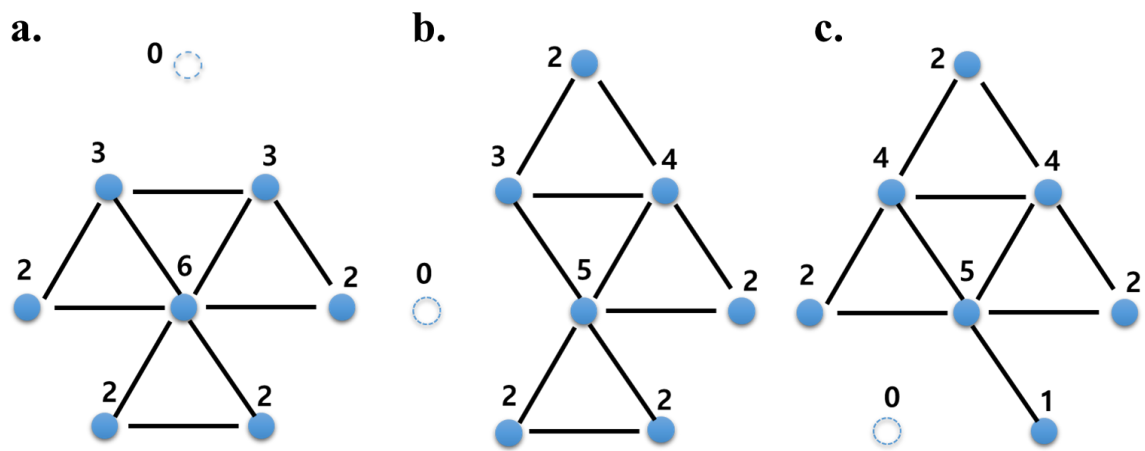

Figure S50. Eulerian graph of pyrene that can make odd number of Kramers doublet and Eulerian trail.